\documentclass[preprint, 10pt]{elsarticle}

\newif\ifInputs
\Inputstrue

\pdfoutput=1
\usepackage{palatino}
\usepackage[usenames]{color}
\usepackage{epsfig}
\usepackage[fleqn,reqno]{amsmath}
\usepackage{amsfonts,amsthm,bm}
\usepackage{amssymb}
\usepackage{mathrsfs}
\usepackage{stmaryrd}
\usepackage{subfigure}
\usepackage[titletoc]{appendix}
\usepackage{tabularx,booktabs}
\usepackage{graphics,caption}
\usepackage{fancybox}
\usepackage[top=1.2in,bottom=1.2in,left=1in, right=1in]{geometry}
\usepackage{array}
\usepackage{multirow}
\usepackage{wrapfig}
\usepackage{algorithmic,algorithm}
\usepackage{paralist}
\usepackage{ifthen}
\usepackage{enumitem}
\usepackage{tikz,pgfplots,filecontents}
\usepackage{float}
\usepackage{mdframed}
\usepackage{todonotes}
\usepackage{xcolor,colortbl}
\usepackage{soul}

\graphicspath{{./figs/}}

\usepackage[pagebackref=false,bookmarks=false]{hyperref} 

\definecolor{dblue}{rgb}{0.03,0.3,0.62}
\definecolor{dorange}{rgb}{1,0.55,0}

\setstcolor{red}

\hypersetup{
  bookmarksnumbered=true,
  bookmarksopen=false,
  hypertexnames=false,     
  breaklinks=true,          
  unicode=false,
  pdffitwindow=true,        
  pdfnewwindow=true,        
  colorlinks=true,         
  linkcolor=dblue,
  anchorcolor=red,
  citecolor=dorange,
  filecolor=magenta,
  urlcolor=dblue,
  pdfstartview = FitH,
  pdfkeywords = {},
  pdfcreator = {LaTeX with hyperref package}
}

\definecolor{sblue}{cmyk}{0.98,0.13,0,0.43} 
\definecolor{sblue}{cmyk}{0.98,0.13,0,0.43} 

\newcommand{\revis}[1]{ #1}

\newcommand{\bigO}{\mathcal{O}}
\newcommand{\Div}{\mathrm{Div}}

\newcommand{\figref}[1]{Figure~\ref{#1}}
\newcommand{\tabref}[1]{Table~\ref{#1}}

\newcommand{\Pen}{\mathrm{Pe}}
\newcommand{\xx}{{\mathbf{x}}}

\newcommand{\yy}{{\mathbf{y}}}

\newcommand{\ff}{{\mathbf{f}}}

\newcommand{\cc}{{\mathbf{c}}}
\newcommand{\uu}{{\mathbf{u}}}
\newcommand{\upi}{\pi}
\newcommand{\UU}{{\mathbf{U}}}

\newcommand{\rr}{{\mathbf{r}}}
\newcommand{\kk}{{\mathbf{k}}}
\newcommand{\nn}{{\mathbf{n}}}

\newcommand{\zzeta}{{\boldsymbol\zeta}}

\newcommand{\ssigma}{{\boldsymbol\sigma}}
\newcommand{\llambda}{{\boldsymbol\lambda}}
\newcommand{\Grad}{{\nabla}}
\newcommand{\Lap}{\Delta}

\newcolumntype{C}{>{\centering\arraybackslash} m{2.5cm}}

\usepackage{lineno}
\usepackage{array}
\newcommand{\mcaption}[2]{\caption{\small \em #1}\label{#2}}
\newcommand{\secref}[1]{Section \ref{#1}}
\newcommand{\appref}[1]{Appendix \ref{#1}}

\begin{document}

\title{Sorting same-size red blood cells in deep deterministic lateral displacement devices}

\author[utMe]{G\"{o}kberk Kabacao\u{g}lu} \ead{gokberk@ices.utexas.edu}
\author[utMe,ut]{George Biros}\ead{gbiros@acm.org}
\address[utMe]{Department of Mechanical Engineering, \\The University 
of Texas at Austin, Austin, TX, 78712, United States}
\address[ut]{Institute for Computational Engineering and Sciences,\\
The University of Texas at Austin, Austin, TX, 78712, United States}

\begin{abstract} 
Microfluidic sorting of deformable particles finds many applications, for example, medical devices for cells. Deterministic lateral displacement (DLD) is 
one of them. Particle sorting via DLD relies only on hydrodynamic forces. For rigid spherical particles, this separation is to a great extend understood and can be attributed to size differences: large particles displace in the lateral direction with respect to the flow while small particles travel in the flow direction with negligible lateral displacement. However, the separation of 
non-spherical deformable particles such as red blood cells (RBCs) is more 
complicated than rigid particles. For example, is it possible to separate deformable particles that have the same size but different mechanical properties?

We study deformability-based sorting of same-size RBCs via DLD using an in-house integral equation 
solver \revis{for vesicle flows in two dimensions}. Our goal is to quantitatively characterize the physical mechanisms
that enable the cell separation. To this end, we systematically investigate 
the effects of the interior fluid viscosity and membrane elasticity of a cell on its behavior. In particular, we consider deep devices in which a cell can show rich dynamics such as 
taking a particular angular orientation depending on its mechanical property. 
We have found out that cells moving with a sufficiently high positive inclination angle with respect to the flow direction displace laterally while those with smaller angles travel with the flow streamlines. Thereby, deformability-based cell sorting is possible. The underlying mechanism here is cell migration due to the cell's positive inclination and the shear gradient. The higher the inclination is, the farther the cell can travel laterally. We also assess the 
efficiency of the technique for dense suspensions. 
It turns out that most of the cells in dense suspensions 
does not displace in the lateral direction no matter what their deformability is. As a result, separating cells using a DLD device becomes harder.

\end{abstract}


\maketitle

\section{Introduction\label{s:intro}}
\begin{figure}
\begin{minipage}{\textwidth}
\centering
\setcounter{subfigure}{0}
\renewcommand*{\thesubfigure}{(a)} 
      \hspace{0cm}\subfigure[Displacement mode]{\scalebox{0.43}{{\includegraphics{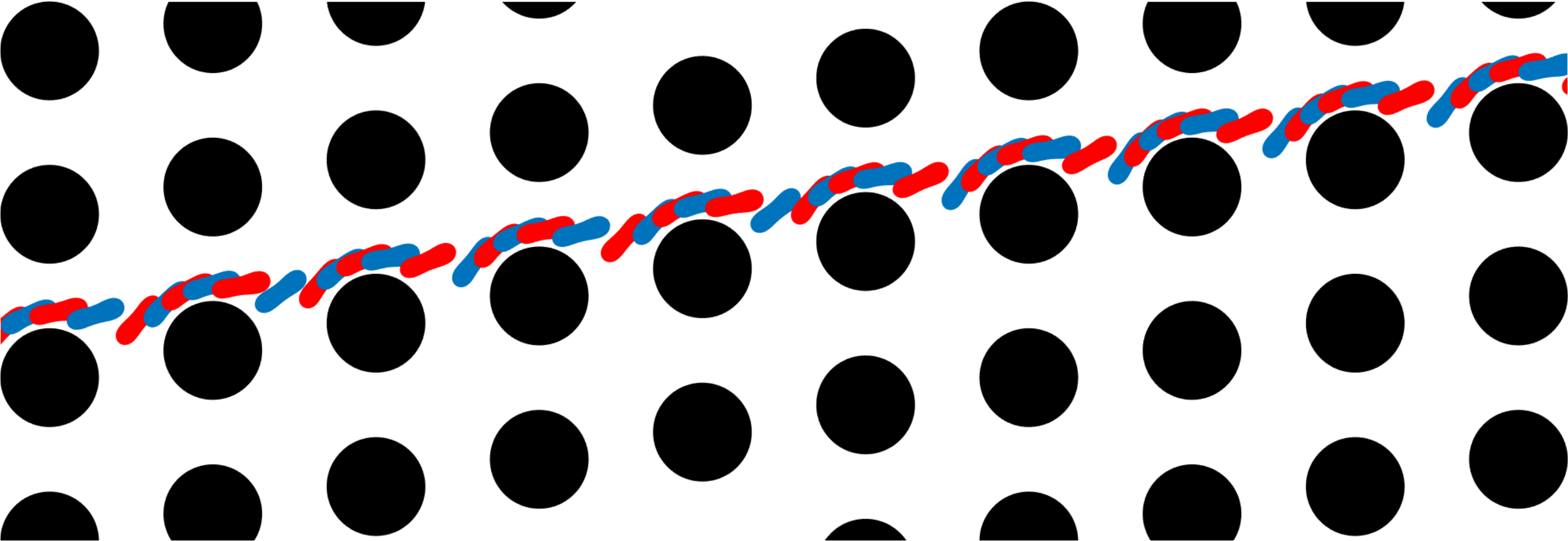}}}
      \label{f:displacementMode}} 
\end{minipage}
\begin{minipage}{\textwidth}
\centering
\setcounter{subfigure}{0}      
\renewcommand*{\thesubfigure}{(b)} 
      \hspace{0cm}\subfigure[Zig-zag mode]{\scalebox{0.43}{{\includegraphics{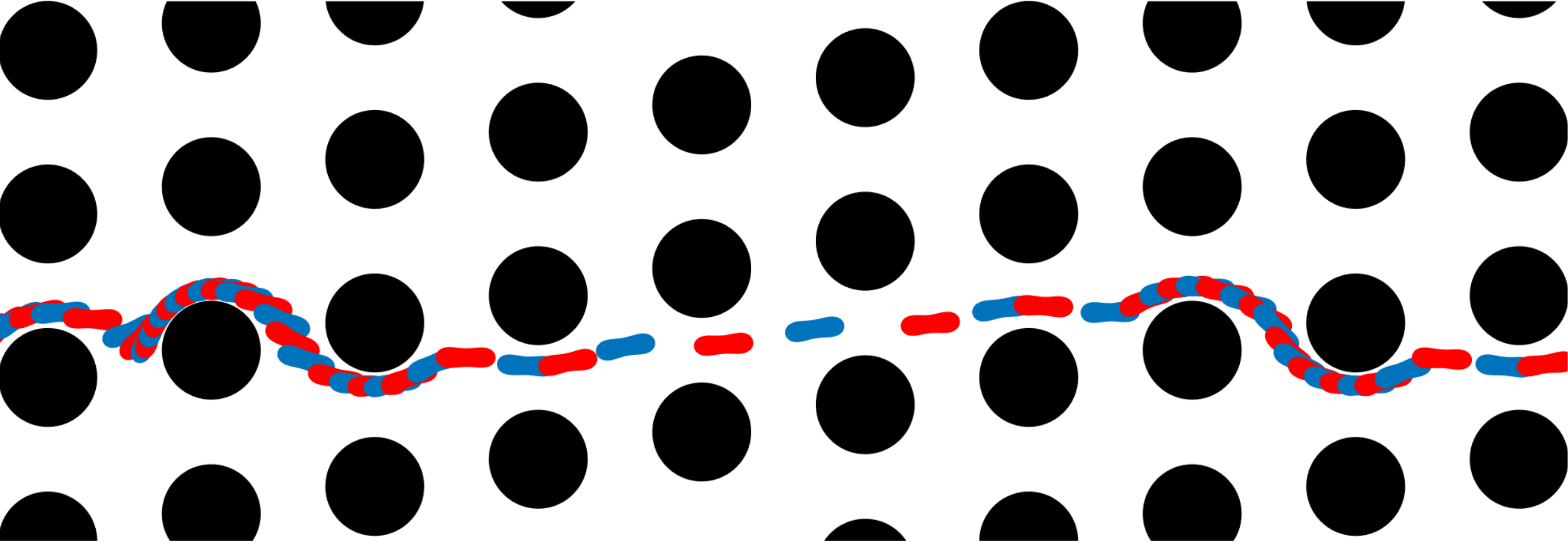}}}
      \label{f:zigzagMode}}
\end{minipage}                    
\mcaption{Transport modes of red blood cells (RBCs) flowing 
in a DLD device. The repeated blue-red RBC shapes indicate different time snapshots from the trajectory of a single cell. The DLD device consists of arrays of
pillars (filled black circles). The flow is from left to
right. Each pillar column is slightly shifted vertically with respect to the previous column. Separation of cells using DLD depends on their orientations,
elasticities and viscosities. Here, the RBC in (a) has a lower
viscosity than the one in (b). The less viscous RBC moves along the inclination of pillars while the more viscous one moves in the flow direction. The former transport mode is called ''{\bf displacement}" and the latter is called ''{\bf zig-zag}".
After multiple interactions between the RBCs and the pillars, the
displacing RBC shows more lateral displacement than the zig-zagging
one. Thereby, they are laterally separated. Here, the tilt angle of the pillar rows is $0.17 \, \mathrm{rad}$.}{f:motionModes}
\end{figure}

Sorting biological cells by their mechanical properties 
(e.g., elasticity, cytoplasmic viscosity) 
is an important process in rapid medical diagnoses and tests using
lab-on-a-chip technology. Therefore, microscale cell separation
techniques have been of great interest. These techniques 
(see~\citet{bhagat-lim-e10, autebert-viovy-e12} for an extensive review of them)
take advantage of various cell fingerprints such as size, shape and
deformability. For example, malaria-infected red blood cells and
metastatic cancer cells can be differentiated based on their
deformabilities as the malaria-infected cells are stiffer
~\citep{suresh-seufferlein-e05} and the metastatic cancer cells
are softer than their healthy counterparts~\citep{guck-mitchell-e05}. The
microscale cell separation techniques are divided into two categories:
active and passive. The former uses an external force field, e.g.,
electrophoresis is an active method using an electric field to
separate cells of different sizes and charges. Passive separation relies mainly on the device geometry and on hydrodynamic interactions between particles
and the device. Thus, passive techniques are cheap and 
readily available. Deterministic lateral displacement (DLD) is a passive
particle separation technique introduced by~\citet{huang-sturm-e04} to
separate rigid particles from their dilute suspensions based on their
sizes. 

A DLD device consists of arrays of pillars
(see~\figref{f:motionModes}). The pillar grid forms a lattice but the lattice vectors are not orthogonal. That is, the pillars are vertically aligned but their ''horizontal" or ''diagonal" alignment direction is at an angle with the $x$-axis, which is also the flow direction axis. This arrangement
determines a critical particle size\citep{davis-austin-e06}. Particles larger than the critical size move along the
direction defined by the pillars (see~\figref{f:displacementMode})
while those smaller than the critical size move with the flow
(see~\figref{f:zigzagMode}). The former transport mode is called
''{\bf displacement}" and the latter is called ''{\bf zig-zag}". After several particle-pillar interactions, the displacing particle
is separated laterally from the zig-zagging one because the latter has
almost zero net lateral displacement. Notice that~\figref{f:motionModes} shows snapshots from simulations of cells, however, these trajectories are characteristic of the transport modes and the same for the rigid particles as well. Analytical DLD theory has been developed for dilute suspensions of rigid spherical particles under the assumption that
the particle-pillar interactions dominate the particle-particle
interactions in determining the particle trajectories. Critical
particle sizes are computed and the devices are designed based on this
assumption~\citep{davis-austin-e06,inglis-sturm-e06,davisPhd14}. However, dilution of
suspensions requires a pre-treatment of the sample, which is time consuming and expensive and high volume fractions of suspensions are needed
for high throughput. Therefore, the performance
of DLD for dense suspensions is also of interest. Additionally, dynamics of non-spherical and deformable particles such as cells flowing in DLD devices needs to be investigated in order to design DLD devices for sorting those particles. So, although the DLD technique has been frequently used and promising, several questions regarding their performance remain open. Given the wide spectrum of possible applications, here we focus on sorting of human red blood cells.

\begin{figure}
\centering
\includegraphics[scale = 0.4]{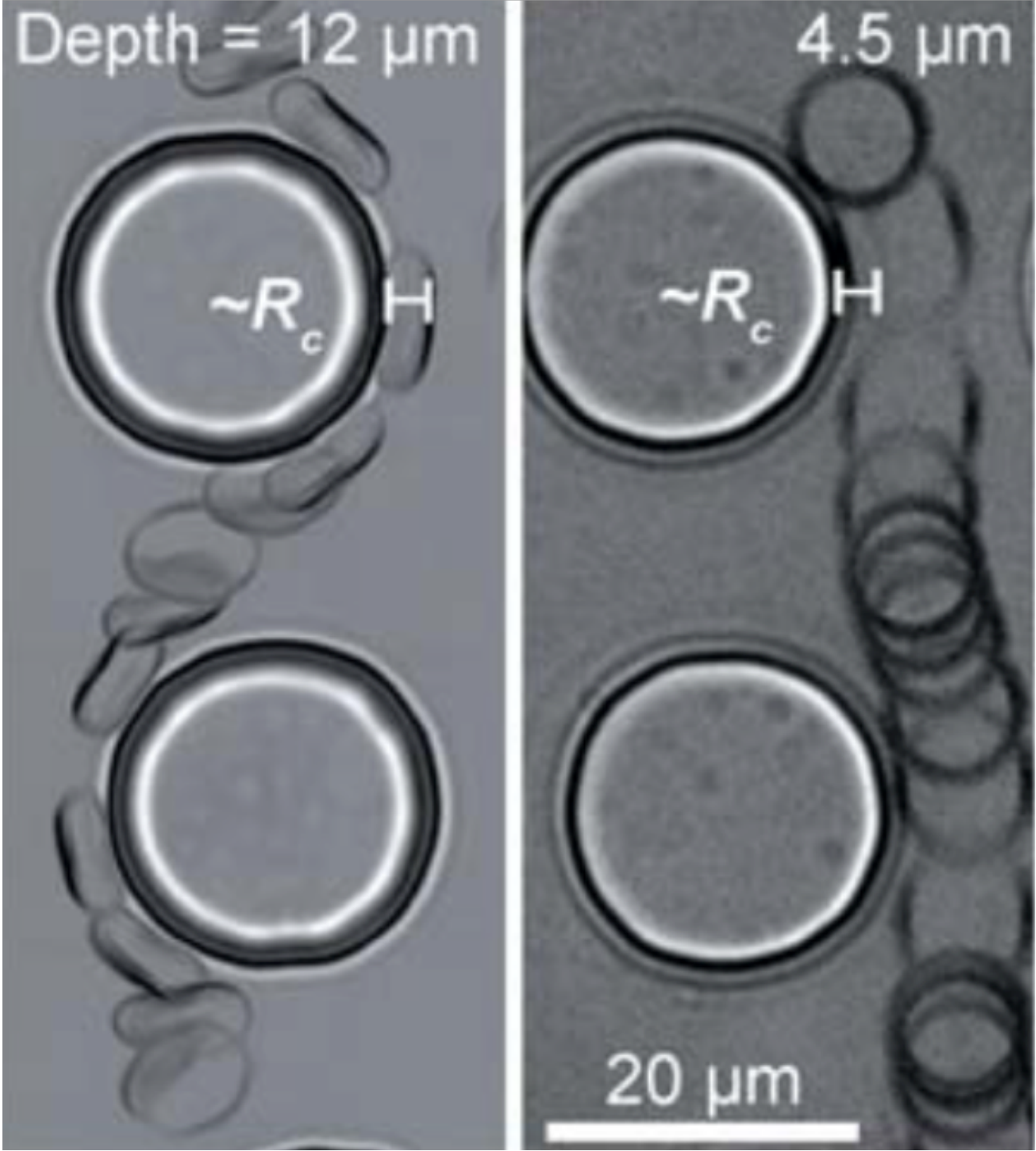}
\mcaption{RBC orientations in a deep (on the left) and a shallow (on the right)
DLD devices (the images show top view and are taken from~\citet{beech-tegenfeldt-e12}). The 
heights of the pillars are $H = 12\mu$m in the deep device and $H = 4.5\mu$m 
in the shallow device. These are superimposed
images of RBCs flowing in actual DLD devices. The fluid flows from
top to bottom. In the deep
device the RBC is orientated in a way that its thickness becomes its
effective size. In the shallow device the confining walls in the out-of-plane direction pushes the
RBC to move parallel to the walls. Therefore what we see is a disc and
the effective size is its diameter. The deep DLD devices are preferable
to shallow ones in practice since they reduce the risk of
clogging and provide higher throughput. We only consider deep devices in our study.}{f:deepShallow}
\end{figure}
Human blood consists of plasma and mainly white and red blood cells and platelets. White blood cells (WBCs) are mostly spherical with a
diameter ranging between $5\mu$m - $20\mu$m, red blood cells (RBCs)
are biconcave with a diameter of $8\mu$m and a thickness of 
$3\mu$m and platelets are nearly rigid discoids (when they are not activated) with a diameter between $1\mu$m - $3\mu$m~\citep{popel-johnson05}.
DLD has been used for fractionation of these components of human blood based
on their sizes~\citep{davis-austin-e06} and also used for separation
of WBCs~\citep{davis-austin-e06}, RBCs~\citep{zeming-zhang-e13}, parasites~\citep{holm-tegenfeldt-e11}
and circulating tumor cells~\citep{loutherback-sturm-e12,karabacak-toner-e14}. 
Diseases such as sickle cell
anemia~\citep{barabino10}, diabetes~\citep{buys-pretorius-e13}, and malaria are responsible for changes in RBCs' deformability
by altering both the cytoplasmic viscosity and the membrane's
viscoelastic properties. Therefore, deformability-based sorting of RBCs could potentially identify and separate abnormal cells from blood. 
These considerations motivate our study of DLD devices. The parameters we used in this study are membrane elasticity and cytoplasmic 
viscosity of RBCs, which determine their deformability. 

A DLD device for RBC separation can be either \emph{shallow} or
\emph{deep}~\figref{f:deepShallow}. A shallow device has short pillars
with a height less than the RBC diameter ($H \approx 4-5\mu$m) while a
deep one has tall pillars with a height much greater than the RBC
diameter ($H \gg D_{\mathrm{RBC}}$). RBCs orient differently in those
devices. A shallow device pushes cells parallel to the confining walls in the out-of-plane direction and results in an effective size (when they are not deformed) equal to the cell diameter. In a deep device they orient themselves in such a way that their effective size becomes their smallest size, i.e., their thickness. In shallow devices the separation of cells is similar to that of rigid spherical
particles. While a suspension of stiff and soft cells flows through a shallow device, soft cells deform more than stiff cells and hence their effective size becomes smaller than that of the stiff cells. This results in zig-zagging soft cells and displacing stiff cells, and hence separation. However, throughput is limited in shallow devices. In contrast, deep devices allow more cells along the pillars and hence higher throughput. Cells can show richer dynamics in deep devices than shallow devices since deep devices do not confine cells as much as shallow ones do. Therefore, the separation in deep devices does not occur by the same means as in shallow devices. It depends on complex dynamics of cells such as moving with a stationary angular orientation (i.e., tank-treading) or varying angular orientation (i.e.,tumbling)~\cite{henry-gompper-e16}. RBCs show these complex dynamics in shear and Poiseuille flows depending on the cytoplasmic viscosity and membrane elasticity~\citep{misbah06,messlinger-gompper-e09,podgorski-misbah-e11,misbah12,fedosov-gompper-e14}. Cell dynamics in these fundamental flows have been studied and well-understood. However, the DLD flow is more complicated than those flows. How do such complex behaviors appear in deep DLD devices? How does sorting depend on them? How can we quantify them and explain sorting? These are the main questions we
aim to address in this article.

\subsection{Contributions}
To the best of our knowledge, this is one of the first studies
investigating the effects of the complex RBC dynamics in deep DLD devices. We show that deformability-based sorting of RBCs is possible. We list our contributions below. 
\begin{itemize}
\item\ We investigate the cell dynamics in DLD flow, i.e., we study cell's inclination angle and lateral velocity in Sections \ref{s:deformSep} and~\ref{s:vcSep}. 

\item\ We compare these cell-in-DLD dynamics with simple flows such as a free shear and a confined Poiseuille flows in Sections~\ref{s:freeShear} and~\ref{s:confPois}. 

\item\ In order to quantify migration, we compute a new quantity in~\secref{s:lift}, which we call the ''pseudo-lift", that turns out to be an indicative measure of migration.

\item\ Lastly, we present phase diagrams for the transport modes in~\secref{s:phase} and investigate the separation in dense RBC suspensions in~\secref{s:dense} for different viscosity contrasts, capillary numbers and device configurations. The efficiency of DLD for dense suspensions in deep devices has not been studied experimentally and numerically. However, it is essential to investigate the dense suspension regime because the deep devices with dense suspensions provide higher throughput than the shallow devices with dilute suspensions.

\end{itemize}

{\em Methodology.}
We use a standard 2D mechanical model for RBCs~\citep{keller-skalak82,noguchi-gompper05,misbah12}. We model an RBC as a vesicle with a reduced area of 0.65 which resists deformation due to bending and tension. It is impermeable to flow and locally inextensible. Since RBCs and vesicles share similar dynamical properties and the differences are minor in 2D, we use the vesicle model for RBCs. We assume that the fluids in the interior and exterior of the vesicle are Newtonian. Since the Reynolds number is $\bigO(10^{-3})$(see \citet{mcgrath-bridle-e14} for a
summary of the experiments reported in the literature), we use a standard quasi-static Stokes approximation scheme to model the flow~\citep{danker-misbah-e09,freund-zhao10,zhao-shaqfeh11a,freund-orescanin11,zhao-shaqfeh13b,zhao-shaqfeh13a}. We formulate the problem as a set of
integro-differential equations using the potential theory~\citep{quaife-biros14,quaife-biros15,quaife-biros16,kabacaoglu-biros-e17a,kabacaoglu-biros-e17b}. Our DLD
model has pillars with a circular cross-section only. Our 2D model can represent deep DLD devices not the shallow ones because RBCs with a reduced area of 0.65 in the 2D model have the same orientation with RBCs in real deep devices (see Figures~\ref{f:motionModes} and~\ref{f:deepShallow}). Additionally, our 2D model does not include the wall effects and the cell-cell interactions in the out-of-plane direction which are negligible in deep devices but important in shallow ones. Unlike other
numerical studies (discussed in~\secref{s:relatedWork}) we include
multiple rows and columns of pillars in our simulation domain (this
results in $\bigO(100)$ pillars). We consider only one free parameter
for the DLD device: the row-shift fraction $\epsilon$ (i.e., the angular orientation of the pillar orientation). We quantify the RBC's elasticity and viscosity with two dimensionless
numbers: the capillary number $C_a$ and viscosity contrast $\nu$, 
respectively. First, we perform several simulations of RBCs with no viscosity
contrast by changing the capillary number $C_a$ in a DLD with the
row-shift fraction $\epsilon = 0.1667$. Second, in the same device we vary the viscosity contrast
$\nu$ and fix the capillary
number to $C_a = 3.41$, which corresponds to a value for the flow of a 
healthy RBC through DLD with an average velocity of $1$mm/s. This flow speed is in the range [1$\mu$m/s, 10mm/s] used in experiments on sorting RBCs depending on their deformability~\citep{mcgrath-bridle-e14}. Additionally, it is not so high that cells do not deform significantly and hence we can observe the effects of viscosity contrast on cell dynamics. Last, we vary the viscosity contrast and capillary number for fixed row-shift fractions $\epsilon = 0.1$ and $\epsilon = 0.1667$. Then we map out the parameter space
for the transport modes as a function of $C_a$, $\nu$, $\epsilon$. We also perform simulations of dense suspensions in DLD and quantify the efficiency of the technique for dense suspensions.

Summary of our findings:
\begin{itemize}
\item\ \emph{Relation to cell migration:} Cell migration is a phenomenon observed in a shear and confined Poiseuille flows. That is, cells having positive inclination with respect to the flow direction migrate towards low shear rate regions due to the parabolic velocity profile. In our numerical studies, we observe that more deformable cells have higher positive inclination angles with respect to the flow direction than the less deformable ones (See \secref{s:confPois}). Since the velocity profile near the pillars is parabolic, the more deformable cells migrate further away from the pillars than the less deformable ones. Thus, the more deformable cells displace while the less deformable ones zig-zag.

\item\ \emph{Effects of complex geometry:} Although we have discovered that cell dynamics in DLD gaps are similar to those in confined Poiseuille flows (See \secref{s:confPois}), the dynamics in the whole device is more complicated than the channel flow because a cell periodically moves from a confined gap to a less confined region between gaps\footnote{By ''DLD" gap, we refer to the flow space between two vertically aligned pillars.}. So it is not possible to estimate the cell dynamics in DLD from a simpler flow such as a channel flow.

\item\ \emph{Quantification of migration:} We have computed cell vertical migration velocity in DLD and demonstrated that they correlate well with the inclination angles (See \secref{s:confPois}). 

\item\ \emph{Pseudo-lift:} By computing the so-called pseudo-lift (a measure of the alignment of local forces to the migration direction) acting for several cells, we have found $5\times$ stronger pseudo-lift on cells with high inclination angles compared to cells with smaller inclination angles which are under either a 
negative or a weak pseudo-lift (See \secref{s:lift}). 

\item\ \emph{Dense suspensions:} Finally, when we assessed the DLD efficiency for dense RBC suspensions, we have observed that most cells zig-zag in dense suspensions no matter what their capillary numbers and viscosity contrasts are. So it is difficult to separate small rigid particles or stiff cells from dense suspensions because those would zig-zag, too. This
result agrees with the numerical~\citep{vernekar-kruger15} and
experimental~\citep{inglis-nordon-e11} studies which use shallow devices (See \secref{s:dense}). 

\end{itemize}

\subsection{Limitations}

Our simulations are in two dimensions. We have
opted to use two-dimensional simulations because three-dimensional simulations of a DLD
device including multiple pillars can be quite
expensive for this purpose~\citep{rossinelli-petros-e15}. Two-dimensional studies have
been proven to be predictive and match actual experiments in many
regimes (e.g motion of RBCs in microchannels~\citep{kaoui-misbah-e11,fedosov-gompper-e14},
 margination of white blood cells in blood
flow~\citep{freund07,zhao-shaqfeh11b,muller-gompper-e14} and sorting
of rigid particles using deterministic lateral displacement
technique~\citep{zhang-fedosov-e15}). The other limitation is
that we do not consider changes in the resting size and shape of an RBC. When an RBC
becomes diseased, not only its elasticity and viscosity change but its
resting size does so. Finally, we do not investigate the effects of the lateral confinement in the gaps on cell sorting. In our study, the confinement in the gaps does not allow a cell to tumble and leads it to move always with a positive inclination angle. In the case of a weaker confinement, the cell can tumble for high viscosity contrasts (i.e., for less deformable cases). We expect tumbling to increase the cell's effective size in those cases and hence result in displacing cells. For example, ~\citet{zeming-zhang-e13,ranjan-zhang-e14} observed that pillars with protrusions cause cells to tumble and tumbling cells start displacing for the same reason.

\subsection{Related work}\label{s:relatedWork}

The study in~\citep{henry-gompper-e16} is very similar to us and the findings are consistent. They study both deep and shallow devices by combining simulations (3D smoothed dissipative particle dynamics) and experiments. They only consider three viscosity contrast values (0.25, 1, 5), whereas we study a much wider range of viscosity contrasts (up to 100). Also, we vary the capillary number (they do not). Going beyond the work in~\citet{henry-gompper-e16}, we conduct a systematic study and map out the parameter space for the transport modes of cells. We investigate the cell dynamics in DLD in detail and compare those dynamics with simpler flows such as confined Poiseuille flow. We claim that cell migration is responsible for cell separation in DLD and quantify migration with migration velocity and pseudo-lift. 
\citet{henry-gompper-e16} shows our 2D simulations can capture the
actual experiments and supports our findings and rationale.

In addition to~\citet{henry-gompper-e16} mentioned above, there have 
been a few numerical studies investigating the separation
of deformable particles using DLD: 

\begin{itemize}
\item\ \citet{quek-chiam-e11} and~\citet{ye-yu-e14} performed two-dimensional simulations of spherical deformable particles with no viscosity contrast in 
order to explain the effects of particle deformability on the separation,

\item\ \citet{kruger-coveney-e14} systematically investigated the effects of the
capillary number $C_a$ on the separation of RBCs with a constant
viscosity contrast of $\nu = 5$ in shallow devices using 
three-dimensional simulations,

\item\ \citet{zhang-fedosov-e15} studied circular, square, diamond and triangular pillar shapes for the separation of rigid particles and RBCs using two-dimensional simulations. 

\end{itemize}
Overall, none of these studies have systematically investigated the separation
of RBCs as a function of the capillary number and viscosity contrast
in deep devices.

Regarding numerical methods, all the studies above used either the immersed boundary method~\citep
{quek-chiam-e11,kruger-coveney-e14,vernekar-kruger15}, 
lattice-Boltzmann method~\citep{kruger-coveney-e14,vernekar-kruger15},
fictitious domain method~\citep{ye-yu-e14} or dissipative particle
dynamics~\citep{zhang-fedosov-e15,henry-gompper-e16}. 
Here, we use our in-house algorithms based on
the boundary integral equation formulation for Stokesian particulate flows~\citep{kabacaoglu-biros-e17b}. 
We refer the reader 
to~\citep{shravan-biros-e09,rahimian-biros-e10,quaife-biros14,quaife-biros15,quaife-biros16} 
for the details of our
scheme and the review of the numerical methods for Stokesian
particulate flows. Additionally, most numerical studies
of DLD mentioned in the previous paragraph reduced the simulation domain to a single pillar and
used periodic boundary conditions. Since actual DLD devices have
walls that result in zero net lateral flow, an artificial force
in the lateral direction needs to be introduced to mimic the wall
effects in the periodic models~\citep{davino13}. Our model contains multiple pillars in the
lateral and flow directions, and also the top and bottom walls.
Therefore, we do not need to add any force mimicking the wall effects.
The only parameters in our scheme are the physical parameters (device geometry, number of cells, viscosity contrast and capillary number) as well as the spatial discretization size. We use an adaptive, semi-implicit time-stepping scheme. No other (non-physical) parameters are necessary. We only have five free parameters in our model (other than the shape of the device): the time-step error tolerance, the points per cell, the discretization size of the rigid walls, the viscosity contrast and the capillary number. We have also performed a convergence study to verify our method in the DLD setting in~\secref{a:appendix1}.

There is only one
numerical study~\citep{vernekar-kruger15} where simulations of dense RBC suspensions with different volume fractions
were performed in shallow devices. The authors found out that the
displacement mode breaks down as the volume fraction increases and
most of the RBCs zig-zags independent of the capillary number. We also 
studied this by performing simulations of dense RBC suspensions with two different
capillary numbers and volume fractions in deep devices. We reached
the same conclusions. In addition, we considered dense RBC
suspensions with different viscosity contrasts. Our results show that
again the displacement mode breaks down.

There are efforts to develop novel pillar shapes in order to improve
the DLD performance. \citet{zeming-zhang-e13}
presented I-shaped pillars for RBC separations and showed that the
I-shape increases the RBC's effective sizes by inducing rotational
motions. \citet{ranjan-zhang-e14} conducted experiments with various
pillar shapes to investigate efficient separation of spherical and
non-spherical bio-particles. They considered pillars with protrusions
and grooves to induce particle rotation and maintain the rotational
movement in the flow direction. We observed that there is a need for
novel shapes to prevent breakdown of the displacement mode for the
dense RBC suspensions. So that small particles (which would zig-zag)
can effectively be separated from dense suspensions. However, this is
beyond the scope of the current study.

\subsection{Outline of the paper}
In~\secref{s:numMeth} we briefly summarize the governing equations for
the RBC flows. We explain the DLD theory in detail and our DLD model
in~\secref{s:dldModel}. After validating our model by comparing our
numerical results for the size-based separation of rigid particles
with the actual experiments~\citep{davisPhd14}
in~\secref{s:validation}, we present our results
in~\secref{s:results}. We show snapshots from the simulations in Sections~\ref{s:deformSep} (for different capillary numbers) and~\ref{s:vcSep} (for different viscosity contrasts). We present the phase diagrams of the transport modes as a function of $C_a$, $\nu$ and $\epsilon$ in~\secref{s:phase}. 
In~\secref{s:dense} we demonstrate the failure rates of the transport
modes in dense suspensions of RBCs. Then, in~\secref{s:discussion} we
quantify the cell dynamics in DLD and compare them with those in shear and confined Poiseuille flows. Finally, we summarize our results
in~\secref{s:conclusions}. In \appref{a:intEqnForm} we present the 
detailed integral equation formulation for the Stokesian particulate flows. 
Then we show that our DLD model is insensitive to the number of 
pillars in the domain when capturing the particle 
trajectories in \appref{a:appendix2} and we show convergence 
studies for verification of our solver in \appref{a:appendix1}.

\subsection{Notation} We summarize the main notation used in
this paper in \tabref{t:notation}.
\begin{table}
\mcaption{List of frequently used notation.}{t:notation}
\centering
\begin{tabular}{l l}

 {\em Symbol} & {\em Definition} \\ 
  \hline
  $C_a$ 
  & Capillary number~\eqref{e:CaNumber}, dimensionless number quantifying RBC elasticity\\

  $\nu$
  & Viscosity contrast~\eqref{e:viscCont}, dimensionless cytoplasmic viscosity\\ 

  $\epsilon$ &
  Row-shift fraction: tangent of the tilt angle of the pillar rows\\

  $v_{\mathrm{mig}}$ & Migration velocity \\

  $F_l$ & Pseudo-lift \\

  $\alpha$ & Cell's inclination angle with respect to the flow \\

  $x$ & Main flow direction \\

  $y$ & Lateral direction, in which separation occurs \\

\end{tabular} 
\end{table} 

\section{Numerical model and geometry\label{s:model}} 
We present the mathematical model for the flow of red blood cells 
in~\secref{s:numMeth} and our DLD model
in~\secref{s:dldModel}. We verify and validate our DLD model by
comparing our numerical results for the separation of rigid spherical
particles with the experimental data available in the literature in
Sections~\ref{s:verification} and~\ref{s:validation}. Finally,
in~\secref{s:dimParams} we describe the dimensionless parameters of
which we will investigate the effects on the RBC separation.

\begin{figure}
\begin{minipage}{\textwidth}
\centering
\setcounter{subfigure}{0}
\renewcommand*{\thesubfigure}{(a)} 
      \hspace{0cm}\subfigure[An RBC suspension in a DLD device]{\scalebox{0.57}{{\includegraphics{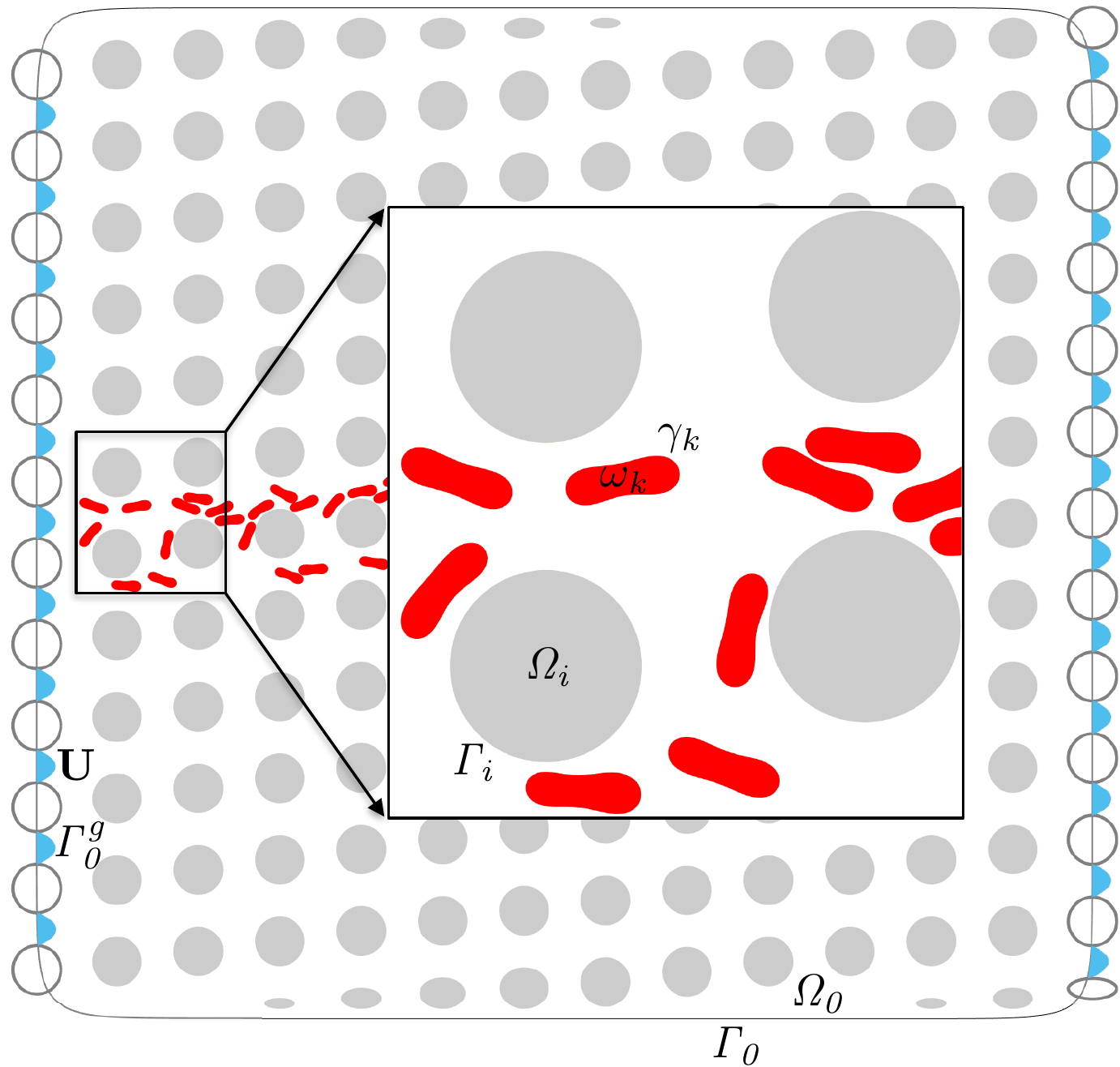}}}
      \label{f:VesInDLD}} 
\setcounter{subfigure}{0}      
\renewcommand*{\thesubfigure}{(b)} 
      \hspace{0cm}\subfigure[DLD parameters]{\scalebox{0.44}{{\includegraphics{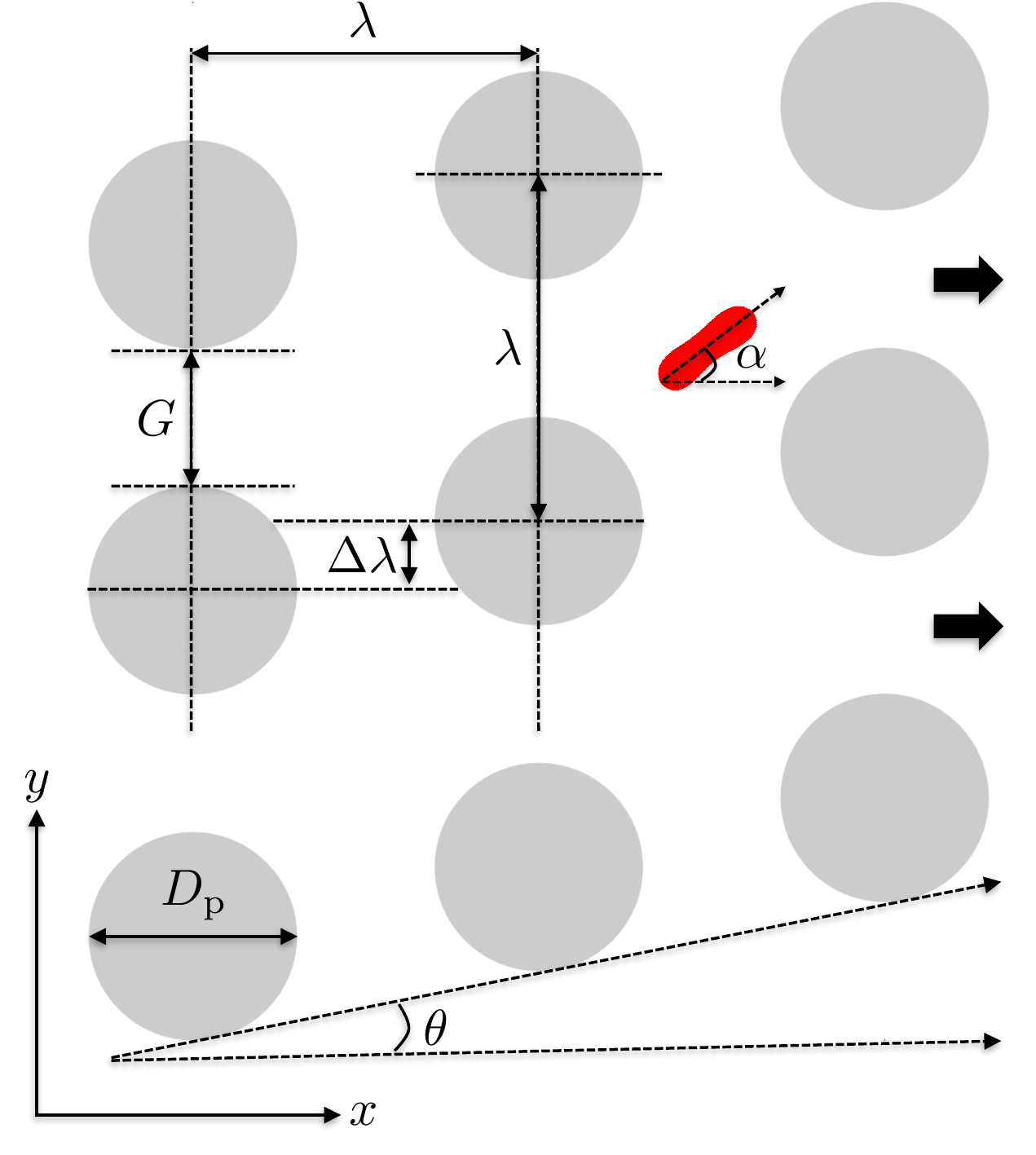}}}
      \label{f:DLDGeom}}
\end{minipage}                    
\mcaption{We illustrate the domain of a cell suspension in a DLD device on
the left. We model RBCs as inextensible vesicles with a reduced area of 0.65. The
interior and boundary of the $i^{th}$ pillar are denoted by $\Omega_i$
and $\Gamma_i$ ($\Omega_0$ and $\Gamma_0$ are the ones of the exterior
wall). Additionally, $\omega_k$ and $\gamma_k$ stand for the interior
and boundary of the $k^{th}$ cell. The empty circles are the pillars in the left and the right imaginary columns. The walls pass through these pillars' centers. The blue arrows show the parabolic velocity $\UU$ imposed as a boundary condition in the gaps $\Gamma_0^g$ between the imaginary pillars. On the right we show the DLD
parameters where $x$ is the flow direction. The geometry is uniquely
defined by the pillar diameter $D_p$, the center-to-center distance
$\lambda$, and the row shift $\Delta \lambda$. Using these parameters
we can determine the gap size $G$, the row-shift fraction 
$\epsilon = \Delta \lambda / \lambda$ and the tilt angle of the pillars (the angle between the flow direction and the alignment direction of the pillars) $\theta = \tan^{-1}(\epsilon)$. The red cell on the right has a positive inclination angle $\alpha$ with respect to the flow direction, which is defined in~\secref{s:numMeth}.}
{f:formulation}
\end{figure}

\subsection{Governing equations}\label{s:numMeth}

We assume that the flow is two-dimensional and there are no external body
forces such as gravity. Both the RBC's interior fluid and the exterior
fluid (plasma) are Newtonian. We assume a Stokesian fluid. See \figref{f:VesInDLD} for the geometry. Let the boundary of each pillar be denoted by
$\Gamma_i$. The pillars are bounded by an exterior wall $\Gamma_0$
which is a mollification of a rectangle. The areas enclosed by the
exterior wall and the $i^{th}$ pillar are denoted with $\Omega_0$ 
and $\Omega_i$, respectively. Therefore, the suspension resides in the domain
\begin{equation*}
\Omega = \Omega_0 \setminus \left( \bigcup_i \Omega_i \right),
\end{equation*}
and its boundary is $\Gamma = \Gamma_0 \cup \left(\bigcup_i \Gamma_i\right)$.

The reduced area, $4\upi A/P^2$, is the ratio between the area of a
cell $A$ and the area of a circle having the same perimeter $P$. 
We model RBCs as inextensible capsules with bending (vesicles)  with the reduced area 0.65. For this reduced area the vesicles exhibit biconcave shape as
RBCs. A cell's inclination angle $\alpha$ (see also the right figure in~\figref{f:formulation}) is the angle between the flow direction and the principal axis corresponding to the smallest principal moment of inertia~\citep{rahimian-biros-e10}. The moment of inertia tensor is 
\begin{equation} \label{e:momInt}
J = \frac{1}{4}\int_{\gamma_k}\rr\cdot\nn\left(|\rr|^2\mathbf{I} - \rr \otimes \rr \right) d\gamma_k.
\end{equation}
Here, $\rr = \xx - \cc$ and $\cc$ is the center of the cell. $\gamma_k$ and $\omega_k$ stand for the boundary and interior of
the $k^{th}$ cell, respectively. 
Let $\gamma = {\bigcup}_k \gamma_k$ and $\omega = {\bigcup}_k \omega_k$, then the equations describing velocity field at an instance are given by 
\begin{equation} \label{e:stokesCont}
  -\eta \Lap \mathit{\uu}(\mathit{\xx}) + \Grad p(\mathit{\xx}) = 0, 
  \quad \text{and} \quad \Div(\mathit{\uu}(\mathit{\xx})) = 0, \quad 
  \mathit{\xx} \in \Omega \setminus \gamma.
\end{equation}

Here, $\eta$ is the viscosity ($\eta_{\mathrm{in}}$ for the interior
and $\eta_{\mathrm{out}}$ for the exterior fluids), $\uu$ is the
velocity and $p$ is the pressure. We impose the velocity on the
external wall and pillars as a Dirichlet boundary condition (we
explain the boundary conditions in detail in~\secref{s:dldModel})
\begin{equation}\label{e:noslipOnWalls}
\uu(\xx,t) = \UU (\xx,t), \quad  \xx \in \Gamma.
\end{equation}
Conservation of mass and the no-slip boundary condition requires velocity continuity on the
interface of cells, i.e.,
\begin{equation}\label{e:velContinuity}
\uu(\xx,t) = \frac{d \xx}{dt}(t), \quad  \xx \in \gamma.
\end{equation}
RBCs are known to be inextensible, i.e., they conserve their area and arc-length in 2D. Inextensibility means that
\begin{equation}\label{e:inexten}
\xx_s \mathbf{\cdot} \uu_s = 0, \quad  \xx \in \gamma,
\end{equation}
where the subscript ''$s$" stands for differentiation with respect to
the arc-length on the boundaries of cells. The last governing equation comes from the
momentum balance on the interface of cells. Since the cells
resist bending and tension, the interface applies an elastic force as
a result of deformation due to them. The momentum balance enforces the
jump in the surface traction to be equal to the net elastic force
applied by the interface,
\begin{equation} \label{e:tracJump}
  [\![ \mathbf{T}\nn(\xx) ]\!] = -\kappa_b \xx_{ssss} + \left(\sigma(\xx,t) \xx_s \right)_s,  \quad
  \xx \in \gamma,
\end{equation}
where $\mathbf{T} = -p\mathbf{I} + \eta (\Grad \uu + \Grad {\uu}^T)$ is the
Cauchy stress tensor, $\nn$ is the outward normal vector on $\gamma$,
$[\![ \cdot ]\!]$ is the jump across the interface. The right-hand
side is the net force applied by the interface onto the fluid. The
first term on the right-hand side is the force due to bending
stiffness $\kappa_b$ and the second term is the force due to tension
$\sigma$, which acts as a Lagrange multiplier enforcing the
inextensibility \citep{shravan-biros-e09}.  Finally, the position of
the boundaries of $M$ cells evolves as
\begin{equation} \label{e:boundaryEvolve}
  \frac{d\xx_i}{dt} = \uu_{\infty}(\xx_i) + \sum_{j=1}^M \uu(\xx_j), \quad i = 1,\ldots,M,
\end{equation}
where $\uu_{\infty}(\xx_i)$ is the background velocity and
$\uu(\xx_j)$ is the velocity due to the $j^{th}$ cell acting on the
$i^{th}$ cell.

The complete set of nonlinear equations
\eqref{e:stokesCont}-\eqref{e:boundaryEvolve} governs the evolution of
the cell interfaces. In line with our previous work
~\citep{shravan-biros-e09, rahimian-biros-e10, quaife-biros14, quaife-biros15, 
quaife-biros16, kabacaoglu-biros-e17b}, we use an integral equation method to
obtain the positions of the cell boundaries 
(see \citet{kabacaoglu-biros-e17b} for the details of the numerical scheme).
The method can naturally handle
the moving geometry, achieves high-order accuracy with a low
computational cost and handle the viscosity contrast between the
interior and exterior fluids.
For completeness, we present the integral equation
formulation in~\appref{a:intEqnForm}.

\subsection{DLD model}\label{s:dldModel}

A DLD device of circular pillars can be uniquely determined by three
parameters: the diameter of a pillar $D_{\mathrm{p}}$, 
the center-to-center distance between two neighboring pillars $\lambda$ and the 
row-shift $\Delta \lambda$. See~\figref{f:DLDGeom} for the DLD geometry.

{\em Geometry.}
Our DLD device consists of circular pillars all of which have the same
diameter $D_{\mathrm{p}} = 15\mu$m. The fluid flows in the $x$
direction. The center-to-center distance between neighboring pillars
is $\lambda = 25\mu$m and the same in both directions. This results in
a gap size of $G = \lambda-D_{\mathrm{p}} = 10\mu$m. Pillars centered
at the same $x$ coordinate form a "column". Each pillar in a neighboring column
is shifted by $\Delta \lambda$ in the $y$-direction with respect to
the previous one, which defines the row-shift fraction 
$\epsilon = \Delta \lambda / \lambda$. Thus, the "rows" are tilted with an angle
$\theta = \tan^{-1}(\epsilon)$. We define {\em "lane"} to be a path between two diagonally parallel rows.

\begin{figure}
\begin{center}
\includegraphics[scale=0.5]{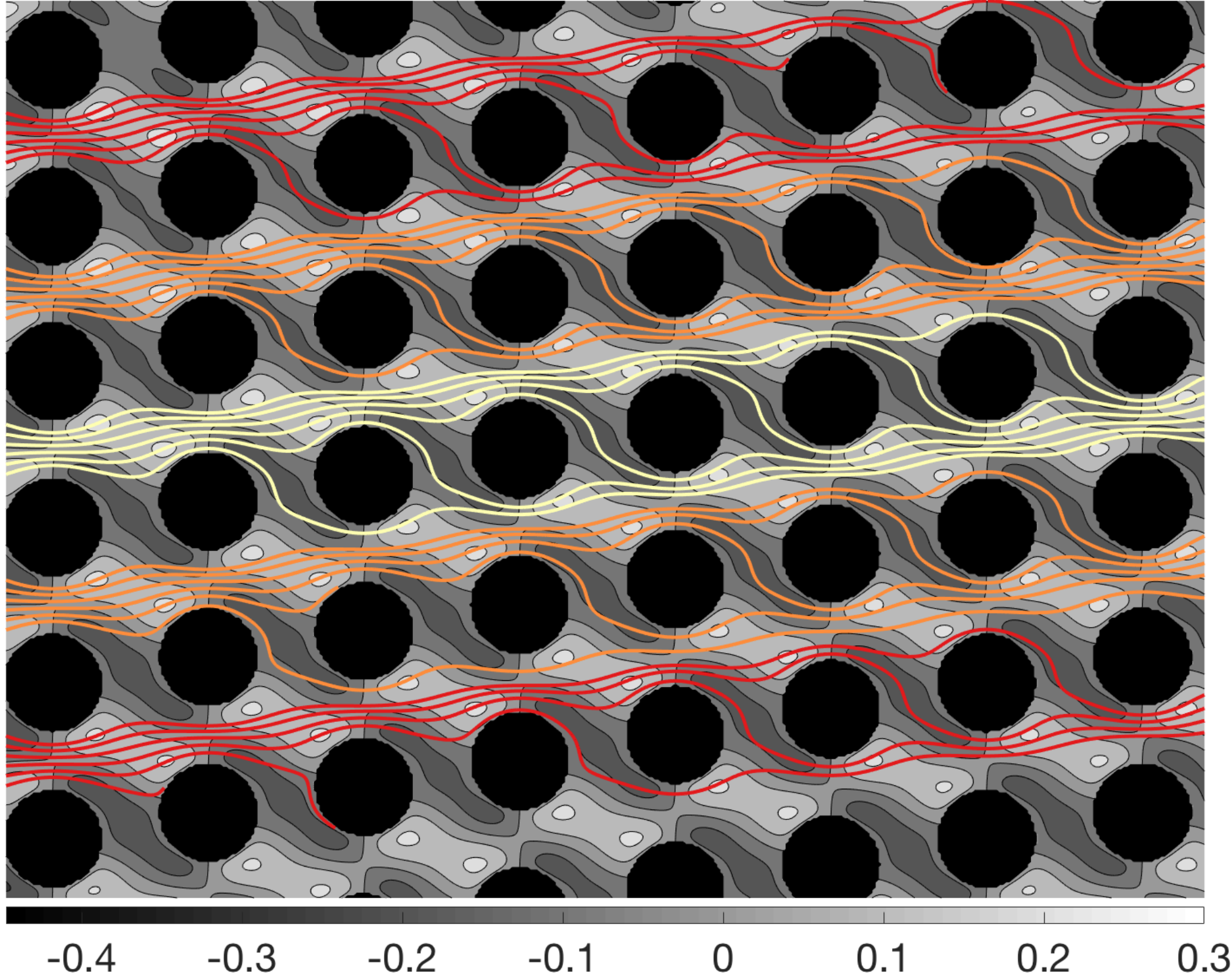} \mcaption{
Velocity in the vertical direction (represented by gray-scale contour colors) 
and streamlines in the absence of particles are 
superimposed. The row-shift fraction is $\epsilon = 0.1667$, which corresponds to the period $n_p = 6$.
The flow is from left to right.
See~\figref{f:N6WholeDLD} for the velocity magnitude in the 
whole device and the velocity field in a gap. Each streamline interacts with a pillar, then 
swaps a lane. The DLD theory assumes that rigid particles do not distort the streamlines 
significantly. Depending on their size, they either follow these streamlines in zig-zag mode 
or cross the streamlines and stay in the same lane 
in displacement mode.}{f:explainStreams}
\end{center}
\end{figure}
{\em DLD theory for rigid spherical particles.}
The row-shift fraction $\epsilon$ sets what is referred to as the period $p$ of the device. ''Period" sets a length-scale in which the column arrangement is exactly repeated. Therefore, we can divide the unperturbed flow in a gap (i.e., in the
absence of any particles) into $n_p$ flow streams of equal mass flux. If we
assume that the unperturbed flow does not change significantly when
particles flow, the width of the stream adjacent to a pillar becomes
the critical particle size~\citep{davis-austin-e06}. Particles small
enough to fit into one of the streams stay in their streams by zig-
zagging, whereas the larger ones cannot. As the tilt angle $\theta$
reduces (i.e., the row-shift fraction $\epsilon$ reduces or the period
$n_p$ increases), the stream width decreases and hence the critical
particle size decreases. This theory is in good agreement with experiments with spherical rigid particles~\citep{davis-austin-e06,davisPhd14}.

{\em Simulation domain.}
The DLD devices usually consist of $\bigO(10)$ rows and $\bigO(100)$
columns, which results in $\bigO(1000)$ pillars in a device.
Performing simulations of cell separation in a whole DLD device is
computationally expensive~\citep{rossinelli-petros-e15}. This renders
its systematic analysis infeasible even in 2D. In
order to evade the computational cost the numerical studies conducted
so far in this area reduced the simulation domain to a single pillar
by assuming periodicity 
\citep{quek-chiam-e11,ye-yu-e14,kruger-coveney-e14, vernekar-kruger15,zhang-fedosov-e15}. However, one pillar with periodic conditions requires imposing an artificial lateral force to enforce zero net lateral flow since in real DLD devices the lateral flow is restricted by the walls.

Due to the high computational costs, most of our numerical experiments are conducted using only one period, i.e., at least
$\left \lceil 1/\epsilon \right \rceil$ pillars in the flow direction instead of using just one pillar. We confine the pillar lattice with an exterior
wall to enforce zero net lateral flow. This raises the questions about the wall effects introducing large errors. We numerically determined that the errors are negligible if we use 12 pillars in the $y$ direction (i.e., 12 rows) and
$\left \lceil 1.5 (1/\epsilon) \right \rceil$ pillars in the $x$
direction (i.e., $\left \lceil 1.5 (1/\epsilon) \right \rceil$ columns)
(see~\figref{f:VesInDLD} and~\figref{f:N6WholeDLD} as an example). We
present the details of this analysis in~\appref{a:appendix2}. The initial locations of the cells are in the middle lane. Since the pillars are shifted
laterally, we end up with pillars crossing the exterior wall at the
top and the bottom. We found out that if we just remove those pillars,
the empty spaces result in less hydraulic resistance and hence induces
a lateral pressure gradient. This breaks the homogeneity of the
flow and introduces significant errors in the cell trajectories~\citep{kulrattanarak-boom-e11,vernekar-inglis-e16}. Therefore, we
decided to replace the circular pillars crossing the walls with
elliptical pillars in such a way that they maintain the same vertical
spacing with the neighboring pillars. Those non-circular pillars near
the top and bottom walls provided a homogeneous flow in the middle
region in our model (see~\figref{f:N6WholeDLD} for our DLD model
and the unperturbed velocity for $n_p = 6$).

\citet{henry-gompper-e16} showed that RBCs might have additional 
zig-zag modes unlike rigid particles, such as zig-zagging within a period or zig-zagging that requires more than a period to take place. In order to capture those zig-zagging modes, we perform simulations of a cell by initializing it at several lateral positions. This is equivalent to simulating the cell for more than one period because in each simulation the cell confronts the first pillar at a different lateral position. We label the cell zig-zagging if it zig-zags in any of these simulations and displacing if it displaces in all of the simulations.
\begin{figure}
\begin{center}
\includegraphics[scale=0.55]{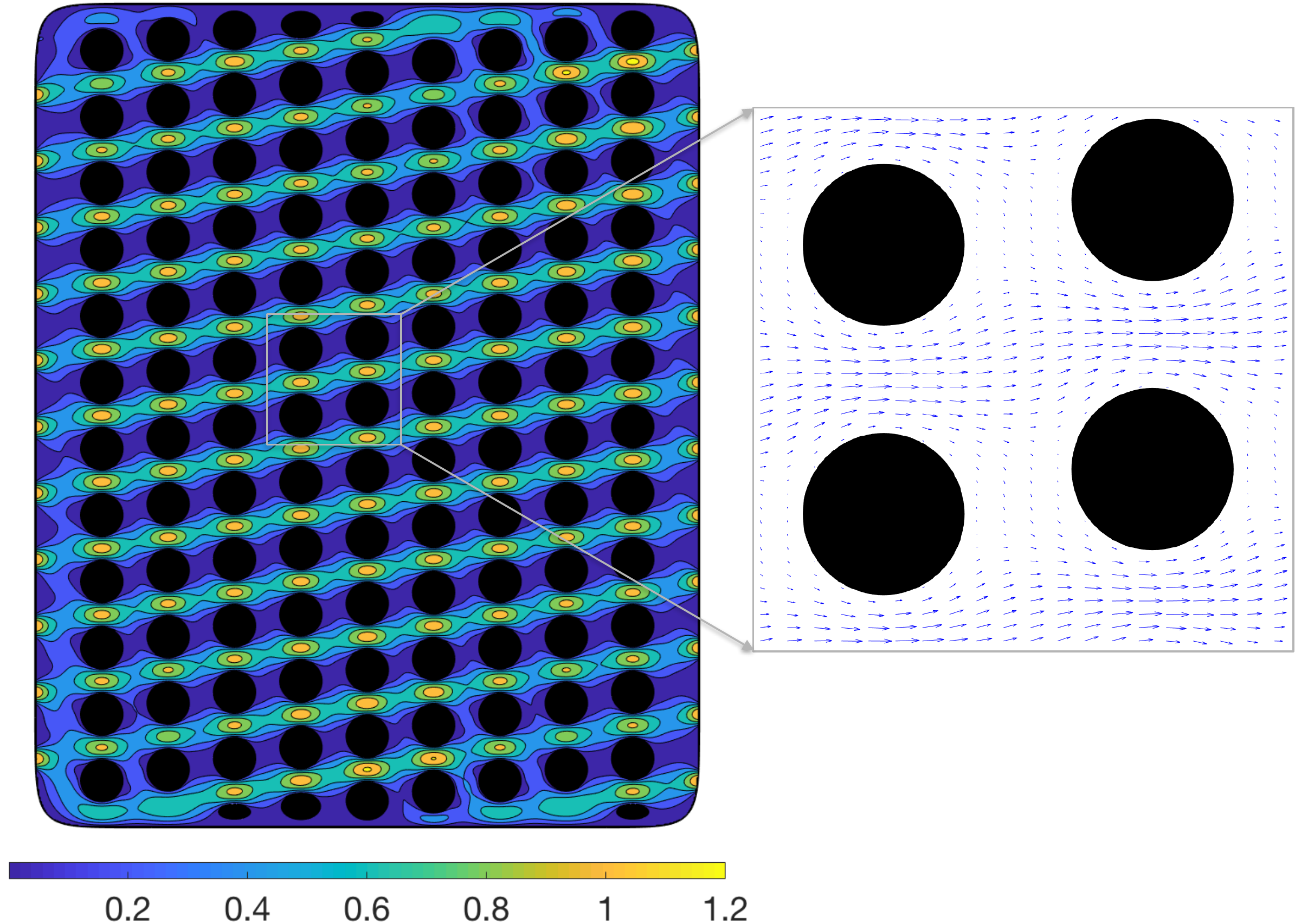} 
\mcaption{Velocity magnitude (on the left) and velocity in a gap (on the right) in
our DLD model for $\epsilon = 0.1667$, i.e., $n_p = 6$. Our DLD model
consists of 12 rows and $\left \lceil 1.5n_p \right \rceil$ columns of
pillars (i.e., 9 columns here). We draw the side walls such that they
pass through the centers of the shifted imaginary columns (not shown).
Then we impose a parabolic velocity profile \eqref{e:uBCs} on the
boundaries which correspond to the gaps between the imaginary pillars
as a boundary condition. We replaced the circular pillars crossing the
top and the bottom walls with the elliptical ones. That leads to a
homogeneous velocity in the device. See~\figref{f:explainStreams} for 
the velocity in the $y$-direction and the streamlines for this 
DLD configuration.}{f:N6WholeDLD}
\end{center}
\end{figure}

{\em Boundary conditions.}
We want to reflect the periodicity of the DLD device on our model in
imposing the velocity as a boundary condition at the intake and the
outtake. For this purpose, we place the side walls as if they pass
through the imaginary columns of the pillars shifted down on the left
and up on the right by $\Delta \lambda$ (the empty circles in the left figure in~\figref{f:formulation} are the pillars in the imaginary columns). Let $\Gamma_{0}^{g}$ be the
boundary on the exterior wall corresponding to the $g^{th}$ gap
between the imaginary pillars. Since the velocity in a gap is
parabolic, we impose the parabolic velocity profile on
$\Gamma_{0}^{g}$ and zero velocity on the rest of the exterior wall
and on the interior pillars as a Dirichlet boundary condition
\begin{equation}\label{e:uBCs}
\UU(\xx) = \begin{cases} 
U_{\max}\left(1-\frac{y^2}{G^2}\right), & \xx \in \Gamma_{0}^{g} \\
0, & \xx \in \left(\Gamma_{0} \setminus \bigcup_{g}\Gamma_{0}^{g}\right) 
\bigcup \left(\bigcup_{i} \Gamma_i\right), \end{cases}
\end{equation}
where $U_{\max}$ sets the velocity scale (see also the left figure in~\figref{f:formulation} for the boundary conditions $\UU$). We demonstrate our model for
a DLD device of the row-shift fraction $\epsilon = 0.1667$
in~\figref{f:N6WholeDLD}. The velocity magnitude far from the top 
and bottom walls seems homogeneous.
Therefore, we expect to capture the behavior of the particles flowing
away from the top and bottom walls accurately in one period using our
model.

\subsection{Verification}\label{s:verification}

We performed a convergence study by considering one zig-zagging and
one displacing RBCs: the zig-zagging RBC with a high viscosity
contrast and the displacing RBC with a low viscosity contrast for the
row-shift fraction $\epsilon = 0.1667$. We refined the spatio-temporal
resolution until the trajectories converged (see~\appref{a:appendix1}
for the details of the convergence study). We used the converged
discretization for the pillars and the converged temporal resolution
in all our simulations. As the row-shift fraction changes, the
number of columns and hence the size of the exterior wall change. In order to 
maintain the same grid quality in all our simulations, we
adjusted the discretization of the exterior wall depending on $\epsilon$. 

We also show that our model is not sensitive to at
what lateral position and with what inclination angle RBCs are
initialized (see~\figref{f:IAsTrajsN6} for the results).

\subsection{Validation}\label{s:validation}

The separation of rigid spherical particles using a DLD device of
circular pillars is well 
understood~\citep{huang-sturm-e04,inglis-sturm-e06,zhang-fedosov-e15}. 
An empirical formula for the critical
particle size is given as a function of the gap size $G$ and row-shift
fraction $\epsilon$~\citep{davisPhd14}
\begin{equation}\label{e:rigidEmpric}
D_{\mathrm{c}} = 1.4G\epsilon^{0.48}
\end{equation}
where $D_{\mathrm{c}}$ is the critical diameter. Particles with
diameters $D>D_c$ displace while those with $D<D_c$ zig-zag. In order
to validate our DLD model we simulated the separation of rigid
spherical particles. We considered four different DLD devices with the
same gap $G = 10\mu$m and different row-shift fractions 
$\epsilon \in \{ 0.05, 0.1, 0.167, 0.25\}$. We had 8 rigid circular particles
with $D = \{2, 3, 4, 5, 6, 7, 8, 8.5\} \mu$m flowing through these DLD
devices. We observed the trajectories of the particles and determined
whether they zig-zag or displace in one period. We present the results
of the validation study in~\figref{f:ValidPhase}. We also plot the
line for the ratio of the critical diameter to the gap
$D_{\mathrm{c}}/G$ given by the empirical
formula~\eqref{e:rigidEmpric}. As expected, the particles with
diameters greater than $D_{\mathrm{c}}$ (above the line in
\figref{f:ValidPhase}) displace while those with smaller diameters
(below the line) zig-zag. Our numerical results agree well with the
experimental results~\citep{inglis-sturm-e06,davisPhd14}. This
validates our 2D model and proves that it can capture the underlying
physics of the particle separation in the DLD devices.
\begin{figure}
\begin{center}
\begin{tikzpicture}

\begin{axis}[
  width = 0.65\textwidth,
  xmin = 0.01,
  xmax = 0.3, 
  ymin = 0.1,
  ymax = 0.9,
  xtick = {0.05,0.1,0.15,0.2,0.25,0.3},
  xticklabels={$0.05$,$0.1$,$0.15$,$0.2$,$0.25$,$0.3$},
  ytick = {0.1,0.2,0.3,0.4,0.5,0.6,0.7,0.8,0.9},
  xlabel = {Row-shift fraction ($\epsilon$)},
  ylabel = {Diameter-to-gap ratio ($D/G$)},
  label style = {font=\normalsize},
  legend style={at={(0.5,1.05)},anchor=south,legend columns=-1,draw=none},
  legend entries = {Zig-zag, Displacement, $D_c/G = 1.4\epsilon^{0.48}$},
  ]

\addplot [color=red,only marks,mark=*,mark size=2.5pt] table{
0.05 0.2
0.05 0.3
0.1 0.2
0.1 0.3
0.1 0.4
0.1667 0.2
0.1667 0.3
0.1667 0.4
0.1667 0.5
0.25 0.2
0.25 0.3
0.25 0.4
0.25 0.5
0.25 0.6
0.25 0.7  
}; 
\addplot [color=blue,only marks,mark=triangle*, mark size=2.5pt] table{
0.05 0.4
0.05 0.5
0.05 0.6
0.05 0.7
0.05 0.8
0.1 0.5
0.1 0.6
0.1 0.7
0.1 0.8  
0.1667 0.6
0.1667 0.7
0.1667 0.8
0.1667 0.85
0.25 0.75
0.25 0.8
0.25 0.85
}; 
\addplot [mark=none,black,line width=1] table{
    0.0100    0.1535
    0.0115    0.1639
    0.0129    0.1736
    0.0144    0.1827
    0.0158    0.1914
    0.0173    0.1996
    0.0187    0.2075
    0.0202    0.2151
    0.0217    0.2224
    0.0231    0.2295
    0.0246    0.2363
    0.0260    0.2430
    0.0275    0.2494
    0.0289    0.2557
    0.0304    0.2618
    0.0319    0.2677
    0.0333    0.2735
    0.0348    0.2792
    0.0362    0.2848
    0.0377    0.2902
    0.0391    0.2955
    0.0406    0.3008
    0.0421    0.3059
    0.0435    0.3109
    0.0450    0.3159
    0.0464    0.3208
    0.0479    0.3256
    0.0493    0.3303
    0.0508    0.3349
    0.0523    0.3395
    0.0537    0.3440
    0.0552    0.3485
    0.0566    0.3529
    0.0581    0.3572
    0.0595    0.3615
    0.0610    0.3657
    0.0625    0.3698
    0.0639    0.3740
    0.0654    0.3780
    0.0668    0.3821
    0.0683    0.3860
    0.0697    0.3900
    0.0712    0.3939
    0.0727    0.3977
    0.0741    0.4015
    0.0756    0.4053
    0.0770    0.4090
    0.0785    0.4127
    0.0799    0.4164
    0.0814    0.4200
    0.0829    0.4236
    0.0843    0.4271
    0.0858    0.4307
    0.0872    0.4342
    0.0887    0.4376
    0.0902    0.4411
    0.0916    0.4445
    0.0931    0.4479
    0.0945    0.4512
    0.0960    0.4545
    0.0974    0.4578
    0.0989    0.4611
    0.1004    0.4644
    0.1018    0.4676
    0.1033    0.4708
    0.1047    0.4740
    0.1062    0.4771
    0.1076    0.4803
    0.1091    0.4834
    0.1106    0.4865
    0.1120    0.4895
    0.1135    0.4926
    0.1149    0.4956
    0.1164    0.4986
    0.1178    0.5016
    0.1193    0.5046
    0.1208    0.5075
    0.1222    0.5104
    0.1237    0.5133
    0.1251    0.5162
    0.1266    0.5191
    0.1280    0.5220
    0.1295    0.5248
    0.1310    0.5277
    0.1324    0.5305
    0.1339    0.5333
    0.1353    0.5360
    0.1368    0.5388
    0.1382    0.5415
    0.1397    0.5443
    0.1412    0.5470
    0.1426    0.5497
    0.1441    0.5524
    0.1455    0.5551
    0.1470    0.5577
    0.1484    0.5604
    0.1499    0.5630
    0.1514    0.5656
    0.1528    0.5682
    0.1543    0.5708
    0.1557    0.5734
    0.1572    0.5760
    0.1586    0.5785
    0.1601    0.5811
    0.1616    0.5836
    0.1630    0.5861
    0.1645    0.5886
    0.1659    0.5911
    0.1674    0.5936
    0.1688    0.5961
    0.1703    0.5986
    0.1718    0.6010
    0.1732    0.6035
    0.1747    0.6059
    0.1761    0.6083
    0.1776    0.6107
    0.1790    0.6131
    0.1805    0.6155
    0.1820    0.6179
    0.1834    0.6203
    0.1849    0.6226
    0.1863    0.6250
    0.1878    0.6273
    0.1892    0.6297
    0.1907    0.6320
    0.1922    0.6343
    0.1936    0.6366
    0.1951    0.6389
    0.1965    0.6412
    0.1980    0.6435
    0.1994    0.6457
    0.2009    0.6480
    0.2024    0.6502
    0.2038    0.6525
    0.2053    0.6547
    0.2067    0.6569
    0.2082    0.6592
    0.2096    0.6614
    0.2111    0.6636
    0.2126    0.6658
    0.2140    0.6680
    0.2155    0.6701
    0.2169    0.6723
    0.2184    0.6745
    0.2198    0.6766
    0.2213    0.6788
    0.2228    0.6809
    0.2242    0.6831
    0.2257    0.6852
    0.2271    0.6873
    0.2286    0.6894
    0.2301    0.6915
    0.2315    0.6936
    0.2330    0.6957
    0.2344    0.6978
    0.2359    0.6999
    0.2373    0.7019
    0.2388    0.7040
    0.2403    0.7061
    0.2417    0.7081
    0.2432    0.7102
    0.2446    0.7122
    0.2461    0.7142
    0.2475    0.7163
    0.2490    0.7183
    0.2505    0.7203
    0.2519    0.7223
    0.2534    0.7243
    0.2548    0.7263
    0.2563    0.7283
    0.2577    0.7303
    0.2592    0.7323
    0.2607    0.7342
    0.2621    0.7362
    0.2636    0.7382
    0.2650    0.7401
    0.2665    0.7421
    0.2679    0.7440
    0.2694    0.7460
    0.2709    0.7479
    0.2723    0.7498
    0.2738    0.7517
    0.2752    0.7537
    0.2767    0.7556
    0.2781    0.7575
    0.2796    0.7594
    0.2811    0.7613
    0.2825    0.7632
    0.2840    0.7651
    0.2854    0.7669
    0.2869    0.7688
    0.2883    0.7707
    0.2898    0.7726
    0.2913    0.7744
    0.2927    0.7763
    0.2942    0.7781
    0.2956    0.7800
    0.2971    0.7818
    0.2985    0.7837
    0.3000    0.7855
};

\end{axis}

\end{tikzpicture}
\mcaption{Phase diagram for rigid  spherical particles in DLD devices of
circular pillars as a function of the row-shift fraction $\epsilon$
and the particle diameter-to-gap ratio $D/G$. We also plot the
critical particle size given by the empirical
formula~\eqref{e:rigidEmpric} (solid black line). Particles having
greater diameters than the critical diameter displace while those with
smaller diameters zig-zag. We show the zig-zagging and displacing
particles with filled red circles and blue triangles, respectively.
Our numerical results agree well with experimental 
results~\citep{inglis-sturm-e06,davisPhd14}.}{f:ValidPhase}
\end{center}
\end{figure}

{\em Remark.}
In our simulations we consider the transport of the particles at high
Peclet numbers (i.e., $\Pen \gg 1$, the transport is advection dominated). 
Therefore, the mixed mode, which
is an irregular alternation between zig-zag and displacement
modes~\citep{huang-sturm-e04,kulrattanarak-boom-e11,zhang-fedosov-e15}, 
is not observed in our study. This mode occurs as a
result of diffusion of particles between different streamlines.

\subsection{Dimensionless numbers}\label{s:dimParams}

In this section, we summarize the parameter values that we will use in our experiments in~\secref{s:discussion}. The parameter selection is related to separating normal and abnormal RBCs based on differences in their deformability. The deformability of a cell depends on the cell's membrane elasticity, cytoplasmic viscosity and the imposed flow. These can be combined to a single non-dimensional number, the capillary number:
\begin{equation}\label{e:CaNumber}
C_a = \frac{\eta_{\mathrm{out}} R_{\mathrm{eff}}^3}{\kappa_b} \frac{U_{\max}}{ G/2} 
\end{equation}
where $\eta_{\mathrm{out}}$ is the viscosity of the exterior fluid,
$R_{\mathrm{eff}}$ is the effective radius of the cell ($R_{\mathrm{eff}} = \sqrt{A/\pi}$ in which $A$ is the area enclosed by the cell), $U_{\max}$ is the maximum velocity of the unperturbed
Poiseuille flow in a gap~\eqref{e:uBCs}, $\kappa_b$ is the bending
stiffness and $G$ is the gap width. The higher the capillary number is, the more deformable the cell is. Typically, the bending stiffness
and cell thickness (corresponding to its effective size in deep devices) for a healthy red blood cell are $\kappa_b = $ 1E-
19Nm~\citep{suresh-seufferlein-e05,tomaiuolo14} and $R_{\mathrm{eff}}
= 2.5 \mu$m~\citep{popel-johnson05}, respectively. The viscosity of the
blood plasma is $\eta_{\mathrm{out}} = 1.2$mPas. As far as the
reported experiments~\citep{mcgrath-bridle-e14}, the order of $U_{\max}$ ranges
between $10\mu$m/s and $10$mm/s in the DLD devices used to separate the
blood components. In those devices the gap width is in the order of $G
\bigO(1\mu$m), which is also the size in our study. Based on those typical
values, the capillary number for a healthy RBC is 
$C_a \in [0.0375, \, 375]$. For the diseased RBC the capillary number
decreases to its 1/10 since the stiffness increases 
10-fold~\citep{suresh-seufferlein-e05}. Therefore, we consider the capillary numbers
in the range of $[0.0038, \, 37.5]$. We set the capillary number by
adjusting the bending stiffness $\kappa_b$ only and fix all other parameters.

The dimensionless number quantifying the cytoplasmic viscosity of an 
RBC~$\eta_{\mathrm{in}}$ is the viscosity contrast
\begin{equation}\label{e:viscCont}
\nu = \frac{\eta_{\mathrm{in}}}{\eta_{\mathrm{out}}}.
\end{equation}
RBC's cytoplasmic viscosity is a nonlinear function of
the mean corpuscular hemoglobin concentration (MCHC), which is the
concentration of the hemoglobin per unit volume of 
an RBC~\citep{aouane-misbah-e14}. The viscosity varies from one cell to the other even
within the same organism due to age because the MCHC increases as RBC
gets older. For a young red blood cell MCHC is around $32$g/dl, which
results in $\eta_{\mathrm{in}} = 5-7$mPas, i.e., the viscosity contrast
is $\nu = 4-6$. If the MCHC increases to
$40$g/dl, the viscosity increases almost four-fold~\citep{chien87,mohandas-gallagher08}. Viscosity contrast is inversely proportional to cell's deformability. So, the smaller the viscosity contrast is, the more deformable the cell is. We consider $\nu \in [0.1, \, 100]$ in our study and set the viscosity contrast by changing the
interior viscosity $\eta_{\mathrm{in}}$. 

Following~\citet{vlahovska-gracia07} we define a time scale as the deformation time scale driven by the imposed flow, which is inverse of the shear gradient in a gap
\begin{equation}\label{e:timeScale}
\tau = \frac{G/2}{U_{\max}}.
\end{equation}

We have already presented the geometry of our DLD model
in~\secref{s:dldModel}. The third dimensionless number is the row-shift 
fraction $\epsilon$ of the DLD device. We adjust $\epsilon$ by
changing the row-shift $\Delta \lambda$ and the other DLD parameters
remain the same.

\section{Results\label{s:results}} 
First, we investigate the
effects of the capillary number in~\secref{s:deformSep} and the
viscosity contrast in~\secref{s:vcSep} on the dynamics of red blood cells in 
deep devices. Then in~\secref{s:phase}, we present phase diagrams for the 
transport modes as a function of the capillary number $C_a$, the viscosity 
contrast $\nu$ and 
the row-shift fraction $\epsilon$. Finally,
we study the cell separation in the flow of dense RBC suspensions through the DLD devices in~\secref{s:dense}.

\subsection{Effects of capillary number}\label{s:deformSep}

\begin{figure}
 \begin{minipage}{\textwidth}
\centering
\setcounter{subfigure}{0}
\renewcommand*{\thesubfigure}{(a)} 
      \hspace{0cm}\subfigure[$C_a = 0.034$ (red cell) and $C_a = 0.34$ (blue cell)]{\scalebox{0.69}{{\includegraphics{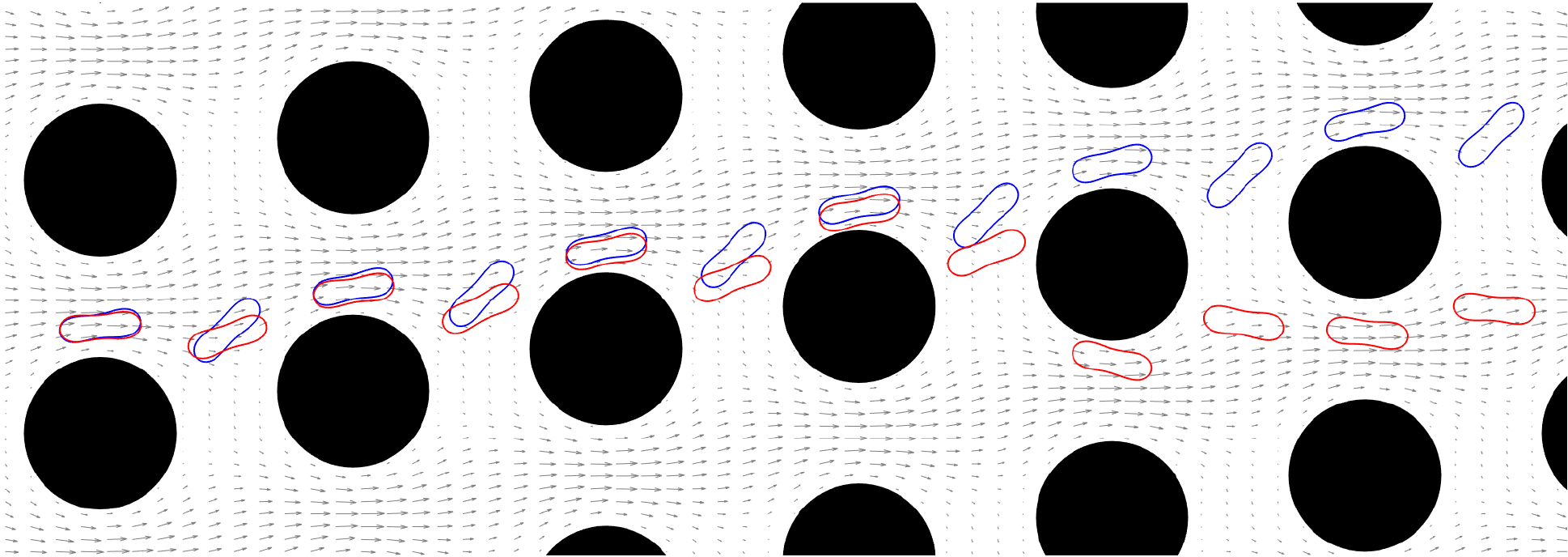}}}
      \label{f:N6CaLowTrajs}} 
\end{minipage}
\begin{minipage}{\textwidth}
\centering
\setcounter{subfigure}{0}      
\renewcommand*{\thesubfigure}{(b)} 
      \hspace{0cm}\subfigure[$C_a = 0.034$ (red cell) and $C_a = 34$ (blue cell)]{\scalebox{0.69}{{\includegraphics{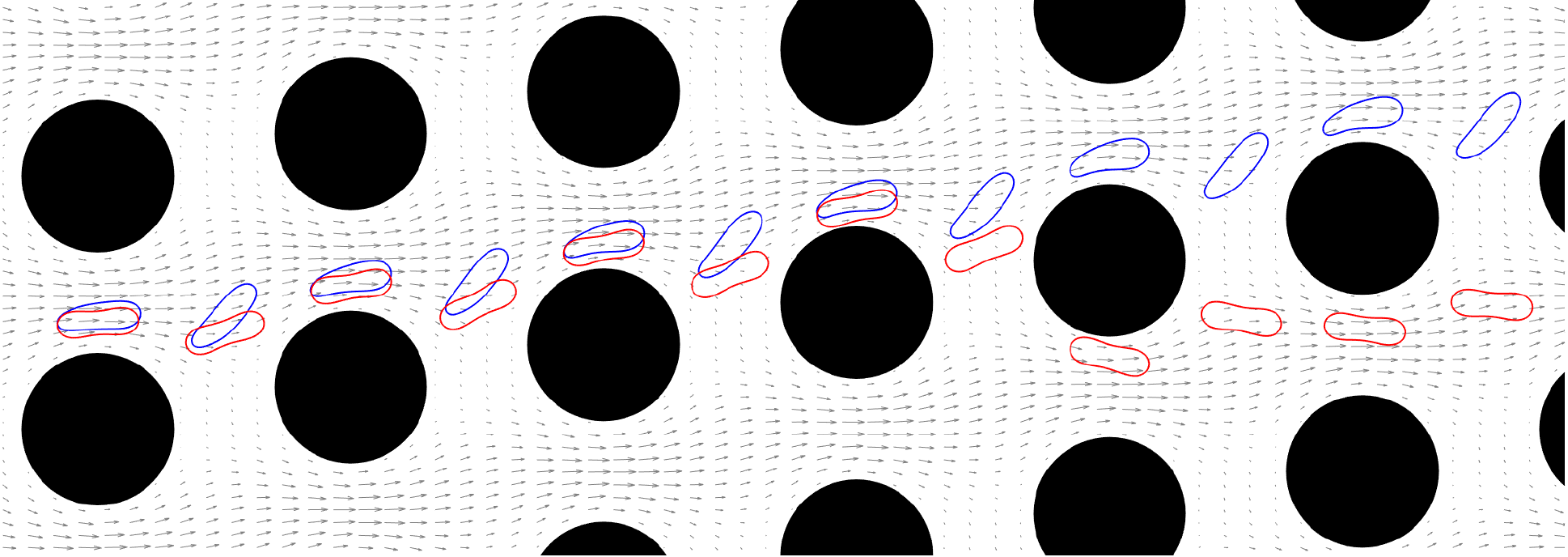}}}
      \label{f:N6CaHighTrajs}}
\end{minipage}                    
\mcaption{Superimposed snapshots from the simulations of RBCs having different capillary numbers $C_a$ and the unperturbed velocity field (shown with gray arrows). Here, we initialize the cells at the same position and simulate them separately. We, then, superimpose the snapshots taken when the cells are in the gaps and between two consecutive gaps. The viscosity contrast and the row-shift fraction in these simulations are the same, $\nu = 1$ and $\epsilon = 0.1667$, respectively. At the top, we compare the dynamics of two stiff cells with $C_a = 0.034$ (red) and $C_a = 0.34$ (blue). While the stiffer cell (red) zig-zags, the softer one (blue) displaces. At the bottom, we remain the stiffest cell ($C_a = 0.034$, red one) but replace the softer one with even a softer cell ($C_a = 34$, blue one). Again, the softer cell (blue) displaces.}{f:CaTrajs}
\end{figure}

We performed simulations of a single RBC flowing through DLD with the capillary numbers $C_a = (0.034, 0.34, 34)$. The viscosity contrast for these cells is $\nu = 1$. We considered only one device with row-shift fraction $\epsilon = 0.1667$. We present the superimposed snapshots from these simulations in~\figref{f:CaTrajs} together with the unperturbed velocity field (shown with gray arrows). The snapshots are taken when the cells are in the gaps and between the consecutive gaps. The cells are initialized at the same location and have effective radius $R_{\mathrm{eff}} = 2.4\mu$m. 

At the top in~\figref{f:CaTrajs} we compare two stiff cells, the red one 
($C_a = 0.034$) is stiffer than the blue one ($C_a = 0.34$). 
While the stiffer one
zig-zags, the softer one displaces. Then, at the bottom we replace the softer
one with even softer cell having $C_a = 34$. The softer cell again
displaces. We, now, discuss our observations from these simulations. In the
gaps, a stiff cell (low $C_a$) does not deform as much as the soft one does
(high $C_a$). The stiff cell behaves like a rigid particle and moves closer to the pillars than the softer cells do. A rigid spherical particle with a diameter
less than $6\mu$m zig-zags in a device with $\epsilon = 0.1667$ as we shown
in~\secref{s:validation}. Since the cell's effective size is less than this
critical size, it zig-zags. There are several other differences in cells'
behaviors depending on their stiffness. The softer cells have an asymmetry in
their shape while they are passing through a gap. They have a thick head and a
thin tail in the gaps (this is more apparent for $C_a = 34$) while the stiff
cell maintains its symmetric relaxed shape. The asymmetry in the soft cells'
shapes and the shear gradient in the gap lead the soft cells to have a higher
inclination angle with respect to the main flow direction than the stiff ones
while they are moving towards the region between two consecutive gaps. 

The positive inclination angle with
respect to the flow direction and the shear gradient together result in a
phenomenon called cell migration, which has been extensively studied in the
literature
~\citep{goldsmith-mason61,olla97,vlahovska-gracia07,coupier-misbah-e08,danker-misbah-e09,messlinger-gompper-e09,ghigliotti-misbah-e10}. A cell placed
near a wall migrates away from the wall in confined Poiseuille flow or shear
flow due to its positive inclination with respect to the wall and the shear
gradient. Here, the flow in DLD resembles the Poiseuille flow (confined flow in
the gaps and almost free flow between the gaps). Therefore, the cell with
positive inclination moves away from the pillars and if it is sufficiently away
from the pillars (i.e., it is out of the adjacent stream swapping a lane), it
displaces. As the inclination increases, it moves even farther away from the
pillars. \figref{f:CaTrajs} shows that as the cell becomes softer, its inclination angle increases. That is, the cells with $C_a = (0.34, 34)$ have higher inclination angles right after the first gap than
the one with $C_a = 0.034$. The softer cells maintain this positive
inclination and stay away from the pillars. Additionally, the cell with $C_a = 34$ has higher inclination angle than the one with $C_a = 0.34$ and moves
farther away from the pillars than the one with $C_a = 0.34$ (see the gaps in
both figures). ~\cite{zhang-fedosov-e15} also observed that the RBCs stay away
from the pillars with square cross-sections and attributed this behavior to the
fact that the flow resembles a confined Poiseuille flow and hence the cells
migrate. We discuss the similarities between the flow in DLD and confined
Poiseuille flow in~\secref{s:discussion} and compare the cells' inclination angles
for various capillary numbers and viscosity contrasts.

\subsection{Effects of viscosity contrast}\label{s:vcSep}

\begin{figure}
 \begin{minipage}{\textwidth}
\centering
\setcounter{subfigure}{0}
\renewcommand*{\thesubfigure}{(a)} 
      \hspace{0cm}\subfigure[$\nu = 1$ (blue cell) and $\nu = 2$ (red cell)]{\scalebox{0.69}{{\includegraphics{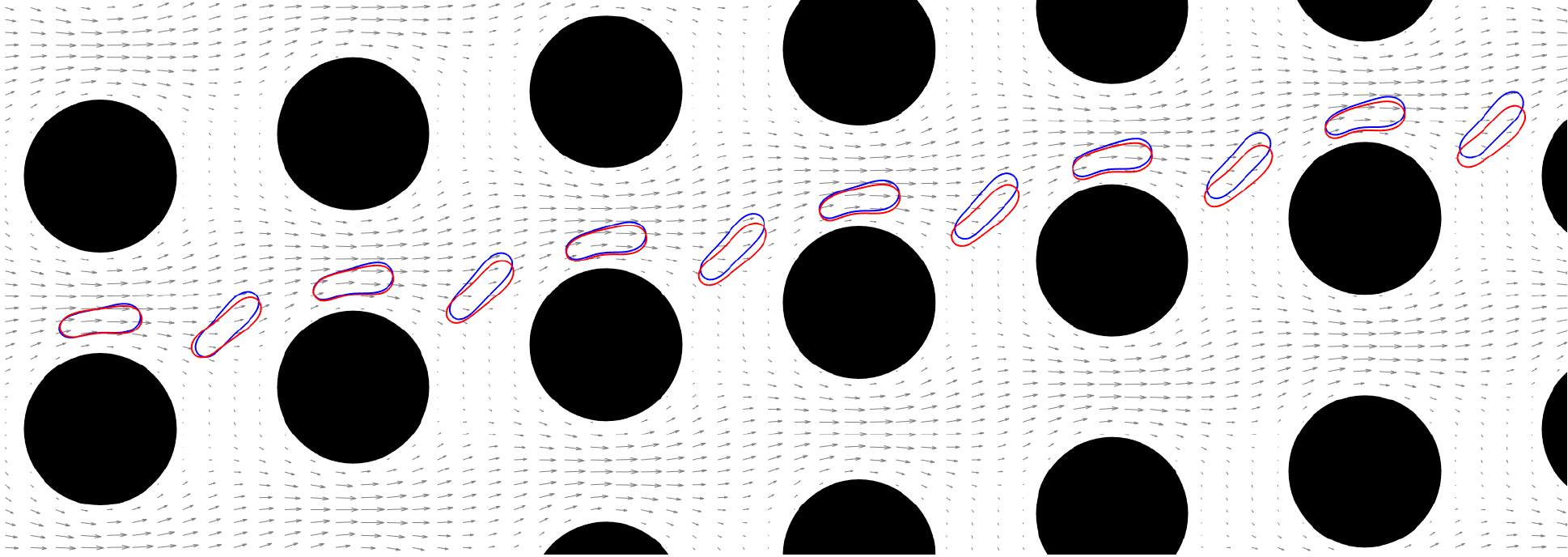}}}
      \label{f:N6VCLowTrajs}} 
\end{minipage}
\begin{minipage}{\textwidth}
\centering
\setcounter{subfigure}{0}      
\renewcommand*{\thesubfigure}{(b)} 
      \hspace{0cm}\subfigure[$\nu = 1$ (blue cell) and $\nu = 5$ (red cell)]{\scalebox{0.69}{{\includegraphics{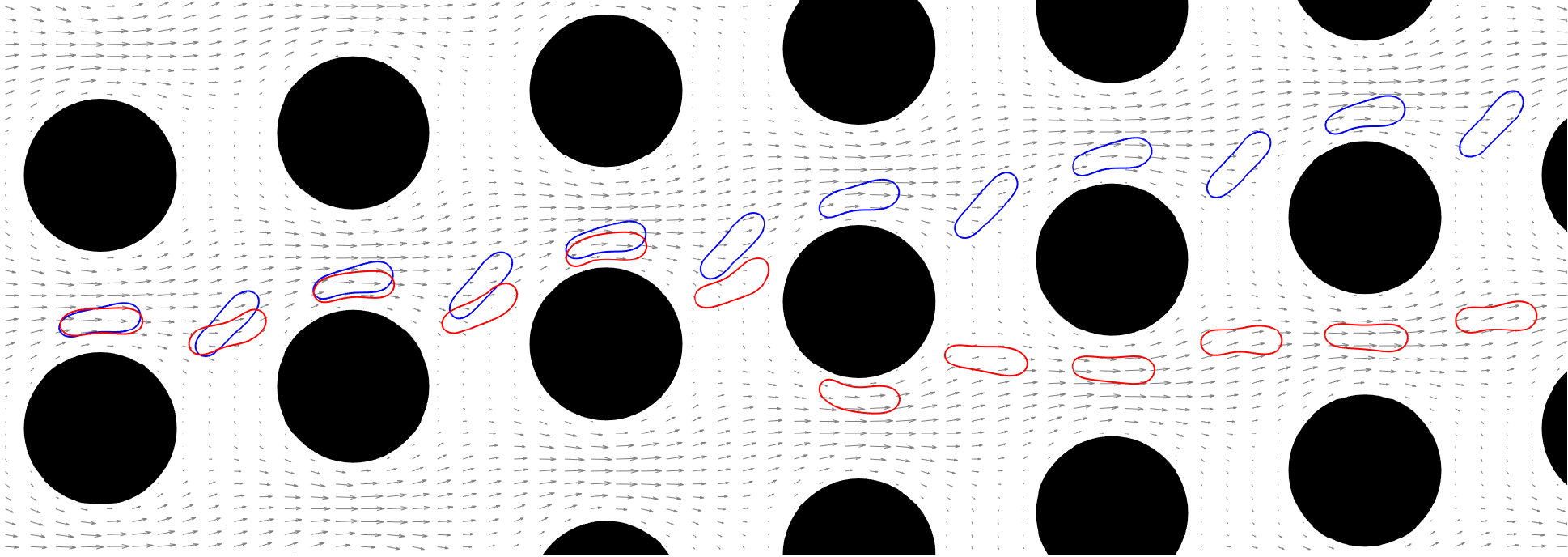}}}
      \label{f:N6VCHighTrajs}}
\end{minipage}                    
\mcaption{Superimposed snapshots from the simulations of RBCs having different viscosity contrasts $\nu$ and the unperturbed velocity field (shown with gray arrows). Here, we initialize the cells at the same position and simulate them separately. We, then, superimpose the snapshots taken when the cells are in the gaps and between two consecutive gaps. The capillary number and the row-shift fraction in these simulations are $C_a = 3.41$ and $\epsilon = 0.1667$, respectively. At the top, we compare the dynamics of the less viscous cells with $\nu = 1$ (blue) and $\nu = 2$ (red). Both of these cells displace. At the bottom, we compare the cell having $\nu = 1$ with the one with $\nu = 5$. The cell with $\nu = 5$ corresponds to a healthy RBC and it zig-zags. }{f:VCTrajs}
\end{figure}

In the previous section, we considered cells with no viscosity contrast. We, now, want to investigate how the cells' dynamics in DLD depend on their viscosity contrast. We performed simulations of a single RBC flowing through DLD with the viscosity contrast values $\nu = (1, 2, 5)$. The capillary number for these cells is $C_a = 3.41$. We considered the same device as in the previous section, i.e., the row-shift fraction is $\epsilon = 0.1667$. We present the superimposed snapshots from these simulations in~\figref{f:VCTrajs} together with the unperturbed velocity field (shown with gray arrows). Here, the snapshots are taken when the cells are passing through the gaps and between two consecutive gaps. The cells are initialized at the same location and have the effective radius $R_{\mathrm{eff}} = 2.4\mu$m.

The RBCs here have a moderate capillary number so they can be considered more deformable than the stiff cells in the previous section. In~\figref{f:N6VCLowTrajs} we compare cells with no viscosity contrast (i.e., $\nu = 1$, blue) and with the viscosity contrast $\nu = 2$ (red). Here, both cells displace. As observed in the previous section, both cells have thick heads and thin tails in the gaps. Additionally, they have positive inclination angle with respect to the flow direction in the gaps. Their inclination angle increases when they are between two consecutive gaps. Then, due to this asymmetry and the shear gradient, they migrate away from the pillars. Since they stay sufficiently away from the pillars, they displace. In~\figref{f:N6VCHighTrajs} we compare the cell having no viscosity contrast with the one having the viscosity contrast $\nu = 5$. The cell with $\nu = 5$ and this capillary number might be considered a healthy cell and it zig-zags in this device. Although the more viscous cell (red one) still maintains a positive inclination with respect to the flow direction, this angle is lower than the one with no viscosity contrast (see the cells right after the first gap). As a result of having smaller inclination angle, the more viscous cell cannot remain as far away from the pillars as the cell with no viscosity contrast does (see the cells in the third gap for instance). So the more viscous cell eventually swaps a lane. 

It is known that the cells show either tank-treading (moving with the same inclination) or tumbling (moving while rotating around its axis) depending on its viscosity contrast and capillary number. In free shear flows, the cells with $\nu = 5$ tumble~\citep{kantsler-steinberg-e06}. So one might expect the same cell to tumble in DLD as well. However, the confinement reduces the inclination angle of a tank-treading cell and also delays the tumbling. It turns out that tumbling does not occur for the confinement level we have in the gaps in this study, which is $2R_{\mathrm{eff}}/G = 0.48$ (See~\cite{kaoui-harting-e12}). That's why, the cells in DLD always maintain positive inclination and the inclination decreases as the viscosity contrast increases. We discuss how the inclination angle varies with viscosity contrast in both confined Poiseuille flows and DLD in detail in~\secref{s:discussion}.

{\em Remark.} 
Another consequence of aging besides increasing viscosity contrast is that RBCs lose surface area (arc-length in 2D)~\citep{mohandas-gallagher08}, so the reduced area increases. In shear and Poiseuille flows, cell's migration velocity and inclination angle decrease as the reduced area increases. In the circular limit (the reduced area of 1), cell does not deform and migrate. Since cell dynamics in DLD are different than those simpler flows, how the transport modes depend on the reduced area needs to be investigated. In order to shed some light on this issue we performed simulations of cells with $C_a = 3.41$ and $\nu = (1, 10)$, and DLD devices with $\epsilon = (0.1, 0.125, 0.167)$. We increased the reduced area from 0.65 to 0.75, 0.85 and 0.95. Recall that the reduced area is 0.65 for the other experiments in the study. We found that the cell with $\nu = 1$ displaces for all the $\epsilon$ values regardless of its reduced area. The cell with $\nu = 10$ zig-zags for these $\epsilon$ values when the reduced area is 0.65. For greater reduced areas the cell tumbles for $\epsilon = (0.1, 0.125)$ and displaces while it still zig-zags for $\epsilon = 0.1667$. Our conclusion is that the cell's transport mode in DLD can change with its reduced area depending on the row-shift fraction $\epsilon$ and the viscosity contrast $\nu$. This shows that a more detailed study of the reduced area effects is required but such a study is beyond the scope of our work.

\subsection{Phase diagrams}\label{s:phase}

We performed an analysis in order to determine how the cells' transport modes depend on their capillary number $C_a$, viscosity contrast $\nu$ and the DLD's row-shift fraction $\epsilon$. We obtained three phase diagrams for the transport modes of a single cell in DLD which are described below.

\begin{enumerate}
\item\ $\epsilon$ vs. $C_a$ (at the top in~\figref{f:CaVCPhaseDiag}): We fixed the viscosity contrast to $\nu = 10$ and performed simulations for $\epsilon \in [0.0625, 2.5]$ and $C_a \in [3.4\times 10^{-3}, 3.4\times 10^{-1}]$. 
\item\ $\epsilon$ vs. $\nu$ (at the bottom in~\figref{f:CaVCPhaseDiag}): We fixed the capillary number to $C_a = 3.4\times 10^{-1}$ and performed simulations for $\epsilon \in [0.0625, 2.5]$ and $\nu \in [1, 10]$. 
\item\ $\nu$ vs. $C_a$ (in~\figref{f:CavsVCPhaseDiag}): We fixed the row-shift fraction to $\epsilon = 0.1667$ (at the top) and $\epsilon = 0.1$ (at the bottom). Then, we performed simulations for $\nu = (1, 2, 5, 8, 10, 100)$ and $C_a = [3.4\times 10^{-2}, 3.4\times 10^{3}]$. 
\end{enumerate}

\begin{figure}
\begin{center}
\begin{tabular}{c}
\begin{tikzpicture}

\begin{axis}[
  width = 0.65\textwidth,
  xmin = 0.0025,
  xmax = 0.5, 
  ymin = 0.05,
  ymax = 0.26,
  xtick = {3.4E-3,3.4E-2,3.4E-1},
  xticklabels = {$3.4$E$-3$,$3.4$E$-2$,$3.4$E$-1$},
  ytick = {0.05,0.1,0.15,0.2,0.25},
  yticklabels = {$0.05$,$0.1$,$0.15$,$0.2$,$0.25$},
  ylabel = {Row-shift fraction ($\epsilon$)},
  xlabel = {Capillary number ($C_a$)},
  xmode = log,
  label style = {font=\normalsize},
  legend style={at={(0.5,1)},anchor=south,legend columns=-1,draw=none},
  legend entries = {Zig-zag, Displacement, $\epsilon = 0.014{(C_a)}^{-0.48}$},
  ]

\addplot [color=red,only marks,mark=*,mark size=2.5pt] table{
0.0034 0.25
0.0034 0.2
0.0034 0.1667
0.0034 0.1429
0.0034 0.125
0.0034 0.1111
0.0034 0.1
0.0034 0.0833
0.0034 0.0625
0.0068 0.25
0.0068 0.2
0.0068 0.1667
0.014 0.25
0.014 0.2
0.014 0.1667
0.014 0.1429
0.014 0.125
0.014 0.1111
0.034 0.25
0.034 0.2
0.034 0.1667
0.034 0.1429
0.034 0.125
0.034 0.1111
0.034 0.1
0.034 0.0833
0.068 0.25
0.068 0.2
0.068 0.1667
0.068 0.1429
0.068 0.125
0.068 0.1111
0.068 0.1
0.068 0.0833
0.068 0.0625
0.17 0.25
0.17 0.2
0.17 0.1667
0.17 0.1429
0.17 0.125
0.17 0.1111
0.17 0.1
0.17 0.0833
0.17 0.0625
0.34 0.25
0.34 0.2
0.34 0.1667
0.34 0.1429
0.34 0.125
0.34 0.1111
0.34 0.1
0.34 0.0833
0.34 0.0625
}; 

\addplot [color=blue,only marks,mark=triangle*, mark size=2.5pt] table{

0.0068 0.1429
0.0068 0.1250
0.0068 0.1111
0.0068 0.1
0.0068 0.0833
0.0068 0.0625
0.014 0.1
0.014 0.0833
0.014 0.0625
0.034 0.0625
}; 
\addplot [mark=none,black,line width=1] table{
       0.0068    0.1536
    0.0071    0.1508
    0.0073    0.1480
    0.0076    0.1454
    0.0079    0.1430
    0.0082    0.1407
    0.0084    0.1384
    0.0087    0.1363
    0.0090    0.1343
    0.0093    0.1324
    0.0095    0.1306
    0.0098    0.1288
    0.0101    0.1271
    0.0104    0.1255
    0.0106    0.1239
    0.0109    0.1224
    0.0112    0.1209
    0.0115    0.1195
    0.0117    0.1182
    0.0120    0.1169
    0.0123    0.1156
    0.0126    0.1144
    0.0128    0.1132
    0.0131    0.1121
    0.0134    0.1110
    0.0137    0.1099
    0.0139    0.1089
    0.0142    0.1078
    0.0145    0.1068
    0.0148    0.1059
    0.0150    0.1050
    0.0153    0.1041
    0.0156    0.1032
    0.0159    0.1023
    0.0161    0.1015
    0.0164    0.1006
    0.0167    0.0998
    0.0170    0.0991
    0.0172    0.0983
    0.0175    0.0976
    0.0178    0.0968
    0.0181    0.0961
    0.0183    0.0954
    0.0186    0.0948
    0.0189    0.0941
    0.0192    0.0934
    0.0194    0.0928
    0.0197    0.0922
    0.0200    0.0916
    0.0203    0.0910
    0.0205    0.0904
    0.0208    0.0898
    0.0211    0.0892
    0.0214    0.0887
    0.0216    0.0882
    0.0219    0.0876
    0.0222    0.0871
    0.0225    0.0866
    0.0227    0.0861
    0.0230    0.0856
    0.0233    0.0851
    0.0236    0.0846
    0.0238    0.0842
    0.0241    0.0837
    0.0244    0.0832
    0.0247    0.0828
    0.0249    0.0824
    0.0252    0.0819
    0.0255    0.0815
    0.0258    0.0811
    0.0260    0.0807
    0.0263    0.0803
    0.0266    0.0799
    0.0269    0.0795
    0.0271    0.0791
    0.0274    0.0787
    0.0277    0.0783
    0.0280    0.0780
    0.0282    0.0776
    0.0285    0.0772
    0.0288    0.0769
    0.0291    0.0765
    0.0293    0.0762
    0.0296    0.0758
    0.0299    0.0755
    0.0302    0.0752
    0.0304    0.0748
    0.0307    0.0745
    0.0310    0.0742
    0.0313    0.0739
    0.0315    0.0736
    0.0318    0.0733
    0.0321    0.0730
    0.0324    0.0727
    0.0326    0.0724
    0.0329    0.0721
    0.0332    0.0718
    0.0335    0.0715
    0.0337    0.0712
    0.0340    0.0710
};

\end{axis}

\end{tikzpicture} \\ 
\begin{tikzpicture}

\begin{axis}[
  width = 0.65\textwidth,
  xmin = 0.5,
  xmax = 10.5, 
  ymin = 0.05,
  ymax = 0.26,
  xtick = {1,2,3,4,5,6,7,8,9,10},
  ytick = {0.05,0.1,0.15,0.2,0.25},
  yticklabels = {$0.05$,$0.1$,$0.15$,$0.2$,$0.25$},
  ylabel = {Row-shift fraction ($\epsilon$)},
  xlabel = {Viscosity contrast ($\nu$)},
  label style = {font=\normalsize},
  legend style={at={(0.5,1)},anchor=south,legend columns=-1,draw=none},
  legend entries = {Zig-zag, Displacement, $\epsilon = -0.29\nu^{0.25}+0.57$},
  ]

\addplot [color=red,only marks,mark=*,mark size=2.5pt] table{
2 0.25
3 0.25
3 0.2
4 0.25
4 0.2
4 0.1667
5 0.25
5 0.2
5 0.1667
5 0.1429
6 0.25
6 0.2
6 0.1667
6 0.1429
6 0.125
7 0.25
7 0.2
7 0.1667
7 0.1429
7 0.125
7 0.1111
8 0.25
8 0.2
8 0.1667
8 0.1429
8 0.125
8 0.1111
8 0.1
8 0.08333
9 0.25
9 0.2
9 0.1667
9 0.1429
9 0.125
9 0.1111
9 0.1
9 0.08333
9 0.0625
10 0.25
10 0.2
10 0.1667
10 0.1429
10 0.125
10 0.1111
10 0.1
10 0.08333
10 0.0625
}; 
\addplot [color=blue,only marks,mark=triangle*, mark size=2.5pt] table{
1 0.0625
1 0.0833
1 0.1
1 0.1111
1 0.125
1 0.1429
1 0.1667
1 0.2 
1 0.25
2 0.0625
2 0.0833
2 0.1
2 0.1111
2 0.125
2 0.1429
2 0.1667
2 0.2
3 0.0625
3 0.0833
3 0.1
3 0.1111
3 0.125
3 0.1429
3 0.1667
4 0.0625
4 0.0833
4 0.1
4 0.1111
4 0.125
4 0.1429
5 0.0625
5 0.0833
5 0.1
5 0.1111
5 0.125
6 0.0625
6 0.0833
6 0.1
6 0.1111
7 0.0625
7 0.0833
7 0.1
8 0.0625

}; 
\addplot [mark=none,black,line width=1] table{
    2.0000    0.2251
    2.1224    0.2200
    2.2449    0.2150
    2.3673    0.2103
    2.4898    0.2057
    2.6122    0.2013
    2.7347    0.1971
    2.8571    0.1930
    2.9796    0.1890
    3.1020    0.1851
    3.2245    0.1814
    3.3469    0.1778
    3.4694    0.1742
    3.5918    0.1708
    3.7143    0.1674
    3.8367    0.1641
    3.9592    0.1609
    4.0816    0.1578
    4.2041    0.1547
    4.3265    0.1518
    4.4490    0.1488
    4.5714    0.1460
    4.6939    0.1431
    4.8163    0.1404
    4.9388    0.1377
    5.0612    0.1350
    5.1837    0.1324
    5.3061    0.1299
    5.4286    0.1273
    5.5510    0.1249
    5.6735    0.1224
    5.7959    0.1200
    5.9184    0.1177
    6.0408    0.1154
    6.1633    0.1131
    6.2857    0.1108
    6.4082    0.1086
    6.5306    0.1064
    6.6531    0.1042
    6.7755    0.1021
    6.8980    0.1000
    7.0204    0.0979
    7.1429    0.0959
    7.2653    0.0939
    7.3878    0.0919
    7.5102    0.0899
    7.6327    0.0880
    7.7551    0.0861
    7.8776    0.0842
    8.0000    0.0823

};

\end{axis}

\end{tikzpicture} \\
\end{tabular}
\end{center}
\mcaption{Phase diagrams for the transport modes as a function of the capillary number $C_a$ and row-shift fraction $\epsilon$ with $\nu = 10$ (at the top) and as a function of the viscosity contrast $\nu$ and $\epsilon$ with $C_a = 0.34$ (at the bottom). Red circles and blue triangles indicate zig-zagging and displacing cells, respectively. The solid line at the top corresponds to
\eqref{e:CaEpsPower} and the one at the bottom corresponds to~\eqref{e:VCEpsPower}. These equations approximate the separation between two
transport modes.}{f:CaVCPhaseDiag}
\end{figure}
We, first, discuss the phase diagrams for $C_a$ vs. $\epsilon$ and $\nu$ vs. $\epsilon$ in~\figref{f:CaVCPhaseDiag}. The top figure shows that as the capillary number increases, the transport mode shifts from zig-zag to displacement and for larger capillary numbers it again shifts to zig-zag. This is not observed at the bottom figure. For low viscosity contrasts the cell displaces and as the viscosity contrast increases, the transport mode shifts to zig-zag. Since rigidity of a cell increases as its viscosity contrast increases or its capillary number decreases, one would expect the same dynamics as a cell becomes rigid. Our results agree with this expectation. The phase diagrams show that a cell has the same transport mode in the limit of low capillary number and high viscosity contrast. This transport mode is zig-zag. This is reasonable because a very rigid cell behaves like a rigid spherical particle with a diameter equal to the cell's thickness $\approx 2.5\mu$m (i.e., it cannot migrate and its effective size is $\approx 2.5\mu$m) and the critical particle size given by~\eqref{e:rigidEmpric} for the row-shift fractions considered here is greater than the cell's thickness (i.e., $D_c = 3.7\mu$m). So the rigid cell zig-zags. The reason why we observe different transport modes depending on the capillary number and viscosity contrast is that a cell's inclination angle depends on these parameters and it migrates some amount which depends on its inclination angle as observed in the previous sections. However, the rigid cell does not migrate. The curves separating the transport modes in these figures are given by a power law~\eqref{e:CaEpsPower} (for the top figure) and~\eqref{e:VCEpsPower} (for the bottom figure). 
\begin{subequations} \label{e:approxCurves}
\begin{alignat}{1}
\epsilon & = 0.014{C_a}^{-0.48}, \label{e:CaEpsPower} \\
\epsilon & = -0.29{\nu}^{0.25}+0.26. \label{e:VCEpsPower}
\end{alignat}%
\end{subequations}
$R^2$ is the coefficient of determination, which indicates the goodness of a
fit. The curve fits~\eqref{e:CaEpsPower} and~\eqref{e:VCEpsPower} have $R^2 = 0.92$ and $R^2 = 0.91$, respectively.

\begin{figure}
\begin{center}
\begin{tabular}{c}
\begin{tikzpicture}

\begin{axis}[
  width = 0.65\textwidth,
  xmin = 1E-2,
  xmax = 5E+3, 
  ymin = 0.5,
  ymax = 10.5,
  xtick = {1E-2,1E-1,1,1E+1,1E+2,1E+3},
  ytick = {1,2,3,4,5,6,7,8,9,10},
  yticklabels = {$1$,$2$,$ $,$ $,$5$,$ $,$ $,$8$,$ $,$10$},
  ylabel = {Viscosity contrast ($\nu$)},
  xlabel = {Capillary number ($C_a$)},
  xmode = log,
  label style = {font=\normalsize},
  legend style={at={(0.5,1)},anchor=south,legend columns=-1,draw=none},
  legend entries = {Zig-zag, Displacement},
  ]
\addplot [color=red,only marks,mark=*,mark size=2.5pt] table{
3.41E+3 5
3.41E+3 8
3.41E+3 10
3.41E+2 5
3.41E+2 8
3.41E+2 10
3.41E+1 5
3.41E+1 8
3.41E+1 10
3.41E+0 5
3.41E+0 8
3.41E+0 10
3.41E-1 5
3.41E-1 8
3.41E-1 10
3.41E-2 1
3.41E-2 2
3.41E-2 5
3.41E-2 8
3.41E-2 10

}; 

\addplot [color=blue,only marks,mark=triangle*, mark size=2.5pt] table{
3.41E+3 1
3.41E+3 2
3.41E+2 1
3.41E+2 2
3.41E+1 1
3.41E+1 2
3.41E+0 1
3.41E+0 2
3.41E-1 1
3.41E-1 2
};

\end{axis}

\end{tikzpicture}\\ 
\begin{tikzpicture}

\begin{axis}[
  width = 0.65\textwidth,
  xmin = 1E-2,
  xmax = 5E+3, 
  ymin = 0.5,
  ymax = 10.5,
  xtick = {1E-2,1E-1,1,1E+1,1E+2,1E+3},
  ytick = {1,2,3,4,5,6,7,8,9,10},
  yticklabels = {$1$,$2$,$ $,$ $,$5$,$ $,$ $,$8$,$ $,$10$},
  ylabel = {Viscosity contrast ($\nu$)},
  xlabel = {Capillary number ($C_a$)},
  xmode = log,
  label style = {font=\normalsize},
  legend style={at={(0.5,1)},anchor=south,legend columns=-1,draw=none},
  legend entries = {Zig-zag, Displacement},
  ]
\addplot [color=red,only marks,mark=*,mark size=2.5pt] table{
3.41E+3 8
3.41E+3 10
3.41E+2 8
3.41E+2 10
3.41E+1 8
3.41E+1 10
3.41E+0 5
3.41E+0 8
3.41E+0 10
3.41E-1 8
3.41E-1 10
6.83E-2 8
6.83E-2 10
3.41E-2 5
3.41E-2 8
3.41E-2 10

}; 

\addplot [color=blue,only marks,mark=triangle*, mark size=2.5pt] table{
3.41E+3 1
3.41E+3 2
3.41E+3 5
3.41E+2 1
3.41E+2 2
3.41E+2 5
3.41E+1 1
3.41E+1 2
3.41E+1 5
3.41E+0 1
3.41E+0 2
3.41E-1 1
3.41E-1 2
3.41E-1 5
6.83E-2 1
6.83E-2 2
6.83E-2 5
3.41E-2 1
3.41E-2 2
};

\end{axis}

\end{tikzpicture} \\
\end{tabular}
\end{center}
\mcaption{Phase diagrams for the transport modes as a function of the viscosity contrast $\nu$ and capillary number $C_a$ for the row-shift fractions $\epsilon = 0.1667$ (at the top) and $\epsilon = 0.1$ (at the bottom). Red circles and blue triangles indicate zig-zagging and displacing cells, respectively.}{f:CavsVCPhaseDiag}
\end{figure}
\figref{f:CavsVCPhaseDiag} indicates how the transport mode depends on the capillary number and viscosity contrast for a fixed row-shift fraction. Here, the top figure is for the row-shift fraction $\epsilon = 0.1667$ and the bottom one is for $\epsilon = 0.1$. We observe that in DLD with higher row-shift fraction cells zig-zag more. That is, under the same conditions ($\nu$ and $C_a$) a cell zig-zags for high $\epsilon$ and displaces for low $\epsilon$. For example, the mode is zig-zag in $\epsilon = 0.1667$ and displacement in $\epsilon = 0.1$ for $\nu = 5$ and $C_a = 3.4 \times 10^{-1}$. The reason is as the row-shift fraction increases, the width of the adjacent stream (and hence the critical size) increases. In order for a cell to displace for that case higher migration velocity (i.e. higher inclination angle) is required. That's why, for the same viscosity contrasts and capillary numbers the top figure ($\epsilon = 0.1667$) in~\figref{f:CavsVCPhaseDiag} has more zig-zagging cases than the bottom figure ($\epsilon = 0.1$). For $\epsilon = 0.1667$, only the cells with the viscosity contrasts $\nu = (1,2)$ displace if their capillary number is greater than $3.4\times10^{-2}$. So the viscosity contrast based separation is possible for these cells. However, the capillary number based separation is only possible for viscosity contrasts $\nu = (1, 2)$ and for very low capillary numbers. For $\epsilon = 0.1$, the transport mode shifts from zig-zag to displacement and then to zig-zag again as the capillary number increases for the viscosity contrast $\nu \geq 5$. So, the separation based on the capillary number is possible for higher capillary numbers than it is for $\epsilon = 0.1667$. 

\subsection{Breakdown of DLD efficiency in dense suspensions}\label{s:dense}

\begin{figure}
 \begin{minipage}{\textwidth}
 \centering
\setcounter{subfigure}{0}
\renewcommand*{\thesubfigure}{(a)} 
      \hspace{-0.3cm}\subfigure[$t = 71.4\tau$]{\scalebox{0.325}{{\includegraphics{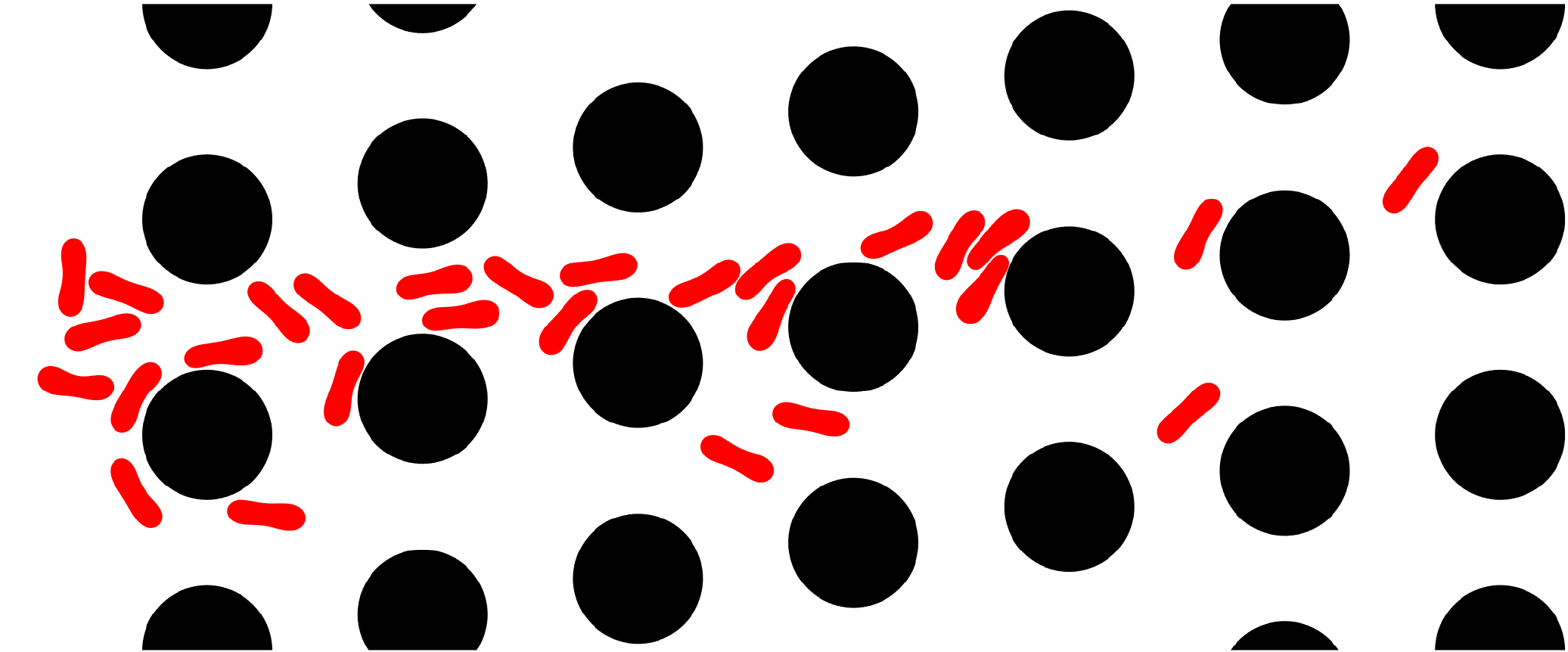}}}
      \label{f:streamt30}} 
\setcounter{subfigure}{0}      
\renewcommand*{\thesubfigure}{(b)} 
      \hspace{-0.3cm}\subfigure[$t = 95.2\tau$]{\scalebox{0.325}{{\includegraphics{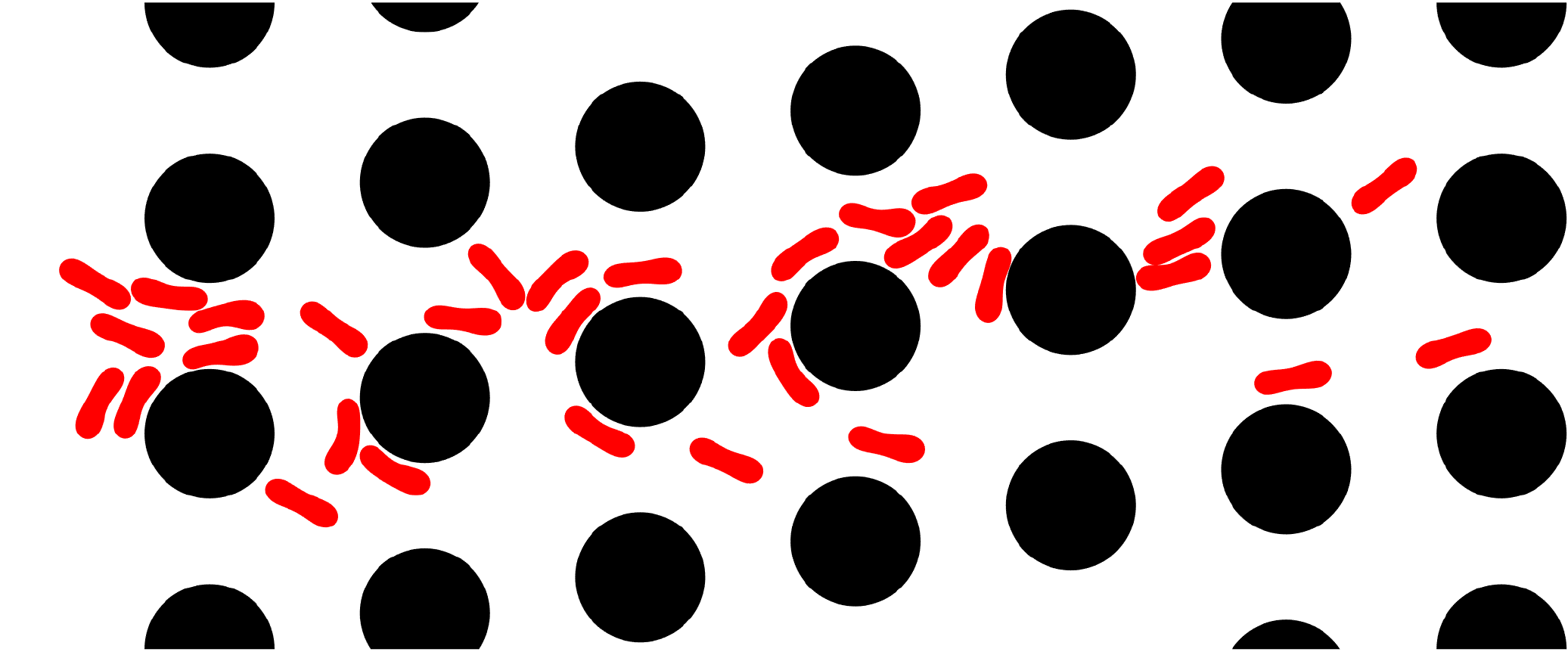}}}
      \label{f:streamt40}}
\end{minipage}          
\begin{minipage}{\textwidth}
\centering
\setcounter{subfigure}{0}
\renewcommand*{\thesubfigure}{(c)} 
      \hspace{-0.3cm}\subfigure[$t = 107.1\tau$]{\scalebox{0.325}{{\includegraphics{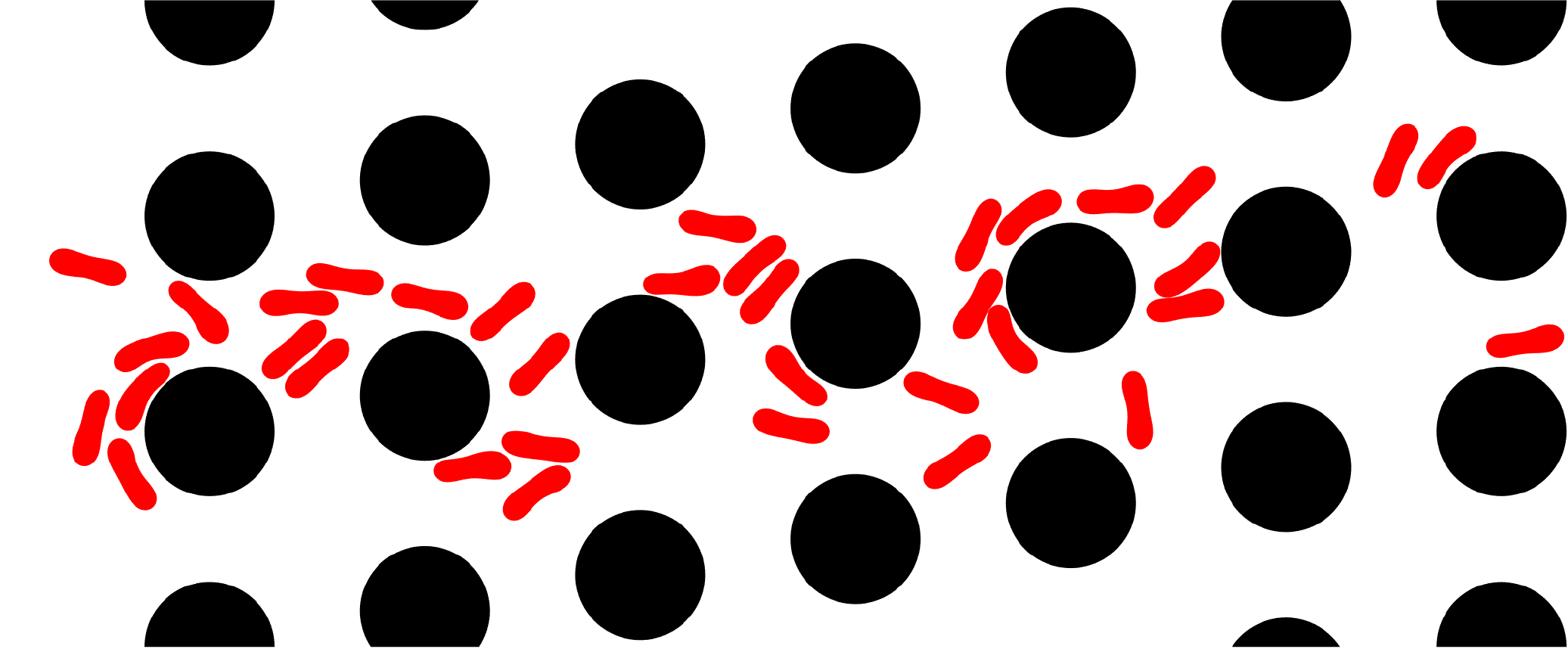}}}
      \label{f:streamt45}}
\setcounter{subfigure}{0}
\renewcommand*{\thesubfigure}{(d)} 
      \hspace{-0.3cm}\subfigure[$t = 119.1\tau$]{\scalebox{0.325}{{\includegraphics{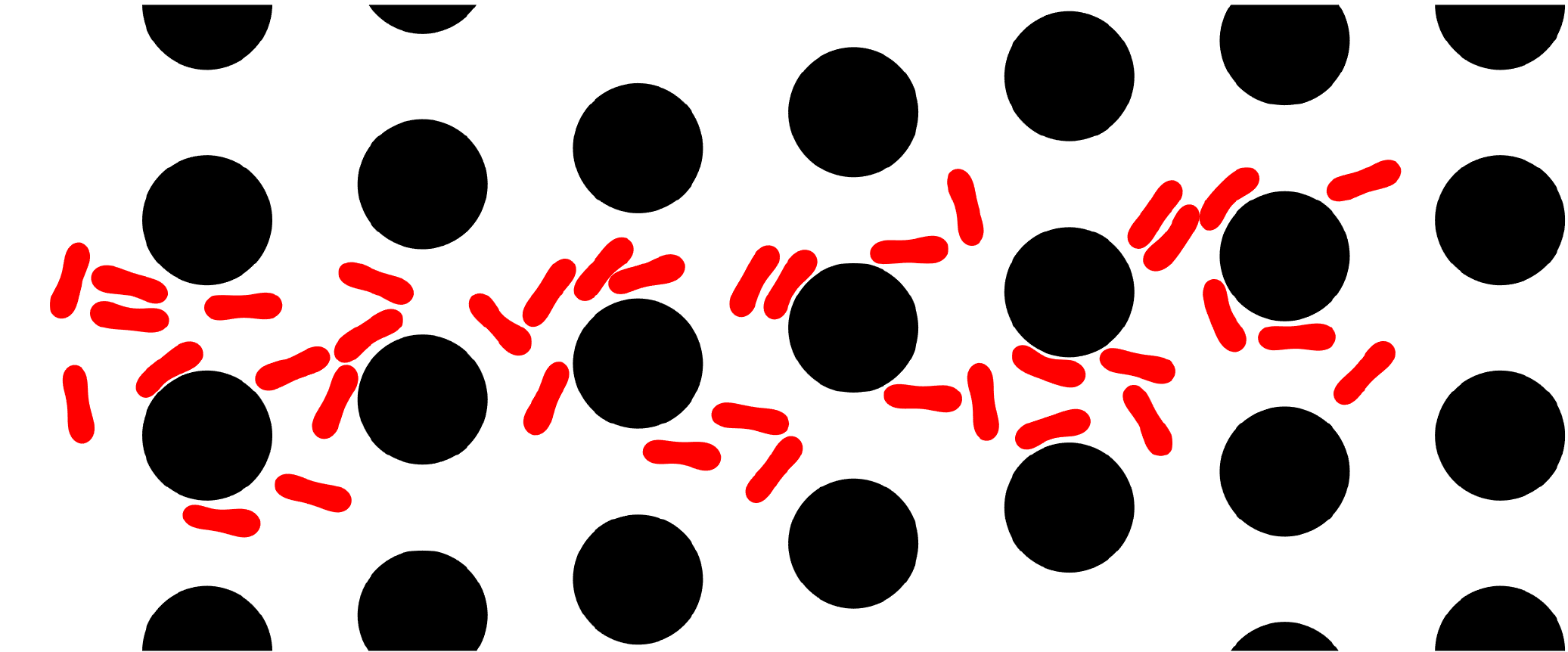}}}
      \label{f:streamt50}}
\end{minipage}  
\begin{minipage}{\textwidth}
\centering
\setcounter{subfigure}{0}
\renewcommand*{\thesubfigure}{(e)} 
      \hspace{-0.3cm}\subfigure[$t = 131\tau$]{\scalebox{0.325}{{\includegraphics{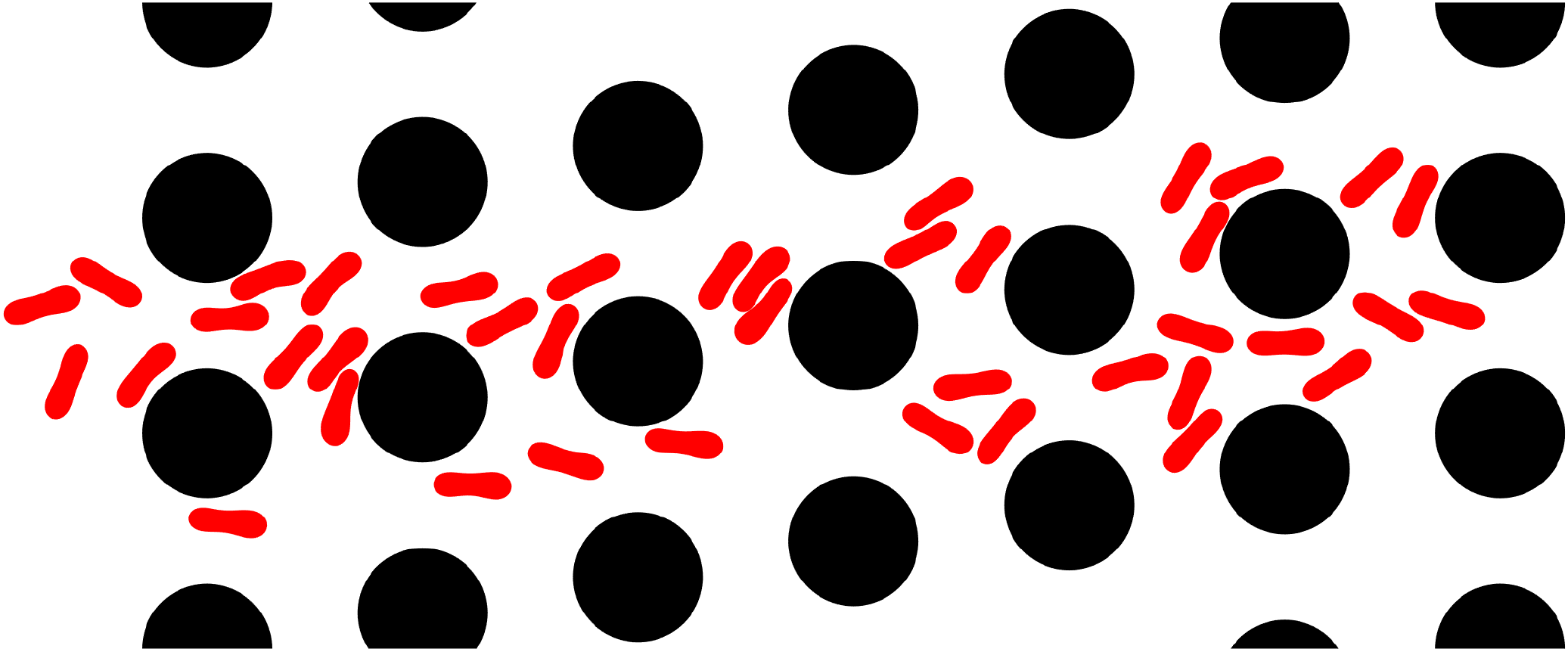}}}
      \label{f:streamt55}}
\setcounter{subfigure}{0}
\renewcommand*{\thesubfigure}{(f)} 
      \hspace{-0.3cm}\subfigure[$t = 142.9\tau$]{\scalebox{0.325}{{\includegraphics{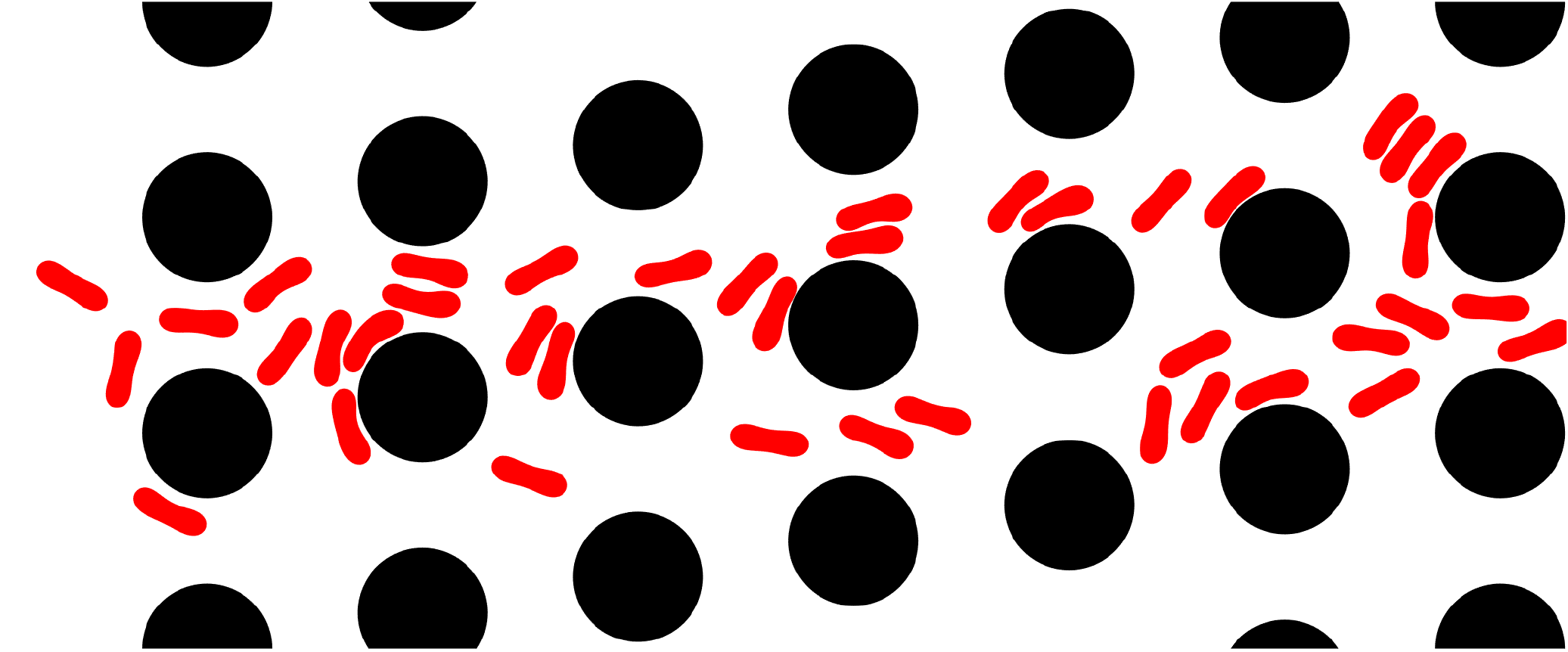}}}
      \label{f:streamt60}}   
 \end{minipage}    
          
\mcaption{Frames from a simulation of a dense RBC suspension for the row shift
fraction $\epsilon = 0.1667$. The suspension has a volume fraction of
15\%. All RBCs in the suspension have the same capillary number $C_a = 0.34$ 
and viscosity contrast $\nu = 1$. The phase diagram
in~\figref{f:CaVCPhaseDiag} shows that a single RBC with these
parameters displaces. In its dense suspension, however, only 30\% of
the RBCs displace. $\tau$ is the time scale defined as the inverse of
the shear gradient in a gap, i.e., $\tau = G/2U_{\max}$.
See~\figref{f:streamFails} for the statistics of the dense
suspensions.}{f:N6Stream3VC1}
\end{figure}

The DLD theory is developed for dilute suspensions assuming that the
unperturbed velocity field (in the absence of particles) is not
distorted when the particles are present. So that the particle-pillar
interactions result in deterministic particle trajectories and hence
separation. Additionally, the phase diagrams for deformability-based
RBC separation in our study and the literature~\citep{quek-chiam-e11,kruger-coveney-e14} 
are obtained simulating a single particle in a
DLD device. As the suspensions become denser, the particle-particle
interactions dominate the particle-pillar interactions and separation
may not occur anymore because the particles of the same properties are
not in the same transport modes~\citep{vernekar-kruger15}. The
DLD efficiency for dense suspensions is of interest because the use
of dense suspensions would reduce the operation time and remove the
pre-treatment (such as dilution) requirement. 
Inspired by~\citet{vernekar-kruger15} we
are interested in failure of the transport modes in dense suspensions
of RBCs in deep devices. In order to explain the failure let's
consider a single RBC with a certain viscosity contrast and capillary
number and suppose that this RBC displaces in its dilute suspension
for a certain row-shift fraction. When a dense suspension of this RBC
flows in the same device, not all of the cells displace. This is called
failure in the displacement mode. Similarly we define failure in the
zig-zag mode, too. \citet{vernekar-kruger15} performed 3D simulations
of the flow of dense RBC suspensions in shallow devices for various
capillary numbers, row-shift fractions and volume fractions under a
constant viscosity contrast of $\nu = 5$. The key findings of their
study are:

\begin{enumerate}
\item\ As the volume fraction increases for the same row-shift fraction and
capillary number, more RBCs zig-zag independent of the transport mode
of a single RBC under the same conditions. So the displacement mode is
more prone to failure than the zig-zag mode. Consequently, it is
easier to separate large particles such as white blood cells from
dense suspensions of RBCs than the small particles (e.g., platelets),
since the small particles also zig-zag like the RBCs.
 
\item\ At the same volume fraction and capillary number increasing the row-
shift fraction reduces the failure rate of the displacement mode in
one period of the device. However, a DLD device with small row-shift
fraction must be long enough to have significant lateral separation
between displacing and zig-zagging particles at the end. This requires
using many periods of the device, which increases the failure rate in
the overall device. Therefore, the use of small row-shift fractions
does not prevent the breakdown.
\end{enumerate}
\citet{inglis-nordon-e11} also made the same observations in their 
experimental study.

Here, we performed numerical simulations of dense suspensions in deep
devices. We, first, considered RBCs with two different capillary
numbers $C_a = (0.0034,0.34)$ but a fixed viscosity contrast of $\nu = 10$ 
for the row-shift fractions 
$\epsilon = (0.0833, 0.125, 0.1667)$. 
Secondly, we conducted the same experiments with two
different viscosity contrasts $\nu = (1, 10)$ but a fixed capillary
number of $C_a = 0.34$. As in the previous sections, we observed the
transport mode in one period of the device. We filled the middle lane
of the device by randomly initialized RBCs. Then as the simulation
went on we initialized new RBCs at a random lateral position at the
leftmost of the device randomly in time. We ran the simulations until
100 RBCs in total traveled to the end of the device. We define and calculate the volume fraction as follows. Considering a circle centered at each cell's center with diameter equal to the cell's arc-length, we compute 
the so-called local volume fraction which is the ratio of the area occupied by 
the cells in the circle to the area of the circle. Finally, taking the average 
of the local volume fraction over time and cells delivers the volume fraction 
of the simulation. In~\figref{f:N6Stream3VC1} we present the frames from the simulation
of RBCs with $C_a = 0.34$ and $\nu = 1$ for the row-shift fraction
$\epsilon = 0.1667$. We report the numbers of zig-zagging and
displacing RBCs in the dense suspensions in~\figref{f:streamFails}.
The dense suspensions in our simulations had volume fractions of
15\%-20\%. We found out that one RBC interacts with approximately two
to three other RBCs in average throughout its motion in our
simulations. The statistics in~\figref{f:streamFails} agree well with
the findings of~\citet{inglis-nordon-e11,vernekar-kruger15}. We
summarize our results:

\begin{enumerate}
\item\ The displacement mode due to either viscosity contrast or capillary number is more susceptible to failure. That is, more
than 50\% of the RBCs in a dense suspension zig-zags for the two
largest row-shift fractions (the first two columns
in~\figref{f:streamFails}) while the transport mode of the RBC with
the same properties is displacement in its dilute suspension(see the
first and the last rows). Whereas only 15\%-20\% of the zig-zagging
RBCs fails to zig-zag (see the second row).

\item\ For the smaller row-shift fractions (from left to right in the same
figure), the breakdown is less pronounced. Only 25\% of the displacing
RBCs zig-zag in a dense suspension for the smallest row-shift fraction
(see the last column).
\end{enumerate}

\begin{figure}
\begin{center}
\begin{tabular}{c| c | c | c |c}
   & $\epsilon = 0.167$ & $\epsilon = 0.125$ & $\epsilon = 0.083$ \\
  \hline 
  {\rotatebox[origin=lB]{90}{{\hspace{0.8cm} $C_a = 0.0034$}}} & \definecolor{zigzag}{HTML}{D7191C}
\definecolor{disp}{HTML}{2B83BA}
\begin{tikzpicture}
\pgfplotsset{every tick label/.append style={font=\small}}
\pgfplotsset{width=7cm,compat=1.9}
\begin{axis}[
  width = 0.35\textwidth,
  ybar stacked,
    bar width = 10mm,
    x=15mm,
    nodes near coords,
    every node near coord/.append,
  legend style={at={(0.5,1)},
    anchor=south, legend columns=-1, draw=none},
  ytick=0,
  ytick style={draw=none},
  ylabel={\%},
  xtick={1,2},
  xticklabels={Dilute, Dense},
  xtick style={draw=none},
  axis line style={draw=none},
  axis y line=none,
  axis x line*=bottom,
  ytick style={draw=none},
  x tick label style={rotate=0,anchor=north},
  enlarge x limits={abs=0.5},
  enlarge y limits=false,
  xbar legend,
  ymin=0,
  ymax=100,
  area legend,
  ]

  \addplot+[ybar]  plot coordinates {(1,100) (2,33)};
  \addplot+[ybar]  plot coordinates {(1,0) (2,67)};
  \legend{\strut Displacement, \strut Zig-zag}
\end{axis}
\end{tikzpicture} &
  \definecolor{zigzag}{HTML}{D7191C}
\definecolor{disp}{HTML}{2B83BA}
\begin{tikzpicture}
\pgfplotsset{every tick label/.append style={font=\small}}
\pgfplotsset{width=7cm,compat=1.9}
\begin{axis}[
  width = 0.35\textwidth,
  ybar stacked,
    bar width = 10mm,
    x=15mm,
    nodes near coords,
    every node near coord/.append,
  legend style={at={(0.5,1)},
    anchor=south, legend columns=-1, draw=none},
  ytick=0,
  ytick style={draw=none},
  ylabel={\%},
  xtick={1,2},
  xticklabels={Dilute, Dense},
  xtick style={draw=none},
  axis line style={draw=none},
  axis y line=none,
  axis x line*=bottom,
  ytick style={draw=none},
  x tick label style={rotate=0,anchor=north},
  enlarge x limits={abs=0.5},
  enlarge y limits=false,
  xbar legend,
  ymin=0,
  ymax=100,
  area legend,
  ]

  \addplot+[ybar]  plot coordinates {(1,100) (2,45)};
  \addplot+[ybar]  plot coordinates {(1,0) (2,55)};
\end{axis}
\end{tikzpicture} &
  \definecolor{zigzag}{HTML}{D7191C}
\definecolor{disp}{HTML}{2B83BA}
\begin{tikzpicture}
\pgfplotsset{every tick label/.append style={font=\small}}
\pgfplotsset{width=7cm,compat=1.9}
\begin{axis}[
  width = 0.35\textwidth,
  ybar stacked,
    bar width = 10mm,
    x=15mm,
    nodes near coords,
    every node near coord/.append,
  legend style={at={(0.5,1)},
    anchor=south, legend columns=-1, draw=none},
  ytick=0,
  ytick style={draw=none},
  ylabel={\%},
  xtick={1,2},
  xticklabels={Dilute, Dense},
  xtick style={draw=none},
  axis line style={draw=none},
  axis y line=none,
  axis x line*=bottom,
  ytick style={draw=none},
  x tick label style={rotate=0,anchor=north},
  enlarge x limits={abs=0.5},
  enlarge y limits=false,
  xbar legend,
  ymin=0,
  ymax=100,
  area legend,
  ]

  \addplot+[ybar]  plot coordinates {(1,100) (2,76)};
  \addplot+[ybar]  plot coordinates {(1,0) (2,24)};
\end{axis}
\end{tikzpicture} & {\rotatebox[origin=lB]{90}{{\hspace{1.3cm} $\nu = 10$}}} \\
  \hline
  {\rotatebox[origin=lB]{90}{{\hspace{0.75cm} $C_a = 0.34$}}} & \definecolor{zigzag}{HTML}{D7191C}
\definecolor{disp}{HTML}{2B83BA}
\begin{tikzpicture}
\pgfplotsset{every tick label/.append style={font=\small}}
\pgfplotsset{width=7cm,compat=1.9}
\begin{axis}[
  width = 0.35\textwidth,
  ybar stacked,
    bar width = 10mm,
    x=15mm,
    nodes near coords,
    every node near coord/.append,
  legend style={at={(0.5,1)},
    anchor=south, legend columns=-1, draw=none},
  ytick=0,
  ytick style={draw=none},
  ylabel={\%},
  xtick={1,2},
  xticklabels={Dilute, Dense},
  xtick style={draw=none},
  axis line style={draw=none},
  axis y line=none,
  axis x line*=bottom,
  ytick style={draw=none},
  x tick label style={rotate=0,anchor=north},
  enlarge x limits={abs=0.5},
  enlarge y limits=false,
  xbar legend,
  ymin=0,
  ymax=100,
  area legend,
  ]

  \addplot+[ybar]  plot coordinates {(1,0) (2,17)};
  \addplot+[ybar]  plot coordinates {(1,100) (2,83)};
  \legend{\strut Displacement, \strut Zig-zag}
\end{axis}
\end{tikzpicture} &
  \definecolor{zigzag}{HTML}{D7191C}
\definecolor{disp}{HTML}{2B83BA}
\begin{tikzpicture}
\pgfplotsset{every tick label/.append style={font=\small}}
\pgfplotsset{width=7cm,compat=1.9}
\begin{axis}[
  width = 0.35\textwidth,
  ybar stacked,
    bar width = 10mm,
    x=15mm,
    nodes near coords,
    every node near coord/.append,
  legend style={at={(0.5,1)},
    anchor=south, legend columns=-1, draw=none},
  ytick=0,
  ytick style={draw=none},
  ylabel={\%},
  xtick={1,2},
  xticklabels={Dilute, Dense},
  xtick style={draw=none},
  axis line style={draw=none},
  axis y line=none,
  axis x line*=bottom,
  ytick style={draw=none},
  x tick label style={rotate=0,anchor=north},
  enlarge x limits={abs=0.5},
  enlarge y limits=false,
  xbar legend,
  ymin=0,
  ymax=100,
  area legend,
  ]

  \addplot+[ybar]  plot coordinates {(1,0) (2,21)};
  \addplot+[ybar]  plot coordinates {(1,100) (2,79)};
\end{axis}
\end{tikzpicture} &
  \definecolor{zigzag}{HTML}{D7191C}
\definecolor{disp}{HTML}{2B83BA}
\begin{tikzpicture}
\pgfplotsset{every tick label/.append style={font=\small}}
\pgfplotsset{width=7cm,compat=1.9}
\begin{axis}[
  width = 0.35\textwidth,
  ybar stacked,
    bar width = 10mm,
    x=15mm,
    nodes near coords,
    every node near coord/.append,
  legend style={at={(0.5,1)},
    anchor=south, legend columns=-1, draw=none},
  ytick=0,
  ytick style={draw=none},
  ylabel={\%},
  xtick={1,2},
  xticklabels={Dilute, Dense},
  xtick style={draw=none},
  axis line style={draw=none},
  axis y line=none,
  axis x line*=bottom,
  ytick style={draw=none},
  x tick label style={rotate=0,anchor=north},
  enlarge x limits={abs=0.5},
  enlarge y limits=false,
  xbar legend,
  ymin=0,
  ymax=100,
  area legend,
  ]

  \addplot+[ybar]  plot coordinates {(1,0) (2,37)};
  \addplot+[ybar]  plot coordinates {(1,100) (2,63)};
\end{axis}
\end{tikzpicture} & {\rotatebox[origin=lB]{90}{{\hspace{1.3cm} $\nu = 10$}}}\\
  \hline
  {\rotatebox[origin=lB]{90}{{\hspace{0.75cm} $C_a = 0.34$}}} & \definecolor{zigzag}{HTML}{D7191C}
\definecolor{disp}{HTML}{2B83BA}
\begin{tikzpicture}
\pgfplotsset{every tick label/.append style={font=\small}}
\pgfplotsset{width=7cm,compat=1.9}
\begin{axis}[
  width = 0.35\textwidth,
  ybar stacked,
    bar width = 10mm,
    x=15mm,
    nodes near coords,
    every node near coord/.append,
  legend style={at={(0.5,1)},
    anchor=south, legend columns=-1, draw=none, font=\small},
  ytick=0,
  ytick style={draw=none},
  ylabel={\%},
  xtick={1,2},
  xticklabels={Dilute, Dense},
  xtick style={draw=none},
  axis line style={draw=none},
  axis y line=none,
  axis x line*=bottom,
  ytick style={draw=none},
  x tick label style={rotate=0,anchor=north},
  enlarge x limits={abs=0.5},
  enlarge y limits=false,
  xbar legend,
  ymin=0,
  ymax=100,
  area legend,
  ]

  \addplot+[ybar]  plot coordinates {(1,100) (2,36)};
  \addplot+[ybar]  plot coordinates {(1,0) (2,64)};
  \legend{\strut Displacement, \strut Zig-zag}
\end{axis}
\end{tikzpicture} &
  \definecolor{zigzag}{HTML}{D7191C}
\definecolor{disp}{HTML}{2B83BA}
\begin{tikzpicture}
\pgfplotsset{every tick label/.append style={font=\small}}
\pgfplotsset{width=7cm,compat=1.9}
\begin{axis}[
  width = 0.35\textwidth,
  ybar stacked,
    bar width = 10mm,
    x=15mm,
    nodes near coords,
    every node near coord/.append,
  legend style={at={(0.5,1)},
    anchor=south, legend columns=-1, draw=none},
  ytick=0,
  ytick style={draw=none},
  ylabel={\%},
  xtick={1,2},
  xticklabels={Dilute, Dense},
  xtick style={draw=none},
  axis line style={draw=none},
  axis y line=none,
  axis x line*=bottom,
  ytick style={draw=none},
  x tick label style={rotate=0,anchor=north},
  enlarge x limits={abs=0.5},
  enlarge y limits=false,
  xbar legend,
  ymin=0,
  ymax=100,
  area legend,
  ]

  \addplot+[ybar]  plot coordinates {(1,100) (2,48)};
  \addplot+[ybar]  plot coordinates {(1,0) (2,52)};
\end{axis}
\end{tikzpicture} &
  \definecolor{zigzag}{HTML}{D7191C}
\definecolor{disp}{HTML}{2B83BA}
\begin{tikzpicture}
\pgfplotsset{every tick label/.append style={font=\small}}
\pgfplotsset{width=7cm,compat=1.9}
\begin{axis}[
  width = 0.35\textwidth,
  ybar stacked,
    bar width = 10mm,
    x=15mm,
    nodes near coords,
    every node near coord/.append,
  legend style={at={(0.5,1)},
    anchor=south, legend columns=-1, draw=none},
  ytick=0,
  ytick style={draw=none},
  ylabel={\%},
  xtick={1,2},
  xticklabels={Dilute, Dense},
  xtick style={draw=none},
  axis line style={draw=none},
  axis y line=none,
  axis x line*=bottom,
  ytick style={draw=none},
  x tick label style={rotate=0,anchor=north},
  enlarge x limits={abs=0.5},
  enlarge y limits=false,
  xbar legend,
  ymin=0,
  ymax=100,
  area legend,
  ]

  \addplot+[ybar]  plot coordinates {(1,100) (2,84)};
  \addplot+[ybar]  plot coordinates {(1,0) (2,16)};
\end{axis}
\end{tikzpicture} & {\rotatebox[origin=lB]{90}{{\hspace{1.25cm} $\nu = 1$}}}\\ 
\end{tabular}
\end{center}
\mcaption{Breakdown of the transport modes in dense RBC suspensions. We performed
simulations of a single RBC (dilute suspension) and multiple RBCs
(dense suspension) for the row-shift fractions $\epsilon = (0.083,0.125, 0.1667)$. 
We considered RBCs with the capillary numbers 
$C_a = (0.0034, 0.34)$ and viscosity contrasts $\nu = (1,10)$. The dense
suspensions have volume fractions of 15\% to 20\%. We present the
histogram plots showing the numbers of zig-zagging and displacing RBCs
(out of 100). In the dilute suspensions, a single RBC either zig-zags
or displaces. Therefore, we observe either 100\% zig-zag or
displacement in the dilute suspensions. Whereas in dense
suspensions the same RBCs are not in the same transport modes. Results
at each column are for different row-shift fractions (decreasing from left to right) 
and those at each row have different combinations of $C_a$ and $\nu$.}{f:streamFails}
\end{figure}

\section{Discussion\label{s:discussion}}
In this section, we investigate the underlying mechanism for the RBC separation depending on the capillary number and viscosity contrast in deep DLD devices. Since the flow in DLD resembles free shear flow (between the vertical gaps) and confined Poiseuille flow (in vertical gaps), we compare cell dynamics in DLD with those in these simpler flows. We present the similarities and seek if these simpler setups can be used to predict whether a cell displaces or zig-zags in a particular DLD device. 

\subsection{Free shear flow}\label{s:freeShear}

The results in~\secref{s:results} show that cells having high positive inclination angles with respect to the flow direction displace and those having lower inclination angles zig-zag. This is a result of a positive inclination angle and shear gradient which lead a cell to migrate away from the pillars. A cell's inclination angle depends on its capillary number and viscosity contrast. In free shear flow, cells tilt to a certain angle and move with that angular orientation for low capillary numbers and viscosity contrasts. This motion is called tank-treading. For high capillary numbers and viscosity contrasts, cells do not have such an angular orientation and tumble~\citep{kantsler-steinberg-e06,misbah06}. This motion is called tumbling. In free shear flow, tank-treading cells migrate from high shear rate regions to low shear rate regions while tumbling cells do not migrate~\citep{olla97,vlahovska-gracia07,messlinger-gompper-e09}.~\cite{henry-gompper-e16} investigated sorting cells depending on those cell dynamics and concluded that tank-treading at the viscosity contrast $\nu = 1$ results in displacement and tumbling usually leads to zig-zag. They observed tumbling between the gaps right before a cell zig-zags which we also observed for some cases. It must be noted that tumbling does not occur in the gaps in DLD due to the confinement~\citep{kaoui-harting-e12}. 

We investigate whether the conditions (the capillary number and the viscosity contrast) causing cells to tank-tread in free shear flow lead cells to displace in DLD for any row-shift fraction. In other words, we investigate whether the conditions causing cells to tumble in free shear flow lead cells to zig-zag in DLD for any row-shift fraction. If this is true, then we can estimate cell dynamics in DLD by simulating cells in a simpler free shear flow. However, as we can see in~\figref{f:CavsVCPhaseDiag}, a cell tank-treading in a free shear flow does not necessarily displace in DLD flow for the same capillary number and the viscosity contrast. For example, a cell that tumbles for $\nu = 5$ and $C_a = \bigO(10)$ in free shear flow displaces in the DLD for $\epsilon = 0.1$ but zig-zags for $\epsilon = 0.1667$. The main reason for that the dynamics in free shear flow cannot predict the transport mode in DLD is that the row-shift fraction information does not enter free shear flow. Additionally, the confinement in the gaps in DLD also affects these dynamics~\citep{kaoui-harting-e12}.

\subsection{Confined Poiseuille flow}\label{s:confPois}

\begin{figure}
\begin{center}
\begin{tabular}{c c}
\begin{tikzpicture}

\begin{axis}[
  width = 0.45\textwidth,
  xmin = 1E-2,
  xmax = 1E+4, 
  ymin = 0.05,
  ymax = 0.35,
  xtick = {1E-2,1E-1,1,1E+1,1E+2,1E+3,1E+4},
  ytick = {0.05,0.1,0.15,0.2,0.25,0.3,0.35},
  ylabel = {$\overline{\alpha}$ [rad]},
  xlabel = {$C_a$},
  xmode = log,
  label style={font=\normalsize},
  legend style={at={(0.5,1.05)},anchor=south,draw=none},
  legend columns=3,
  legend entries={$\nu = 1\quad\,\,$,$\nu = 2\quad\,\,$,$\nu = 5\quad\,\,$,$\nu = 8\quad\,\,$,$\nu = 10\,\,\,\,$,$\nu = 1000$},
  ]

\addplot [mark=*,lightgray,line width=1, mark size=2.5pt] table{
3.4141e-02   1.9209e-01
3.4141e-01   2.2921e-01
3.4141e+00   2.6125e-01
3.4141e+01   2.9228e-01
3.4141e+02   3.0071e-01
3.4141e+03   3.0197e-01
};

\addplot [mark=*,gray,line width=1, mark size=2.5pt] table{
3.4141e-02   1.9093e-01
3.4141e-01   2.0793e-01
3.4141e+00   2.3367e-01
3.4141e+01   2.6254e-01
3.4141e+02   2.7063e-01
3.4141e+03   2.7176e-01
};

\addplot [mark=*,brown,line width=1, mark size=2.5pt] table{
3.4141e-02   1.8064e-01
3.4141e-01   1.7273e-01
3.4141e+00   1.8811e-01
3.4141e+01   2.1356e-01
3.4141e+02   2.2047e-01
3.4141e+03   2.2123e-01
};

\addplot [mark=*,green,line width=1, mark size=2.5pt] table{
3.4141e-02   1.7494e-01
3.4141e-01   1.5383e-01
3.4141e+00   1.6293e-01
3.4141e+01   1.8522e-01
3.4141e+02   1.9103e-01
3.4141e+03   1.9225e-01
};

\addplot [mark=*,cyan,line width=1, mark size=2.5pt] table{
3.4141e-02   1.6851e-01
3.4141e-01   1.4551e-01
3.4141e+00   1.5084e-01
3.4141e+01   1.7123e-01
3.4141e+02   1.7714e-01
3.4141e+03   1.7794e-01
};

\addplot [mark=*,orange,line width=1, mark size=2.5pt] table{
3.4141e-02   6.8047e-02
3.4141e-01   7.5641e-02
3.4141e+00   6.8500e-02
3.4141e+01   6.5784e-02
3.4141e+02   6.5616e-02
3.4141e+03   6.5597e-02
};

\end{axis}

\end{tikzpicture} & \begin{tikzpicture}

\begin{axis}[
  width = 0.45\textwidth,
  xmin = 1E-2,
  xmax = 1E+4, 
  ymin = 0.01,
  ymax = 0.045,
  xtick = {1E-2,1E-1,1,1E+1,1E+2,1E+3,1E+4},
  ytick = {0.01,0.015,0.02,0.025,0.03,0.035,0.04,0.045},
  ylabel = {$\overline{v}_{\mathrm{mig}}$},
  xlabel = {$C_a$},
  label style={font=\normalsize},
  xmode = log,
  ]

\addplot [mark=*,lightgray,line width=1, mark size=2.5pt] table{
3.4141e-02   2.6790e-02
3.4141e-01   4.0827e-02
3.4141e+00   4.2905e-02
3.4141e+01   4.3533e-02
3.4141e+02   4.3444e-02
3.4141e+03   4.3380e-02
};

\addplot [mark=*,gray,line width=1, mark size=2.5pt] table{
3.4141e-02   2.6392e-02
3.4141e-01   3.5231e-02
3.4141e+00   3.5687e-02
3.4141e+01   3.6010e-02
3.4141e+02   3.5909e-02
3.4141e+03   3.5870e-02
};

\addplot [mark=*,brown,line width=1, mark size=2.5pt] table{
3.4141e-02   2.5558e-02
3.4141e-01   2.7387e-02
3.4141e+00   2.5653e-02
3.4141e+01   2.6125e-02
3.4141e+02   2.6249e-02
3.4141e+03   2.6252e-02
};

\addplot [mark=*,green,line width=1, mark size=2.5pt] table{
3.4141e-02   2.4866e-02
3.4141e-01   2.3689e-02
3.4141e+00   2.1236e-02
3.4141e+01   2.2031e-02
3.4141e+02   2.2296e-02
3.4141e+03   2.2307e-02
};

\addplot [mark=*,cyan,line width=1, mark size=2.5pt] table{
3.4141e-02   2.4417e-02
3.4141e-01   2.2192e-02
3.4141e+00   1.9450e-02
3.4141e+01   2.0413e-02
3.4141e+02   2.0721e-02
3.4141e+03   2.0736e-02
};

\addplot [mark=*,orange,line width=1, mark size=2.5pt] table{
3.4141e-02   1.0392e-02
3.4141e-01   1.1645e-02
3.4141e+00   1.1942e-02
3.4141e+01   1.1803e-02
3.4141e+02   1.1800e-02
3.4141e+03   1.1799e-02
};

\end{axis}

\end{tikzpicture}
\end{tabular}
\end{center}
\mcaption{The time-averaged inclination angles $\overline{\alpha}$ (on the left) and the migration velocities $\overline{v}_{\mathrm{mig}}$ (on the right) for a red blood cell in confined Poiseuille flow for various capillary numbers $C_a$ and viscosity contrasts $\nu$. We considered a channel having the same confinement as the gaps in our DLD simulations. We initialized a cell below the centerline and let the cell to migrate towards the center.}{f:IAandMigInTube}
\end{figure}
Confining a cell by walls delays the cell's transition from tank-treading to tumbling and can even avoid tumbling~\citep{kaoui-harting-e12}. In order to incorporate the confinement effects into a simpler model we, now, consider confined Poiseuille flow, to which the flow in a DLD gap resembles, and we investigate the cell dynamics. We imposed a parabolic velocity similar to the flow in DLD gaps on the side walls of a channel confined by two parallel walls. The width of this channel is the same as the width of the gap in our DLD simulations. Thus, the confinement is $\chi = 2R_{\mathrm{eff}}/G$. The channel is centered at the origin and the cell is initialized at $y_i = -3G/10$. We performed simulations of a cell for the same capillary numbers and viscosity contrasts we considered in our DLD simulations, i.e., $C_a \in [3.41\times10^{-2}, 3.41\times10^3]$ and $\nu \in [1, 10]$. We let the cell migrate to $y_f = -G/10$. We do not allow the cell to reach the center because there it shows complex shape changes and equilibrium dynamics at the center for this confinement~\citep{aouane-misbah-e14}. Additionally, the cell in DLD usually stays near the pillars not at the center of the gaps. We measured the cell's migration velocity $v_{\mathrm{mig}}$ (i.e., velocity in the direction perpendicular to the flow direction) and inclination angle $\alpha$ with respect to the walls at various lateral positions. We present the time-averaged inclination angles $\overline{\alpha}$ and migration velocities $\overline{v}_{\mathrm{mig}}$ for various capillary numbers $C_a$ and viscosity contrasts $\nu$ in~\figref{f:IAandMigInTube}. The figure on the left shows that the inclination angle varies less with the viscosity contrast for the lowest capillary number than higher capillary numbers. As the capillary number increases, the inclination angle increases. It also increases as the viscosity contrast decreases. So for lower viscosity contrasts and higher capillary numbers the inclination angle is higher. For $C_a \geq 10^2$ the inclination angle does not change significantly with $C_a$. Additionally, the inclination angle increases monotonously with the capillary number for the viscosity contrasts $\nu = (1, 2)$ but it does not change monotonously with the capillary number for the viscosity contrasts $\nu \geq 5$. The figure on the right indicates that the average migration velocity parallels the average inclination angle. A cell migrates faster for lower viscosity contrasts. The migration velocity monotonously increases with the capillary number for the viscosity contrasts $\nu = (1, 2)$ like the average inclination angle, however, this behavior vanishes for viscosity contrasts $\nu \geq 5$. Although the inclination angle still changes with $C_a$ for $C_a = \bigO(1)$, the migration velocity does not change significantly with $C_a$ anymore for $C_a = \bigO(1)$. The reason might be the fact that the bending relaxation time scale dominates the time scale for shear flow for $C_a > 1$. Another observation is that a very rigid cell (i.e., the viscosity contrast $\nu = 1000$) also has a positive inclination with respect to the walls but the angle is an order of magnitude smaller than that for softer cells. This results in a very small migration velocity for the stiff cell. Overall, similar to the DLD flow examples we do not observe tumbling for this confinement in confined Poiseuille flow, which agrees with~\cite{kaoui-harting-e12}.  

Based on these results we expect that the transport mode in DLD depends on whether the degree of cell's inclination results in enough migration to keep the cell far away from the pillars. We, now, discuss how the inclination angle of a cell depends on $C_a$, $\nu$ and $\epsilon$ in DLD flows and compare it with the above results for confined Poiseuille flows.

\begin{figure}
\begin{center}
\begin{tabular}{c c}
\input{figs/N16DiffVCAngles.tikz} & \input{figs/VC1DiffNsAngles.tikz} \\
\end{tabular}
\end{center}
\mcaption{Inclination angles $\alpha$ of RBCs for various viscosity contrasts
$\nu$ at a fixed row-shift fraction $\epsilon = 0.0625$ on the left
and for various row-shift fractions at a fixed viscosity contrast 
$\nu = 1$ on the right. The capillary number is the same for all the
results, $C_a = 0.34$. The RBCs with $\nu = 1, 2, 5$ displace and the
one with $\nu = 10$ zig-zags. The displacing RBCs have positive
inclination angles alternating between two values (see the figure on
the left). The minimum angle is attained when the cell is in the gap and the cell reaches the maximum angle when it is between two gaps. 
The minimum and maximum angles reduce as the viscosity
contrast increases. The maximum angle depends also on the row-shift 
fraction $\epsilon$. As the row-
shift fraction increases the maximum angle increases, however, the
minimum angle does not change significantly (see the figure on the
right). The $x$-coordinates are normalized by the lengths of the DLD
devices $L$.}{f:IAsAnalysisVC}
\end{figure}
We demonstrate how the inclination angle of a cell changes as the cell flows through a DLD device in~\figref{f:IAsAnalysisVC}. On the left, we vary the viscosity contrast $\nu$ by setting the capillary number to $C_a = 0.34$ and row-shift fraction to $\epsilon = 0.0625$. On the right we vary the row-shift fraction for a fixed capillary number $C_a = 0.34$ and viscosity contrast $\nu = 1$. All these cells displace except the one with $\nu = 10$ on the left figure. The displacing cells attain their minimum angles in the gaps and tilt to their maximum angles between two consecutive gaps. These maximum angles are $0.58$rad for $\nu = 1$, $0.5$rad for $\nu = 2$ and $0.37$rad for $\nu = 5$. These values are
close to the steady inclination angles of the RBCs having these
viscosity contrasts in an unbounded shear 
flow~\citep{beaucourt-misbah-e04,kantsler-steinberg-e06}. However, the displacing RBCs'
maximum inclination angles depend also on the row-shift fraction. The figure on the right indicates that as the row-shift fraction
increases, the maximum inclination angle increases. The reason for that
is as the row-shift fraction increases, the angle between the direction 
of the flow between the gaps and the main flow direction increases. This
results in a higher maximum inclination angle. Additionally, the
figure on the right shows that the minimum
angle does not change significantly with the row-shift fraction. This is 
because the flow in the gap is always in the main flow
direction no matter what the row-shift fraction is(see~\figref{f:N6WholeDLD}). 
The minimum inclination angle
depends only on the viscosity contrast and the capillary number.

\begin{figure}
\begin{center}
\begin{tabular}{c c}
\begin{tikzpicture}

\begin{axis}[
  width = 0.45\textwidth,
  title = {$\epsilon = 0.1667$},
  xmin = 1E-2,
  xmax = 1E+4, 
  ymin = 0,
  ymax = 0.25,
  xtick = {1E-2,1E-1,1,1E+1,1E+2,1E+3,1E+4},
  ytick = {0,0.05,0.1,0.15,0.2,0.25},
  ylabel = {$\overline{\alpha}$ [rad]},
  xlabel = {$C_a$},
  xmode = log,
  label style={font=\normalsize},
  legend style={at={(0.5,1.05)},anchor=south,draw=none},
  legend columns=3,
legend entries={$\nu = 1\quad$, $\nu = 2\quad$, $\nu = 5\quad$,$\nu = 8\quad$,$\nu = 10\,\,$},
  ]

\addplot [mark=none,lightgray,line width=1.5] table{
3.4141e-02   1.6046e-01
3.4141e-01   2.1536e-01
3.4141e+00   2.2115e-01
3.4141e+01   2.2260e-01
3.4141e+02   2.2933e-01
3.4141e+03   2.3050e-01
};

\addplot [mark=none,gray,line width=1.5] table{
3.4141e-02   1.3319e-01
3.4141e-01   1.9094e-01
3.4141e+00   1.8759e-01
3.4141e+01   1.8888e-01
3.4141e+02   1.9526e-01
3.4141e+03   1.9579e-01
};

\addplot [mark=none,brown,line width=1.5] table{
3.4141e-02   1.1729e-01
3.4141e-01   1.2557e-01
3.4141e+00   1.0094e-01
3.4141e+01   1.0737e-01
3.4141e+02   1.1360e-01
3.4141e+03   1.1261e-01
};

\addplot [mark=none,green,line width=1.5] table{
3.4141e-02   6.2047e-02
3.4141e-01   4.4293e-02
3.4141e+00   4.0226e-02
3.4141e+01   7.3143e-02
3.4141e+02   8.0734e-02
3.4141e+03   8.1571e-02
};

\addplot [mark=none,cyan,line width=1.5] table{
3.4141e-02   4.6579e-02
3.4141e-01   2.0946e-02
3.4141e+00   1.6544e-02
3.4141e+01   5.6663e-02
3.4141e+02   5.9345e-02
3.4141e+03   5.8323e-02
};

\addplot [color=red,only marks,mark=*,mark size=2.5pt] table{
3.4141E+3 1.1261e-01
3.4141E+3 8.1571e-02
3.4141E+3 5.8323e-02
3.4141E+2 1.1360e-01
3.4141E+2 8.0734e-02
3.4141E+2 5.9345e-02
3.4141E+1 1.0737e-01
3.4141E+1 7.3143e-02
3.4141E+1 5.6663e-02
3.4141E+0 1.0094e-01
3.4141E+0 4.0226e-02
3.4141E+0 1.6544e-02
3.4141E-1 1.2557e-01
3.4141E-1 4.4293e-02
3.4141E-1 2.0946e-02
3.4141E-2 1.6046e-01
3.4141E-2 1.3319e-01
3.4141E-2 1.1729e-01
3.4141E-2 6.2047e-02
3.4141E-2 4.6579e-02

}; 

\addplot [color=blue,only marks,mark=triangle*, mark size=2.5pt] table{
3.4141E+3 2.3050e-01
3.4141E+3 1.9579e-01
3.4141E+2 2.2933e-01
3.4141E+2 1.9526e-01
3.4141E+1 2.2260e-01
3.4141E+1 1.8888e-01
3.4141E+0 2.2115e-01
3.4141E+0 1.8759e-01
3.4141E-1 2.1536e-01
3.4141E-1 1.9094e-01
};

\end{axis}

\end{tikzpicture} & \begin{tikzpicture}

\begin{axis}[
  width = 0.45\textwidth,
  xmin = 1E-2,
  xmax = 1E+4, 
  ymin = -0.1,
  ymax = 0.9,
  xtick = {1E-2,1E-1,1,1E+1,1E+2,1E+3,1E+4},
  ytick = {-0.1,0.1,0.3,0.5,0.7,0.9},
  ylabel = {$\overline{\alpha}$ [rad]},
  xlabel = {$C_a$},
  xmode = log,
  label style={font=\normalsize},
  legend style={at={(0.5,1.05)},anchor=south,draw=none},
  legend columns=2,
  legend entries={Zig-zag,Displacement},
  ]

\addplot [color=red,only marks,mark=*,mark size=2.5pt] table{
   3.4141e+03   4.5126e-01
   3.4141e+03   2.6236e-01
   3.4141e+03   1.3877e-01
   3.4141e+02   4.4622e-01
   3.4141e+02   2.5289e-01
   3.4141e+02   1.2273e-01
   3.4141e+01   4.1532e-01
   3.4141e+01   2.1203e-01
   3.4141e+01   7.4114e-02
   3.4141e+00   4.2133e-01
   3.4141e+00   9.9132e-02
   3.4141e+00  -3.1823e-02
   3.4141e-01   4.5863e-01
   3.4141e-01   1.1981e-01
   3.4141e-01   4.6025e-02
3.4141e-02   3.8341e-01
3.4141e-02   3.4139e-01
 3.4141e-02   2.0388e-01
3.4141e-02   1.2661e-01
   3.4141e-02   5.4352e-02

}; 

\addplot [color=blue,only marks,mark=triangle*, mark size=2.5pt] table{
3.4141E+3 8.3923e-01
3.4141E+3 7.3450e-01
3.4141E+2 8.3745e-01
3.4141E+2 7.3157e-01
3.4141E+1 8.2637e-01
3.4141E+1 7.1470e-01
3.4141E+0 8.1356e-01
3.4141E+0 7.0731e-01
3.4141E-1 7.7637e-01
3.4141E-1 6.9192e-01
}; 

\addplot [mark=none,lightgray,line width=1.5] table{
3.4141e-02   3.8341e-01
   3.4141e-01   7.7637e-01
   3.4141e+00   8.1356e-01
   3.4141e+01   8.2637e-01
   3.4141e+02   8.3745e-01
   3.4141e+03   8.3923e-01
};

\addplot [mark=none,gray,line width=1.5] table{
3.4141e-02   3.4139e-01
   3.4141e-01   6.9192e-01
   3.4141e+00   7.0731e-01
   3.4141e+01   7.1470e-01
   3.4141e+02   7.3157e-01
   3.4141e+03   7.3450e-01
};

\addplot [mark=none,brown,line width=1.5] table{
 3.4141e-02   2.0388e-01
   3.4141e-01   4.5863e-01
   3.4141e+00   4.2133e-01
   3.4141e+01   4.1532e-01
   3.4141e+02   4.4622e-01
   3.4141e+03   4.5126e-01
};

\addplot [mark=none,green,line width=1.5] table{
3.4141e-02   1.2661e-01
   3.4141e-01   1.1981e-01
   3.4141e+00   9.9132e-02
   3.4141e+01   2.1203e-01
   3.4141e+02   2.5289e-01
   3.4141e+03   2.6236e-01
};

\addplot [mark=none,cyan,line width=1.5] table{
   3.4141e-02   5.4352e-02
   3.4141e-01   4.6025e-02
   3.4141e+00  -3.1823e-02
   3.4141e+01   7.4114e-02
   3.4141e+02   1.2273e-01
   3.4141e+03   1.3877e-01
};

\end{axis}

\end{tikzpicture}
\end{tabular}
\end{center}
\mcaption{The average inclination angles $\overline{\alpha}$ of RBCs in the gaps (on the left) and between the gaps (on the right) for the row-shift fraction $\epsilon = 0.1667$. The results are from the simulations we performed for the phase diagram at the top in~\figref{f:CavsVCPhaseDiag}. Red circles and blue triangles indicate the zig-zagging and displacing cells, respectively. The displacing cells have higher inclination angles than the zig-zagging ones. That is, the displacing cells have $\alpha > 0.16$rad in the gaps and $\alpha > 0.6$rad between the gaps. Whereas the zig-zagging ones have $\alpha < 0.16$rad in the gaps and $\alpha < 0.5$rad between the gaps.}{f:IAgapP6}
\end{figure}
\begin{figure}
\begin{center}
\begin{tabular}{c c}
\begin{tikzpicture}

\begin{axis}[
  width = 0.45\textwidth,
  xmin = 1E-2,
  xmax = 1E+4, 
  ymin = 0,
  ymax = 0.25,
  xtick = {1E-2,1E-1,1,1E+1,1E+2,1E+3,1E+4},
  ytick = {0,0.05,0.1,0.15,0.2,0.25},
  ylabel = {$\overline{\alpha}$ [rad]},
  xlabel = {$C_a$},
  xmode = log,
  label style={font=\normalsize},
  legend style={at={(0.5,1.05)},anchor=south,draw=none},
  legend columns=3,
  legend entries={$\nu = 1\quad$, $\nu = 2\quad$, $\nu = 5\quad$,$\nu = 8\quad$,$\nu = 10\,\,$},
  ]

\addplot [mark=none,lightgray,line width=1.5] table{
3.4141e-02   1.5851e-01
3.4141e-01   1.9402e-01
3.4141e+00   2.0379e-01
3.4141e+01   2.1632e-01
3.4141e+02   2.2209e-01
3.4141e+03   2.2326e-01
};

\addplot [mark=none,gray,line width=1.5] table{
3.4141e-02   1.5236e-01
3.4141e-01   1.7019e-01
3.4141e+00   1.7234e-01
3.4141e+01   1.8731e-01
3.4141e+02   1.9444e-01
3.4141e+03   1.9558e-01
};

\addplot [mark=none,brown,line width=1.5] table{
3.4141e-02   1.2526e-01
3.4141e-01   1.2386e-01
3.4141e+00   1.2169e-01
3.4141e+01   1.4972e-01
3.4141e+02   1.5678e-01
3.4141e+03   1.5731e-01
};

\addplot [mark=none,green,line width=1.5] table{
3.4141e-02   9.3063e-02
3.4141e-01   9.7931e-02
3.4141e+00   8.4559e-02
3.4141e+01   1.1664e-01
3.4141e+02   1.2023e-01
3.4141e+03   1.2090e-01
};

\addplot [mark=none,cyan,line width=1.5] table{
3.4141e-02   7.5645e-02
3.4141e-01   5.5766e-02
3.4141e+00   5.8960e-02
3.4141e+01   8.7867e-02
3.4141e+02   9.7038e-02
3.4141e+03   9.5931e-02
};

\addplot [color=red,only marks,mark=*,mark size=2.5pt] table{
3.4141E+3 1.2090e-01
3.4141E+3 9.5931e-02
3.4141E+2 1.2023e-01
3.4141E+2 9.7038e-02
3.4141E+1 1.1664e-01
3.4141E+1 8.7867e-02
3.4141E+0 1.2169e-01
3.4141E+0 8.4559e-02
3.4141E+0 5.8960e-02
3.4141E-1 9.7931e-02
3.4141E-1 5.5766e-02
3.4141E-2 1.2526e-01
3.4141E-2 9.3063e-02
3.4141E-2 7.5645e-02

}; 

\addplot [color=blue,only marks,mark=triangle*, mark size=2.5
pt] table{
3.4141E+3 2.2326e-01
3.4141E+3 1.9558e-01
3.4141E+3 1.5731e-01
3.4141E+2 2.2209e-01
3.4141E+2 1.9444e-01
3.4141E+2 1.5678e-01
3.4141E+1 2.1632e-01
3.4141E+1 1.8731e-01
3.4141E+1 1.4972e-01
3.4141E+0 2.0379e-01
3.4141E+0 1.7234e-01
3.4141E-1 1.9402e-01
3.4141E-1 1.7019e-01
3.4141E-1 1.2386e-01
3.4141E-2 1.5851e-01
3.4141E-2 1.5236e-01
};

\end{axis}

\end{tikzpicture} & \begin{tikzpicture}

\begin{axis}[
  width = 0.45\textwidth,
  xmin = 1E-2,
  xmax = 1E+4, 
  ymin = -0.1,
  ymax = 0.9,
  xtick = {1E-2,1E-1,1,1E+1,1E+2,1E+3,1E+4},
  ytick = {-0.1,0.1,0.3,0.5,0.7,0.9},
  ylabel = {$\overline{\alpha}$ [rad]},
  xlabel = {$C_a$},
  xmode = log,
  label style={font=\normalsize},
  legend style={at={(0.5,1.05)},anchor=south,draw=none},
  legend columns=2,
  legend entries={Zig-zag,Displacement},
  ]

\addplot [color=red,only marks,mark=*,mark size=2.5pt] table{
   3.4141e+03   2.6860e-01
   3.4141e+03   2.0713e-01
   3.4141e+02   2.6242e-01
   3.4141e+02   2.0492e-01
   3.4141e+01   2.5153e-01
   3.4141e+01   1.7031e-01
   3.4141e+00   3.9150e-01
   3.4141e+00   1.9698e-01
   3.4141e+00  -1.2856e-02
   3.4141e-01   2.6890e-01
   3.4141e-01   1.0745e-01
   3.4141e-02   2.3534e-01
   3.4141e-02   1.3651e-01
   3.4141e-02   9.0599e-02

}; 

\addplot [color=blue,only marks,mark=triangle*, mark size=2.5pt] table{
   3.4141e+03   7.4093e-01
   3.4141e+03   6.4317e-01
   3.4141e+03   4.5038e-01
   3.4141e+02   7.3959e-01
   3.4141e+02   6.4097e-01
   3.4141e+02   4.4722e-01
   3.4141e+01   7.2897e-01
   3.4141e+01   6.2631e-01
   3.4141e+01   4.3300e-01
   3.4141e+00   7.0861e-01
   3.4141e+00   6.0066e-01
   3.4141e-01   6.6532e-01
   3.4141e-01   5.8181e-01
   3.4141e-01   4.1148e-01
  3.4141e-02   3.6665e-01
  3.4141e-02   3.3624e-01
}; 

\addplot [mark=none,lightgray,line width=1.5] table{
  3.4141e-02   3.6665e-01
   3.4141e-01   6.6532e-01
   3.4141e+00   7.0861e-01
   3.4141e+01   7.2897e-01
   3.4141e+02   7.3959e-01
   3.4141e+03   7.4093e-01
};

\addplot [mark=none,gray,line width=1.5] table{
  3.4141e-02   3.3624e-01
   3.4141e-01   5.8181e-01
   3.4141e+00   6.0066e-01
   3.4141e+01   6.2631e-01
   3.4141e+02   6.4097e-01
   3.4141e+03   6.4317e-01
};

\addplot [mark=none,brown,line width=1.5] table{
   3.4141e-02   2.3534e-01
   3.4141e-01   4.1148e-01
   3.4141e+00   3.9150e-01
   3.4141e+01   4.3300e-01
   3.4141e+02   4.4722e-01
   3.4141e+03   4.5038e-01
};

\addplot [mark=none,green,line width=1.5] table{
   3.4141e-02   1.3651e-01
   3.4141e-01   2.6890e-01
   3.4141e+00   1.9698e-01
   3.4141e+01   2.5153e-01
   3.4141e+02   2.6242e-01
   3.4141e+03   2.6860e-01
};

\addplot [mark=none,cyan,line width=1.5] table{
   3.4141e-02   9.0599e-02
   3.4141e-01   1.0745e-01
   3.4141e+00  -1.2856e-02
   3.4141e+01   1.7031e-01
   3.4141e+02   2.0492e-01
   3.4141e+03   2.0713e-01
};

\end{axis}

\end{tikzpicture}
\end{tabular}
\end{center}
\mcaption{The average inclination angles $\overline{\alpha}$ of RBCs in the gaps (on the left) and between the gaps (on the right) for the row-shift fraction $\epsilon = 0.1$. The results are from the simulations we performed for the phase diagram at the top in~\figref{f:CavsVCPhaseDiag}. Red circles and blue triangles indicate the zig-zagging and displacing cells, respectively. }{f:IAgapP10}
\end{figure}
Since the inclination angle of a cell is much higher when the cell is between the gaps than when the cell is in the gaps, we compute the average inclination angles in the gaps and between the gaps, separately. We present these average inclination angles $\overline{\alpha}$ as a function of the capillary number and viscosity contrast for the row-shift fractions $\epsilon = 0.1667$ in~\figref{f:IAgapP6} and for $\epsilon = 0.1$ in~\figref{f:IAgapP10}. The figures on the left demonstrate the average angles in the gaps and those on the right show the angles between the gaps. We indicate the zig-zagging and displacing cells with red circles and blue triangles, respectively. The results for $\epsilon = 0.1667$ (See~\figref{f:IAgapP6}) show that the average inclination angle is higher for low viscosity contrasts and increases with the capillary number for $\nu = (1,2)$. The average inclination angle does not monotonously change with the capillary number for $\nu \geq 5$. In this sense, the cell dynamics in DLD are similar to those in the confined Poiseuille flow (See~\figref{f:IAandMigInTube}). While the average angle is less in the gap than in the channel flow, the average angle between the gaps is much higher than the angle in the confined Poiseuille flow due to a weaker confinement between the gaps. So, since a cell periodically moves from strong confinement in the gaps than to weak confinement between the consecutive gaps, it shows more complicated dynamics in DLD than in the confined Poiseuille flow. Additionally, since the row-shift fraction information does not appear in the confined Poiseuille flow, the transport modes of cells in DLD cannot be captured by this simpler setup. For example, while the cells having average inclination angles in the gaps greater than 0.16rad zig-zag for $\epsilon = 0.1667$, those with the average angles less than 0.16rad displace for $\epsilon = 0.1$. 

We made another observation that supports the complexity of the flow in DLD. \figref{f:IAgapP6} shows that the displacing cells have higher inclination angles than the zig-zagging ones both in the gaps and between the gaps for $\epsilon = 0.1667$. Then, one might generalize and state that the displacing cells always have higher inclination angles than the zig-zagging ones. However, the results for $\epsilon = 0.1$ in~\figref{f:IAgapP10} present a counter-example. That is, the higher inclination angle in the gaps does not guarantee that the cell displaces. For example, the average inclination angle in the gaps for $\nu = 5$ and $C_a = 0.034$ (the lowest capillary number) is slightly higher than the angle for the same viscosity contrast but the next capillary number $C_a = 0.34$ (see the left figure). However, the cell with the lowest capillary number zig-zags and the one with a higher capillary number displaces. The reason is the cell with a higher capillary number has higher average inclination angle between the gaps than the one with the lowest capillary number (see the right figure). 

Although neither free shear flow or confined Poiseuille flow can capture the transport mode of a cell in DLD devices, it is helpful to understand the cell dynamics in these flows to explain the underlying mechanism for the cell separation in DLD.

\subsection{Pseudo-lift}\label{s:lift}
\begin{figure}
\begin{center}
\begin{tabular}{c c}
 \begin{tikzpicture}

\begin{axis}[
  width = 0.45\textwidth,
  xmin = 1E-2,
  xmax = 5E+2, 
  ymin = -7,
  ymax = 7,
  ylabel = {$\overline{F}_l$},
  xlabel = {$C_a$},
  xmode = log,
  label style={font=\normalsize},
  legend style={at={(0.5,1)},anchor=south,legend columns=2,draw=none},
  legend entries = {Zig-zag, Displacement,$\nu = 1$,$\nu = 2$,$\nu = 5$,$\nu = 8$},
  ]

\addplot [color=red,only marks,mark=*,mark size=2.5pt] table{
3.4141E-2 4.58E-2
3.4141E-2 3.23E-3
3.4141E-2 4.24E-3
3.4141E-2 -4.57E-3
3.4141E+0 -2.45E-1
3.4141E+0 -1.20
3.4141E+2 -2.72E-1
3.4141E+2 -6.82
};

\addplot [color=blue,only marks,mark=triangle*, mark size=2.5pt] table{
3.4141E+0 4.28
3.4141E+2 6.55
3.4141E+0 2.39
3.4141E+2 4.42
};

\addplot [mark=none,lightgray,line width=1.5] table{
3.4141E-2 4.58E-2
3.4141E+0 4.28
3.4141E+2 6.55
};

\addplot [mark=none,gray,line width=1.5] table{
3.4141E-2 3.23E-3
3.4141E+0 2.39
3.4141E+2 4.42
};

\addplot [mark=none,brown,line width=1.5] table{
3.4141E-2 4.24E-3
3.4141E+0 -2.45E-1
3.4141E+2 -2.72E-1
};

\addplot [mark=none,green,line width=1.5] table{
3.4141E-2 -4.57E-3
3.4141E+0 -1.2
3.4141E+2 -6.82
};

\end{axis}

\end{tikzpicture} & \begin{tikzpicture}

\begin{axis}[
  width = 0.45\textwidth,
  xmin = 1E-2,
  xmax = 5E+2, 
  ymin = -0.1,
  ymax = 0.1,
  ylabel = {$\overline{v}_{\mathrm{mig}}$},
  xlabel = {$C_a$},
  xmode = log,
  label style={font=\normalsize},
  legend style={at={(0.5,1)},anchor=south,legend columns=2,draw=none},
  legend entries = {Zig-zag, Displacement,$\nu = 1$,$\nu = 2$,$\nu = 5$,$\nu = 8$},
  ]

\addplot [color=red,only marks,mark=*,mark size=2.5pt] table{
3.4141E-2 1.68E-2
3.4141E-2 6.74E-3
3.4141E-2 7.12E-3
3.4141E-2 -2.21E-2
3.4141E+0 7.36E-3
3.4141E+0 -2.83E-2
3.4141E+2 1.66E-2
3.4141E+2 -1.80E-2
};

\addplot [color=blue,only marks,mark=triangle*, mark size=2.5pt] table{
3.4141E+0 8.30E-2
3.4141E+0 6.82E-2
3.4141E+2 8.02E-2
3.4141E+2 6.43E-2
};

\addplot [mark=none,lightgray,line width=1.5] table{
3.4141E-2 1.68E-2
3.4141E+0 8.30E-2
3.4141E+2 8.02E-2
};

\addplot [mark=none,gray,line width=1.5] table{
3.4141E-2 6.74E-3
3.4141E+0 6.82E-2
3.4141E+2 6.43E-2
};

\addplot [mark=none,brown,line width=1.5] table{
3.4141E-2 7.12E-3
3.4141E+0 7.36E-3
3.4141E+2 1.66E-2
};

\addplot [mark=none,green,line width=1.5] table{
3.4141E-2 -2.21E-2
3.4141E+0 -2.83E-2
3.4141E+2 -1.80E-2
};

\end{axis}

\end{tikzpicture}\\
\end{tabular}
\end{center}
\mcaption{The time-averaged pseudo-lift $\overline{F}_l$~\eqref{e:liftForce} 
(on the left) and migration velocities $\overline{v}_{\mathrm{mig}}$ (on the right) for RBCs as a function of the capillary number $C_a$ and viscosity contrast $\nu$ for the row-shift fraction $\epsilon = 0.1667$. While the $x$-axis corresponds to $C_a$, lines with different colors correspond to $\nu$. Red circles and blue triangles indicate zig-zagging and displacing cells, respectively.}{f:liftForN6}
\end{figure}

In inertial flows (i.e., $Re \gg 1$) the net force in the direction perpendicular to the free-stream velocity causes a body to drift in that direction. This force is called lift $F_l$ and given by
\begin{equation} \label{e:liftForce}
  F_l(t) = \int_{\gamma} (\mathbf{T}\nn) \mathbf{\cdot} \kk d\gamma,
\end{equation}
where $\gamma$ is the surface of the body, $\mathbf{T}$ is total hydrodynamic stress on the body, $\nn$ is the normal direction on the surface, and $\kk$ is a unit vector perpendicular to the free-stream velocity. In non-inertial flows (i.e., $Re \ll 1$) such as the flow of RBCs in DLD in our study, the net lift force on a body given by~\eqref{e:liftForce} turns out to be zero. By modifying the term $\kk$ in~\eqref{e:liftForce} we define the so-called {\em pseudo-lift} for the flow of cells to quantify the cell migration in DLD. We compute the pseudo-lift $F_l(t)$ at the time $t$ using the same equation~\eqref{e:liftForce}. Instead of a constant $\kk$ along the boundary, we use a $\kk$ varying along the boundary which is a unit vector perpendicular to the velocity on $\gamma$. The other terms in~\eqref{e:liftForce} remain the same. That is, $\mathbf{T}$ is the total hydrodynamic stress on the cell's boundary $\gamma$ (see \appref{a:intEqnForm} for the integral equation formulation to compute $\mathbf{T}$) and $\nn$ is the unit normal vector on $\gamma$. Since $\kk$ varies along the boundary $\gamma$, there is a net pseudo-lift on a migrating cell. We, then, define the time-averaged pseudo-lift $\overline{F}_l$
\begin{equation} \label{e:timeAveliftForce}
  \overline{F}_l = \frac{1}{T}\int_{t_i}^{t_f} F_l(t) dt,
\end{equation}
where $T = t_f - t_i$. Here, we integrate between the time $t_i$ 
when the cell is passing through the second gap and the time $t_f$ which is 
either when a displacing RBC is at the end of a period or when a 
zig-zagging RBC switches a lane. Since $\mathbf{T}\nn$ scales with $\kappa_b R_{\mathrm{eff}}$ (see \eqref{e:tracJump}), we normalize $\overline{F}_l$ with $\kappa_b R_{\mathrm{eff}}$. On the left figure in~\figref{f:liftForN6} we present the time-averaged pseudo-lift on the cells with the capillary numbers  $C_a = (3.41\times 10^{-2}, 3.41, 3.41\times 10^2)$ and viscosity contrasts $\nu = (1, 2, 5, 8)$ for the row-shift fraction $\epsilon = 0.1667$. On the right figure in~\figref{f:liftForN6} we also show the time-averaged migration velocity $\overline{v}_{\mathrm{mig}}$ for these cells. We computed the velocity of a cell's center in the $y$-direction at each time step and averaged the velocity as we did it for the pseudo-lift in~\eqref{e:timeAveliftForce}. The time-averaged inclination angles corresponding to these cells are already presented in~\figref{f:IAgapP6}. In these figures we also marked the zig-zagging and displacing cells with red circles and blue triangles, respectively. The zig-zagging cells have pseudo-lift and migration velocity either very close to zero or less than zero. Whereas, the displacing cells have an order of magnitude greater pseudo-lift and migration velocity. The average migration velocity seems to depend on the capillary number and viscosity contrast as the average inclination angle does so (See~\figref{f:IAgapP6}). That is, the average migration velocity monotonously increases and reaches a plateau for $\nu = (1,2)$, and it decreases then increases for $\nu = (5,8)$. Although the pseudo-lift does not exactly follow this behavior, it can be used as a measure of the cell migration.

{\em Remark.} The normal stress difference (i.e. $N = \langle T_{xx}\rangle - \langle T_{yy}\rangle$), where the angle bracket $\langle \cdot \rangle$ denotes volume average, has been used to quantify cell migration in shear and Poiseuille flows~\citep{ghigliotti-misbah-e10b,ghigliotti-misbah-e10}. In these flows, the normal stress difference is positive during migration of a tank-treading cell and becomes zero when migration ends. So, the positive normal stress difference indicates cell migration. A tumbling cell's inclination angle oscillates periodically, which results in a nonlinear behavior of $N$ in time. In DLD flows, a cell's inclination angle oscillates like a tumbling cell (see~\figref{f:IAsAnalysisVC}). As a result of that, the normal stress difference shows a nonlinear behavior in time in DLD flows. In order to investigate if the normal stress difference indicates whether a cell displaces or zig-zags, we computed the its time average for several cases: cells with $C_a = (0.034,3.4,340)$ and $\nu = (1,2,5,8)$ for $\epsilon = 0.1667$. We found positive average normal stress difference for the displacing cells (those with $(C_a,\nu) = \{(3.4,1),(3.4,2),(340,1),(340,2)\}$). However, the average normal stress difference turned out to be not negative or nearly zero always for zig-zagging cells. The reason is although a zig-zagging cell maintains positive inclination angle most of the time (hence positive average normal stress difference), this may not be able to generate sufficient migration to displace. Overall, we have not observed any stronger correlation between cell's transport mode and the average normal stress difference than the one between the mode and the pseudo-lift.

\section{Conclusion\label{s:conclusions}}
Using an in-house integral equation solver, we have studied 
deformability-based red blood cell (RBC) separation in deep deterministic 
lateral displacement (DLD) devices. Deformability of an RBC is determined by 
the RBC's membrane stiffness and interior fluid viscosity, which correspond to 
the dimensionless numbers: capillary number $C_a$ and viscosity contrast $\nu$
, respectively. We have performed a systematic study to map out the parameter
space for the transport modes as a function of $C_a$, $\nu$ and 
the row-shift fraction $\epsilon$ describing the DLD geometry. 
We have observed that an RBC is either in the displacement mode with 
a steady angular orientation or in the zig-zag mode depending on its $C_a$ 
and $\nu$. This leads RBCs to have a net non-zero or almost zero lateral displacement as they leave a DLD device. Thereby, the cells can be separated laterally. 

We have discussed the mechanism enabling the separation in deep
devices. As already known, RBCs either tank-tread with a steady 
inclination angle or tumble in confined Poiseuille flows. Lateral confinement in the gaps in DLD, however, restricts tumbling and all the cells move with a positive inclination with respect to the flow direction. The degree of the inclination angle depends on the cells' capillary number and viscosity contrast. Positive inclination with respect to the flow and the shear gradient results in cell migration towards low shear gradient region. Cells having higher inclination angles stay farther away from the pillars than those having smaller inclination angles. Since the RBC dynamics in DLD resemble those in confined Poiseuille flows,
we have compared the cell dynamics in these flows. Although there are several similarities between the cell dynamics in these flows (such as the variation of inclination angles with the capillary number and the viscosity contrast), the DLD flow is more complicated. Thus, the cell dynamics in DLD cannot be estimated using simpler flows such as confined Poiseuille flows. Finally, we define the so-called pseudo-lift to quantify the degree of migration. Letting the direction of the row-shift be the positive lateral direction, we have found out 
that strong positive pseudo-lift acts on the displacing RBCs 
while either weak or negative lift force acts on the zig-zagging RBCs. 
Furthermore, 
the magnitude of the pseudo-lift depends on the capillary number and 
viscosity contrast.

We have also performed simulations with dense RBC suspensions for
various capillary numbers, viscosity contrasts and row-shift
fractions. Our findings agree well with the 
numerical~\citep{vernekar-kruger15} and experimental~\citep{inglis-nordon-e11} studies:
\begin{enumerate}
\item\ While a single RBC with a certain viscosity contrast and 
capillary number displaces in its dilute suspension, more than 50\% 
of the same RBCs zig-zags in their dense suspension for the same device. 
\item\ This is not observed for the RBCs that zig-zag in their dilute 
suspensions. The majority of RBCs each of which zig-zags alone still zig-zags
in their dense suspensions. In other words, while the displacement mode 
breaks down as the volume fraction increases, the zig-zag mode seems more robust.
\end{enumerate}

Overall, our study contributes to the understanding of the RBC dynamics in 
deep DLD devices. Thereby, it helps design and optimization of the devices. 
This is important because these devices can be used 
for rapid medical diagnoses and tests as the deformability of a diseased RBC
is evidently different than its healthy counterparts.

\appendix
\section{Integral equation formulation}\label{a:intEqnForm}
We revisit the integral equation formulation
of~\eqref{e:stokesCont}-\eqref{e:tracJump}  in~\citet{kabacaoglu-biros-e17b}. 
We refer the reader to~\figref{f:formulation} for the schematic. Let viscosity contrast be defined as 
$\nu = \eta_{\mathrm{in}}/\eta_{\mathrm{out}}$ between the interior fluid
with viscosity $\eta_{\mathrm{in}}$ and the exterior fluid with
viscosity $\eta_{\mathrm{out}}$. The single and double layer
potentials for Stokes flow (${\mathcal{S}}_{pq}$ and
${\mathcal{D}}_{pq}$, respectively) denote the potential induced by
hydrodynamic densities of  the interfacial force $\ff$  and velocity
$\uu$ on cell $q$ and evaluated on cell $p$:

\begin{subequations} \label{e:potentials}
\begin{alignat}{1}
{\mathcal{S}}_{pq}[\ff](\xx) & := \frac{1}{4\upi \eta_{\mathrm{out}}}
\int_{{\gamma}_q} \left(-\mathbf{I} \log \rho + \frac{\rr \otimes
\rr}{{\rho}^2}\right) \ff(\yy) ds_{\yy}, \quad 
\xx  \in {\gamma}_p, \label{e:singleLayer} \\
{\mathcal{D}}_{pq}[\uu](\xx) & := \frac{1-{\nu}_q}{\upi}
\int_{{\gamma}_q} \frac{\rr \cdot \nn}{{\rho}^2} \frac{\rr \otimes
\rr}{{\rho}^2} \uu(\yy) ds_{\yy}, \quad  \xx \in {\gamma}_p, \label{e:doubleLayer}
\end{alignat}%
\end{subequations}
where $\rr = \xx - \yy$ and $\rho = \|\rr\|_2$. Let ${\mathcal{S}}_p := {\mathcal{S}}_{pp}$ and ${\mathcal{D}}_p := {\mathcal{D}}_{pp}$ denote cell self-interactions. We, then, define
\begin{alignat*}{1}
& {\mathcal{E}}_{pq}[\ff, \uu](\xx) = {\mathcal{S}}_{pq}[\ff](\xx) + {\mathcal{D}}_{pq}[\uu](\xx), \quad  \xx \in {\gamma}_p,\\
& {\mathcal{E}}_p [\ff, \uu](\xx) = \sum_{q=1}^M {\mathcal{E}}_{pq} [\ff, \uu](\xx), \quad \xx \in {\gamma}_p.
\end{alignat*}%
For confined flows such as flows in DLD devices, we use the completed
double layer potential due to a density function $\zzeta$ defined on
the solid walls
\begin{align*}
\mathcal{B}[\zzeta](\xx) = {\mathcal{D}}_{\Gamma}[\zzeta](\xx) + \sum_{q=1}^M R\left[ {\xi}_q(\zzeta), {\cc}_q\right](\xx) + \sum_{q=1}^M S\left[ {\llambda}_q(\zzeta), {\cc}_q\right](\xx), \quad \xx \in \gamma \cup \Gamma.
\end{align*}
The Stokeslets and rotlets are
\begin{align*}
S\left[ {\llambda}_q(\zzeta), {\cc}_q\right](\xx) =
\frac{1}{4\pi\eta_{\mathrm{out}}} \left(-\log \rho + \frac{\rr \otimes \rr}{{\rho}^2}\right){\llambda}_q(\zzeta) \quad \text{and} \quad R\left[ {\xi}_q(\zzeta), {\cc}_q\right](\xx) = \frac{{\xi}_q(\zzeta)}{\eta_{\mathrm{out}}}\frac{{\rr}^{\perp}}{{\rho}^2},
\end{align*}
where ${\cc}_q$ is a point inside ${\omega}_q$, $\rr = \xx -
{\cc}_q$, and ${\rr}^{\perp} = (r_2, -r_1)$. The size of the
Stokeslets and rotlets are 
\begin{align*} 
{\llambda}_{q,i} = \frac{1}{2\upi} \int_{{\gamma}_q} {\zzeta}_i (\yy)
ds_{\yy}, \,\, i = 1,2 \quad \text{and} \quad {\xi}_q =
\frac{1}{2\upi}\int_{{\gamma}_q} {\yy}^{\perp} \cdot \zzeta (\yy) ds_{\yy}.
\end{align*}
If $\xx \in {\Gamma}_0$, we add the rank one modification
${\mathcal{N}}_0[\zzeta](\xx) = \int_{{\Gamma}_0}
\left(\nn(\xx)\otimes\nn(\yy)\right)\zzeta(\yy)ds_{\yy}$ to
$\mathcal{B}$ to remove a one-dimensional null space.  Finally, by
expressing the inextensibility constraint in operator form as 
\begin{align*} 
\mathcal{P}[\uu](\xx) = {\xx}_s \cdot {\uu}_s,
\end{align*}
the integral equation formulation of~\eqref{e:stokesCont}-\eqref{e:tracJump} is
\begin{subequations} \label{e:integEquation}
\begin{alignat}{1}
 &\left(1+{\nu}_p\right) \uu(\xx) = {\mathcal{E}}_p [\ff,\uu](\xx) + {\mathcal{B}}_p [\zzeta](\xx), \quad  \xx  \in {\gamma}_p, \,\, \text{cell evolution,}\\
&\left(1+{\nu}_p\right) \UU(\xx) = -\frac{1}{2} \zzeta(\xx) + {\mathcal{E}}_{\Gamma} [\ff,\uu](\xx) + \mathcal{B}[\zzeta](\xx), \quad  \xx \in \Gamma, \,\, \text{fixed boundaries,}\\
&\mathcal{P}[\uu](\xx)  = 0, \quad \xx \in {\gamma}_p, \,\,  \text{cell inextensibility.}
\end{alignat}%
\end{subequations}
Since the velocity
$\uu = d\xx/dt$ and the interfacial force $\ff$ depend on $\sigma$ and
$\xx$,~\eqref{e:integEquation} is a system of
integro-differential-algebraic equations for $\xx, \sigma$, and $\zzeta$.

{\em Computing total stress on a membrane.}
In order to compute the hydrodynamic lift on a cell's membrane we need to compute the total stress 
$\mathbf{T} = -p\mathbf{I} + \eta (\Grad \uu + \Grad {\uu}^T)$. We revisit the integral 
equation formulation for the stress from \citet{quaife-biros14}. The stress ${\mathbf{T}}^S$ 
of the single-layer potential $\uu (\xx) = \mathcal{S}[\ff](\xx)$ is 
\begin{equation} \label{e:slpStress}
  {\mathbf{T}}^S [\ssigma](\xx) = \frac{1}{\pi}\int_{\gamma}\frac{\rr \mathbf{\cdot} \ssigma}{\rho^2}\frac{\rr \otimes \rr}{\rho^2}\ff ds_{\yy},
\end{equation}
which gives the stress at $\xx$ due to a cell with an interfacial force $\ff$. 
The stress ${\mathbf{T}}^D$ of the double-layer potential $\uu(\xx) = \mathcal{D}[\zzeta](\xx)$ is 
\begin{multline}
  {\mathbf{T}}^D [\ssigma](\xx) = \frac{1}{\pi}\sum_{q = 1}^M\int_{\gamma_q} \left( \frac{\nn\mathbf{\cdot}\zzeta_q}{\rho^2}\ssigma - \frac{8}{\rho^6}(\rr \mathbf{\cdot} \nn)(\rr \mathbf{\cdot} \zzeta_q) (\rr \mathbf{\cdot} \ssigma)\rr + \frac{\rr \mathbf{\cdot} \nn}{\rho^4} (\rr \otimes \zzeta_q + \zzeta_q \otimes \rr)\ssigma \right)\\ 
  + \left(\frac{\rr \mathbf{\cdot}\zzeta_q}{\rho^4}(\rr \otimes \nn + \nn \otimes \rr)\ssigma \right) ds_{\yy},
\end{multline}
which is the stress at $\xx$ due to the pillars with density $\zzeta$. 
Total stress is the summation of the stresses due to a cell, pillars and an exterior wall.

{\em Temporal discretization.}
We linearize~\eqref{e:integEquation} and treat the stiff terms, such as
the bending, implicitly, while treating nonlinear terms, such as the
layer potential kernel, explicitly.  In particular, an approximation
for the position $\xx$ and tension $\sigma$ of cell $p$ at time
$n+1$ is computed by solving
\begin{subequations} \label{e:temporalDiscretization}
\begin{alignat}{1}
& \frac{{\alpha}_p}{\Delta t} \left({\xx}_p^{n+1} - {\xx}_p^n\right)  = {\mathcal{S}}_p^n {\ff}_p^{n+1} + {\mathcal{D}}_p^n{\uu}_p^{n+1} + {\mathcal{B}}_p[{\zzeta}^{n+1}] + \sum_{\substack{q = 1 \\ q \neq p}}^M \, {\mathcal{E}}_{pq}^n [{\ff}_q^{n+1}, {\uu}_q^{n+1}], \,\,  \xx \in {\gamma}_p \\
& \UU^{n+1}(\xx) = -\frac{1}{2} {\zzeta}^{n+1}(\xx) + {\mathcal{E}}_{\Gamma}^n [{\ff}^{n+1},{\uu}^{n+1}](\xx) + \mathcal{B}[{\zzeta}^{n+1}](\xx) + {\mathcal{N}}_0 [{\zzeta}^{n+1}](\xx), \,\, \xx \in \Gamma, \\
& {\mathcal{P}}^n {\xx}_p^{n+1}  = {\mathcal{P}}^n {\xx}_p^n, \,\,  \xx  \in {\gamma}_p, \\
& {\uu}_p^{n+1} = \frac{{\xx}_p^{n+1} - {\xx}_p^n}{\Delta t}, \,\, \xx \in {\gamma}_p,
\end{alignat}
\end{subequations}
where ${\alpha}_p = (1+{\nu}_p)/2$, and operators with a superscript
$n$ are discretized at ${\xx}^n$.

{\em Spatial discretization.}
Let $\xx(\theta)$, $\theta \in (0, 2\upi]$ be a parametrization of the
interface ${\gamma}_p$, and let$\{{\xx(\theta}_k) = 2k\upi/N\}_{k=1}^N$
be $N$ uniformly distributed discretization points. Then, a spectral
representation of the cell membrane is given by
\begin{equation*} \label{e:spatialDiscretization}
  \xx (\theta) = \sum_{k = -N/2 + 1}^{N/2} \hat{\xx}(k)e^{ik\theta}.
\end{equation*}
We use the fast Fourier transform to compute $\hat{\xx}$, and
arc-length derivatives are computed pseudo-spectrally. While we use an interpolation 
scheme~\citep{quaife-biros14} for nearly singular
integrals, we use a Gauss-trapezoid
quadrature rule with accuracy $\mathcal{O}(h^8 \log h)$
to evaluate the single layer potential and the spectrally accurate
trapezoid rule for the double layer potential.

\section{Wall effects\label{a:appendix2}}
Period ($n_p = 1/\epsilon$) of a DLD device is determined by its row-shift
fraction $\epsilon = \Delta \lambda/ \lambda$, where $\Delta \lambda$ is the
amount of the lateral shift and $\lambda$ is the center-to-center distance
between two pillars. Each pillar in the next column is shifted by $\Delta
\lambda$ with respect to the previous one and hence the arrangement of the
pillars in a column is repeated after $n_p$ columns. Since we are interested in
the RBC behavior in a period, we reduce the domain of the DLD simulations to a
single period. Due to the Stokes' paradox, we have to confine the pillars with
an exterior wall. Confining the pillars introduces wall effects which spoil
the underlying physics and hence have to be avoided. Here, we want to find the
sufficient amount of rows and columns that must be included in our domain to
minimize the wall effects.

\begin{figure}
\begin{center}
\begin{tabular}{c c}
 \input{figs/rowIAconvLowVC.tikz} & \input{figs/rowIAconvHighVC.tikz} \\
\input{figs/rowTRconvLowVC.tikz} & \input{figs/rowTRconvHighVC.tikz} \\
\end{tabular}
\end{center}
\mcaption{ Effects of the top and bottom walls on the RBC behavior. We performed
simulations of RBCs with the capillary number $C_a = 0.648$ and the viscosity
contrasts $\nu = 0.1$ (on the left) and $\nu = 10$ (on the right). The DLD
device has $n_{\mathrm{col}} = \left \lceil 1.5p \right \rceil$ number of
columns and its row-shift fraction is $\epsilon = 0.1667$, i.e., the period $n_p = 6$. As increasing the number of rows $n_{\mathrm{row}}$ we plotted the RBCs'
inclination angles (on the top) and trajectories (on the bottom). While the
trajectories and inclination angles are evidently different for the number of
rows $n_{\mathrm{row}} = 8$ and $n_{\mathrm{row}} = 10$, there is no significant
difference between the results given by $n_{\mathrm{row}} = 12$ and
$n_{\mathrm{row}} = 16$. Therefore, we decided to include $n_{\mathrm{row}} = 12$ number of rows in our domain. $x$ and $y$ coordinates are normalized by the
length of the device $L$.}{f:rowConv}
\end{figure}
We considered a DLD device with the row-shift fraction $\epsilon = 0.1667$, i.e.,
$n_p = 6$. We performed simulations of RBCs with the capillary number 
$C_a = 0.648$ and two different viscosity contrasts $\nu = 0.1$ and $\nu = 10$. The RBC
with the lower viscosity contrast displaces and the one with the higher
viscosity contrast zig-zags. We, first, fixed the number of columns to
$n_{\mathrm{col}} = \left \lceil 1.5 p \right \rceil$ and varied the number of
rows $n_{\mathrm{row}} = (8, 10, 12, 16)$. We present the inclination angles 
(at the top) and trajectories (at the bottom) of the RBCs during their motions
in~\figref{f:rowConv}. Here, the results on the left are for $\nu = 0.1$ and
those on the right are for $\nu = 10$. The displacing RBC ($\nu = 0.1$) shows
different transitions in the devices with $n_{\mathrm{row}} = (8, 10)$ than
those with $n_{\mathrm{row}} = (12, 16)$. But the RBCs have the same
trajectories and inclination angles after the transition for all
$n_{\mathrm{row}}$. The zig-zagging RBC ($\nu = 10$) follows similar
trajectories with similar variations in the inclination angles in the devices
with $n_{\mathrm{row}} = (12, 16)$. Since the trajectories and inclination
angles do not change much after $n_{\mathrm{row}} \geq 12$, we decided to
include $n_{\mathrm{row}} = 12$ rows in our DLD domain.

\begin{figure}
\begin{center}
\begin{tabular}{c c}
 \begin{tikzpicture}

\begin{axis}[
  width = 0.45\textwidth,
  title = {$\nu = 0.1$},
  xmin = 0,
  xmax = 2, 
  ymin = -2,
  ymax = 2,
  xtick = {0,0.5,1,1.5,2},
  ytick = {-2,-1.5,-1,-0.5,0,0.5,1,1.5,2},
  ylabel = {Inclination angle ($\alpha$) [rad]},
  xlabel = {$x/L_p$},
  label style={font=\normalsize},
  legend style={anchor=east,draw=none},
  legend entries={$n_{\mathrm{col}} = p\quad\quad\,$,$n_{\mathrm{col}} = \left \lceil 1.5p \right \rceil$,$n_{\mathrm{col}} = 2p\quad\,\,\,$},
  legend pos = south east,
  ]
\addplot [mark=none,black,line width=1.5] table{

         0   -1.5708
    0.0002   -1.5599
    0.0015   -1.5010
    0.0051   -1.3431
    0.0097   -1.1748
    0.0162   -1.0023
    0.0260   -0.8284
    0.0387   -0.6541
    0.0535   -0.4554
    0.0721   -0.2451
    0.1003   -0.0939
    0.1501   -0.0801
    0.1867    0.0090
    0.2070    0.1041
    0.2243    0.1053
    0.2367    0.0748
    0.2494    0.0418
    0.2624    0.0252
    0.2758    0.0491
    0.2880    0.1387
    0.2967    0.2603
    0.3045    0.4078
    0.3115    0.5519
    0.3184    0.6726
    0.3295    0.7698
    0.3426    0.7384
    0.3519    0.6419
    0.3596    0.5416
    0.3672    0.4448
    0.3742    0.3648
    0.3819    0.2942
    0.3900    0.2408
    0.3983    0.2118
    0.4066    0.2131
    0.4146    0.2491
    0.4221    0.3190
    0.4289    0.4161
    0.4351    0.5280
    0.4406    0.6377
    0.4469    0.7463
    0.4557    0.8435
    0.4710    0.8854
    0.4808    0.8387
    0.4884    0.7576
    0.4946    0.6741
    0.5012    0.5773
    0.5067    0.4997
    0.5129    0.4205
    0.5197    0.3467
    0.5261    0.2920
    0.5316    0.2561
    0.5371    0.2337
    0.5427    0.2244
    0.5491    0.2315
    0.5554    0.2613
    0.5615    0.3136
    0.5671    0.3851
    0.5723    0.4698
    0.5779    0.5759
    0.5834    0.6809
    0.5901    0.7800
    0.5998    0.8602
    0.6138    0.8824
    0.6224    0.8381
    0.6292    0.7674
    0.6350    0.6892
    0.6407    0.6058
    0.6460    0.5304
    0.6519    0.4522
    0.6583    0.3773
    0.6632    0.3299
    0.6684    0.2892
    0.6738    0.2578
    0.6793    0.2375
    0.6847    0.2303
    0.6907    0.2389
    0.6969    0.2694
    0.7027    0.3213
    0.7082    0.3913
    0.7135    0.4787
    0.7189    0.5808
    0.7244    0.6834
    0.7310    0.7789
    0.7406    0.8564
    0.7539    0.8849
    0.7623    0.8521
    0.7692    0.7869
    0.7744    0.7206
    0.7799    0.6401
    0.7848    0.5690
    0.7902    0.4934
    0.7962    0.4179
    0.8009    0.3678
    0.8058    0.3223
    0.8110    0.2843
    0.8163    0.2556
    0.8216    0.2382
    0.8269    0.2335
    0.8328    0.2448
    0.8388    0.2772
    0.8446    0.3282
    0.8499    0.3980
    0.8554    0.4888
    0.8606    0.5871
    0.8661    0.6866
    0.8728    0.7795
    0.8822    0.8531
    0.8944    0.8862
    0.9026    0.8621
    0.9087    0.8101
    0.9142    0.7419
    0.9188    0.6772
    0.9240    0.6004
    0.9286    0.5349
    0.9332    0.4728
    0.9382    0.4114
    0.9437    0.3534
    0.9494    0.3018
    0.9554    0.2590
    0.9616    0.2265
    0.9677    0.2068
    0.9737    0.1999
    0.9795    0.2060
    0.9855    0.2263
    0.9916    0.2615
    0.9979    0.3102

};

\addplot [mark=none,cyan,line width=1.5] table{
      
         0   -1.5708
    0.0011   -1.5197
    0.0081   -1.2312
    0.0208   -0.9172
    0.0437   -0.5946
    0.0690   -0.2697
    0.1271   -0.0722
    0.1915    0.0617
    0.2241    0.1337
    0.2461    0.0680
    0.2696    0.0495
    0.2905    0.1998
    0.3030    0.4216
    0.3141    0.6460
    0.3346    0.8030
    0.3517    0.6645
    0.3649    0.4886
    0.3776    0.3426
    0.3922    0.2387
    0.4073    0.2196
    0.4213    0.3104
    0.4334    0.4900
    0.4438    0.6882
    0.4589    0.8588
    0.4810    0.8487
    0.4926    0.7183
    0.5027    0.5716
    0.5133    0.4280
    0.5230    0.3286
    0.5327    0.2617
    0.5427    0.2314
    0.5535    0.2515
    0.5643    0.3415
    0.5741    0.4916
    0.5839    0.6756
    0.5975    0.8376
    0.6176    0.8845
    0.6291    0.7934
    0.6384    0.6657
    0.6474    0.5338
    0.6562    0.4203
    0.6664    0.3199
    0.6773    0.2528
    0.6884    0.2367
    0.6990    0.2819
    0.7092    0.3947
    0.7188    0.5592
    0.7291    0.7352
    0.7448    0.8685
    0.7619    0.8826
    0.7716    0.7916
    0.7797    0.6761
    0.7879    0.5512
    0.7958    0.4459
    0.8050    0.3477
    0.8150    0.2732
    0.8254    0.2375
    0.8355    0.2517
    0.8464    0.3344
    0.8560    0.4732
    0.8653    0.6430
    0.8769    0.7952
    0.8935    0.9004
    0.9059    0.8818
    0.9139    0.7923
    0.9209    0.6875
    0.9283    0.5758
    0.9355    0.4765
    0.9438    0.3815
    0.9530    0.3017
    0.9627    0.2525
    0.9723    0.2448
    0.9825    0.2921
    0.9923    0.4008
    1.0016    0.5571
    1.0114    0.7173
    1.0252    0.8445
    1.0405    0.9126
    1.0496    0.8665
    1.0566    0.7752
    1.0630    0.6773
    1.0699    0.5721
    1.0768    0.4767
    1.0847    0.3836
    1.0936    0.3061
    1.1029    0.2562
    1.1122    0.2443
    1.1220    0.2829
    1.1315    0.3823
    1.1408    0.5344
    1.1503    0.6978
    1.1634    0.8281
    1.1793    0.9069
    1.1894    0.8674
    1.1969    0.7753
    1.2036    0.6736
    1.2110    0.5613
    1.2179    0.4674
    1.2258    0.3768
    1.2345    0.3026
    1.2436    0.2559
    1.2528    0.2466
    1.2627    0.2893
    1.2721    0.3920
    1.2815    0.5512
    1.2912    0.7180
    1.3055    0.8471
    1.3226    0.8903
    1.3323    0.8215
    1.3403    0.7134
    1.3483    0.5933
    1.3562    0.4826
    1.3654    0.3750
    1.3758    0.2894
    1.3866    0.2436
    1.3972    0.2538
    1.4072    0.3265
    1.4171    0.4671
    1.4264    0.6407
    1.4382    0.8012
    1.4577    0.8896
    1.4707    0.8324
    1.4798    0.7179
    1.4887    0.5874
    1.4969    0.4731

};

\addplot [mark=none,red,dashed,line width=1.5] table{
         0   -1.5708
    0.0066   -1.2857
    0.0347   -0.7121
    0.0845   -0.1466
    0.2040    0.1208
    0.2504    0.0544
    0.2908    0.1974
    0.3124    0.6063
    0.3469    0.7192
    0.3710    0.4121
    0.3972    0.2203
    0.4235    0.3340
    0.4437    0.6848
    0.4774    0.8720
    0.4991    0.6244
    0.5176    0.3803
    0.5350    0.2510
    0.5542    0.2550
    0.5728    0.4680
    0.5924    0.7929
    0.6245    0.8439
    0.6413    0.6232
    0.6573    0.4099
    0.6765    0.2569
    0.6964    0.2641
    0.7148    0.4830
    0.7340    0.7893
    0.7634    0.8783
    0.7788    0.6928
    0.7931    0.4831
    0.8098    0.3094
    0.8285    0.2368
    0.8474    0.3441
    0.8644    0.6225
    0.8882    0.8774
    0.9100    0.8503
    0.9229    0.6628
    0.9358    0.4770
    0.9515    0.3153
    0.9690    0.2425
    0.9871    0.3314
    1.0040    0.5919
    1.0255    0.8452
    1.0475    0.8986
    1.0594    0.7435
    1.0711    0.5612
    1.0846    0.3906
    1.1008    0.2672
    1.1174    0.2568
    1.1349    0.4254
    1.1520    0.7118
    1.1772    0.9115
    1.1939    0.8326
    1.2055    0.6558
    1.2187    0.4654
    1.2313    0.3324
    1.2454    0.2545
    1.2615    0.2791
    1.2788    0.4908
    1.2971    0.7761
    1.3245    0.9000
    1.3393    0.7467
    1.3525    0.5462
    1.3674    0.3671
    1.3813    0.2679
    1.3969    0.2567
    1.4149    0.4250
    1.4325    0.7274
    1.4616    0.8993
    1.4787    0.7550
    1.4932    0.5382
    1.5074    0.3665
    1.5236    0.2551
    1.5408    0.2629
    1.5588    0.4666
    1.5772    0.7688
    1.6064    0.8927
    1.6219    0.7248
    1.6358    0.5178
    1.6518    0.3364
    1.6702    0.2403
    1.6891    0.3113
    1.7066    0.5718
    1.7283    0.8400
    1.7519    0.8844
    1.7642    0.7202
    1.7765    0.5353
    1.7910    0.3619
    1.8082    0.2507
    1.8257    0.2714
    1.8432    0.4772
    1.8612    0.7510
    1.8845    0.9347
    1.8978    0.8607
    1.9083    0.6950
    1.9189    0.5331
    1.9319    0.3744
    1.9437    0.2836
    1.9571    0.2476
    1.9738    0.3414
    1.9900    0.5980

};

\end{axis}

\end{tikzpicture} & \begin{tikzpicture}

\begin{axis}[
  width = 0.45\textwidth,
  title = {$\nu = 10$},
  xmin = 0,
  xmax = 2, 
  ymin = -2,
  ymax = 2,
  xtick = {0,0.5,1,1.5,2},
  ytick = {-2,-1.5,-1,-0.5,0,0.5,1,1.5,2},
  ylabel = {Inclination angle ($\alpha$) [rad]},
  xlabel = {$x/L_p$},
  label style={font=\normalsize},
  ]

\addplot [mark=none,black,line width=1.5] table{

         0   -1.5708
    0.0016   -1.5127
    0.0049   -1.4065
    0.0088   -1.3019
    0.0134   -1.1997
    0.0194   -1.0960
    0.0270   -0.9933
    0.0368   -0.8851
    0.0481   -0.7664
    0.0600   -0.6344
    0.0708   -0.5150
    0.0842   -0.3919
    0.1123   -0.2781
    0.1493   -0.3924
    0.1727   -0.5506
    0.1920   -0.6573
    0.2185   -0.6374
    0.2474   -0.6192
    0.2596   -0.7086
    0.2685   -0.8263
    0.2743   -0.9290
    0.2801   -1.0471
    0.2854   -1.1668
    0.2904   -1.2785
    0.2957   -1.3868
    0.3016   -1.4931
    0.3084    1.5427
    0.3157    1.4394
    0.3233    1.3327
    0.3306    1.2226
    0.3373    1.1066
    0.3424    1.0100
    0.3478    0.9027
    0.3537    0.7868
    0.3581    0.7020
    0.3631    0.6110
    0.3687    0.5195
    0.3747    0.4307
    0.3812    0.3484
    0.3882    0.2755
    0.3969    0.2074
    0.4083    0.1544
    0.4324    0.1935
    0.4511    0.3743
    0.4659    0.5504
    0.4829    0.6508
    0.4924    0.5964
    0.4988    0.5245
    0.5047    0.4493
    0.5113    0.3620
    0.5187    0.2715
    0.5267    0.1834
    0.5350    0.1042
    0.5434    0.0389
    0.5521   -0.0123
    0.5636   -0.0449
    0.5861    0.0473
    0.5988    0.2183
    0.6079    0.4131
    0.6156    0.5898
    0.6250    0.7089
    0.6321    0.6811
    0.6368    0.6244
    0.6412    0.5619
    0.6455    0.4964
    0.6504    0.4205
    0.6547    0.3568
    0.6589    0.2964
    0.6634    0.2353
    0.6681    0.1751
    0.6730    0.1170
    0.6779    0.0625
    0.6830    0.0098
    0.6886   -0.0425
    0.6946   -0.0895
    0.7014   -0.1297
    0.7101   -0.1557
    0.7246   -0.1243
    0.7353   -0.0469
    0.7416    0.0409
    0.7450    0.1559
    0.7469    0.2965
    0.7483    0.4384
    0.7494    0.5641
    0.7506    0.6756
    0.7520    0.7867
    0.7535    0.8958
    0.7551    1.0035
    0.7567    1.1108
    0.7584    1.2206
    0.7601    1.3293
    0.7619    1.4489
    0.7639   -1.5676
    0.7659   -1.4425
    0.7679   -1.3205
    0.7700   -1.2051
    0.7718   -1.1181
    0.7738   -1.0277
    0.7762   -0.9341
    0.7791   -0.8375
    0.7821   -0.7515
    0.7850   -0.6784
    0.7883   -0.6039
    0.7921   -0.5278
    0.7963   -0.4507
    0.8010   -0.3732
    0.8062   -0.2962
    0.8117   -0.2212
    0.8175   -0.1496
    0.8234   -0.0829
    0.8293   -0.0221
    0.8350    0.0309
    0.8413    0.0807
    0.8488    0.1226
    0.8623    0.1230
    0.8809   -0.0510
    0.9016   -0.1717
    0.9286   -0.1629
    0.9465   -0.0914
    0.9597   -0.0310
    0.9752    0.0257
        
};

\addplot [mark=none,cyan,line width=1.5] table{

         0   -1.5708
    0.0016   -1.5128
    0.0051   -1.4068
    0.0090   -1.3027
    0.0138   -1.2015
    0.0200   -1.0977
    0.0280   -0.9947
    0.0384   -0.8843
    0.0504   -0.7600
    0.0615   -0.6346
    0.0729   -0.5089
    0.0872   -0.3840
    0.1281   -0.3107
    0.1546   -0.4543
    0.1743   -0.6105
    0.1903   -0.7234
    0.2176   -0.7611
    0.2533   -0.7649
    0.2629   -0.8605
    0.2705   -0.9777
    0.2761   -1.0861
    0.2815   -1.1980
    0.2865   -1.3030
    0.2922   -1.4114
    0.2989   -1.5193
    0.3067    1.5181
    0.3154    1.4150
    0.3243    1.3044
    0.3322    1.1914
    0.3383    1.0885
    0.3445    0.9716
    0.3492    0.8786
    0.3542    0.7788
    0.3598    0.6741
    0.3660    0.5672
    0.3728    0.4621
    0.3803    0.3648
    0.3883    0.2813
    0.3971    0.2133
    0.4090    0.1620
    0.4356    0.2239
    0.4544    0.4169
    0.4695    0.5888
    0.4883    0.6392
    0.4952    0.5759
    0.5013    0.5003
    0.5070    0.4239
    0.5133    0.3422
    0.5203    0.2578
    0.5277    0.1765
    0.5354    0.1033
    0.5431    0.0419
    0.5515   -0.0096
    0.5624   -0.0474
    0.5853    0.0199
    0.5986    0.1843
    0.6064    0.3623
    0.6128    0.5432
    0.6188    0.6986
    0.6276    0.7778
    0.6322    0.7377
    0.6362    0.6810
    0.6398    0.6234
    0.6439    0.5554
    0.6474    0.4962
    0.6513    0.4325
    0.6557    0.3644
    0.6604    0.2938
    0.6655    0.2226
    0.6708    0.1528
    0.6763    0.0863
    0.6817    0.0248
    0.6871   -0.0303
    0.6922   -0.0777
    0.6980   -0.1226
    0.7045   -0.1610
    0.7124   -0.1864
    0.7243   -0.1848
    0.7393   -0.1753
    0.7450   -0.2176
    0.7488   -0.1833
    0.7524   -0.1053
    0.7580   -0.0601
    0.7702   -0.0843
    0.7810   -0.1017
    0.7888   -0.0893
    0.7955   -0.0608
    0.8009   -0.0273
    0.8067    0.0166
    0.8118    0.0608
    0.8162    0.1031
    0.8207    0.1485
    0.8251    0.1950
    0.8293    0.2432
    0.8334    0.2910
    0.8373    0.3367
    0.8415    0.3848
    0.8455    0.4289
    0.8500    0.4697
    0.8551    0.4985
    0.8646    0.4537
    0.8721    0.2979
    0.8778    0.1295
    0.8843   -0.0265
    0.9047   -0.1698
    0.9298   -0.1494
    0.9423   -0.0877
    0.9532   -0.0204
    0.9631    0.0461
    0.9727    0.1135
    0.9821    0.1783
    0.9931    0.2380     

};

\addplot [mark=none,red,dashed,line width=1.5] table{
          0   -1.5708
    0.0043   -1.4257
    0.0118   -1.2367
    0.0228   -1.0510
    0.0407   -0.8548
    0.0618   -0.6259
    0.0842   -0.4016
    0.1429   -0.3828
    0.1780   -0.6534
    0.2103   -0.7995
    0.2542   -0.8360
    0.2689   -1.0340
    0.2808   -1.2819
    0.2914   -1.4908
    0.3048    1.4620
    0.3214    1.2772
    0.3359    1.0809
    0.3471    0.8852
    0.3563    0.7134
    0.3672    0.5280
    0.3803    0.3493
    0.3952    0.2107
    0.4203    0.1349
    0.4584    0.4418
    0.4856    0.6550
    0.4987    0.5381
    0.5093    0.3968
    0.5201    0.2612
    0.5305    0.1476
    0.5432    0.0367
    0.5592   -0.0501
    0.5916    0.0607
    0.6076    0.3890
    0.6174    0.7092
    0.6277    0.8296
    0.6345    0.7429
    0.6407    0.6344
    0.6471    0.5236
    0.6541    0.4052
    0.6607    0.3024
    0.6680    0.1982
    0.6757    0.0983
    0.6835    0.0079
    0.6924   -0.0819
    0.7026   -0.1599
    0.7163   -0.2058
    0.7339   -0.2168
    0.7418   -0.3055
    0.7465   -0.3499
    0.7515   -0.2745
    0.7600   -0.1904
    0.7778   -0.1892
    0.7915   -0.1406
    0.8021   -0.0621
    0.8129    0.0381
    0.8211    0.1260
    0.8288    0.2150
    0.8366    0.3074
    0.8446    0.3983
    0.8537    0.4726
    0.8709    0.3363
    0.8824    0.0545
    0.9115   -0.1550
    0.9432   -0.0722
    0.9667    0.0702
    0.9881    0.2057
    1.0263    0.1782
    1.0915    0.0044
    1.1476    0.1653
    1.2037    0.1083
    1.2717    0.0224
    1.3294    0.2359
    1.3579    0.1136
    1.3822   -0.0196
    1.4188   -0.0978
    1.4566    0.1657
    1.4841    0.2756
    1.4977    0.1673
    1.5117    0.0366
    1.5262   -0.0955
    1.5397   -0.2069
    1.5543   -0.3015
    1.5705   -0.3918
    1.5810   -0.5459
    1.5864   -0.7350
    1.5906   -0.9450
    1.5944   -1.1648
    1.5977   -1.3736
    1.6006    1.5562
    1.6027    1.3997
    1.6038    1.3661
    1.6045    1.3861
    1.6050    1.4383
    1.6057    1.5214
    1.6070   -1.4893
    1.6090   -1.3338
    1.6114   -1.1875
    1.6144   -1.0461
    1.6184   -0.8960
    1.6224   -0.7743
    1.6279   -0.6398
    1.6337   -0.5234
    1.6401   -0.4090
    1.6478   -0.2872
    1.6566   -0.1638
    1.6659   -0.0449
    1.6752    0.0620
    1.6847    0.1525
    1.6971    0.2171
    1.7216    0.0021
    1.7551   -0.1862
    1.7887   -0.0907
    1.8131    0.0290

};

\end{axis}

\end{tikzpicture} \\
\begin{tikzpicture}

\begin{axis}[
  width = 0.45\textwidth,
  title = {$\nu = 0.1$},
  xmin = 0,
  xmax = 2, 
  ymin = -0.1,
  ymax = 0.3,
  xtick = {0,0.5,1,1.5,2},
  ytick = {-0.1,0,0.1,0.2,0.3},
  ylabel = {$y/L_p$},
  xlabel = {$x/L_p$},
  label style={font=\normalsize},
  legend style={anchor=east,draw=none},
  legend pos = south east,
  ]

\addplot [mark=none,black,line width=1.5] table{
         0   -0.0000
    0.0002    0.0000
    0.0015    0.0000
    0.0051    0.0002
    0.0097    0.0005
    0.0162    0.0011
    0.0260    0.0017
    0.0387    0.0019
    0.0535    0.0016
    0.0721    0.0008
    0.1003   -0.0002
    0.1501    0.0002
    0.1867    0.0054
    0.2070    0.0103
    0.2243    0.0138
    0.2367    0.0155
    0.2494    0.0164
    0.2624    0.0165
    0.2758    0.0161
    0.2880    0.0159
    0.2967    0.0160
    0.3045    0.0162
    0.3115    0.0165
    0.3184    0.0170
    0.3295    0.0187
    0.3426    0.0225
    0.3519    0.0262
    0.3596    0.0292
    0.3672    0.0319
    0.3742    0.0339
    0.3819    0.0355
    0.3900    0.0366
    0.3983    0.0373
    0.4066    0.0375
    0.4146    0.0376
    0.4221    0.0376
    0.4289    0.0376
    0.4351    0.0375
    0.4406    0.0372
    0.4469    0.0367
    0.4557    0.0364
    0.4710    0.0382
    0.4808    0.0413
    0.4884    0.0447
    0.4946    0.0476
    0.5012    0.0506
    0.5067    0.0529
    0.5129    0.0550
    0.5197    0.0569
    0.5261    0.0581
    0.5316    0.0589
    0.5371    0.0593
    0.5427    0.0596
    0.5491    0.0598
    0.5554    0.0599
    0.5615    0.0599
    0.5671    0.0600
    0.5723    0.0599
    0.5779    0.0596
    0.5834    0.0592
    0.5901    0.0585
    0.5998    0.0583
    0.6138    0.0605
    0.6224    0.0638
    0.6292    0.0671
    0.6350    0.0701
    0.6407    0.0729
    0.6460    0.0752
    0.6519    0.0775
    0.6583    0.0794
    0.6632    0.0806
    0.6684    0.0815
    0.6738    0.0822
    0.6793    0.0826
    0.6847    0.0829
    0.6907    0.0830
    0.6969    0.0831
    0.7027    0.0831
    0.7082    0.0831
    0.7135    0.0830
    0.7189    0.0826
    0.7244    0.0820
    0.7310    0.0812
    0.7406    0.0808
    0.7539    0.0826
    0.7623    0.0856
    0.7692    0.0889
    0.7744    0.0917
    0.7799    0.0946
    0.7848    0.0970
    0.7902    0.0993
    0.7962    0.1014
    0.8009    0.1027
    0.8058    0.1038
    0.8110    0.1047
    0.8163    0.1053
    0.8216    0.1057
    0.8269    0.1059
    0.8328    0.1061
    0.8388    0.1061
    0.8446    0.1061
    0.8499    0.1061
    0.8554    0.1058
    0.8606    0.1054
    0.8661    0.1047
    0.8728    0.1038
    0.8822    0.1032
    0.8944    0.1047
    0.9026    0.1076
    0.9087    0.1107
    0.9142    0.1138
    0.9188    0.1164
    0.9240    0.1191
    0.9286    0.1213
    0.9332    0.1232
    0.9382    0.1249
    0.9437    0.1263
    0.9494    0.1275
    0.9554    0.1282
    0.9616    0.1287
    0.9677    0.1288
    0.9737    0.1287
    0.9795    0.1284
    0.9855    0.1279
    0.9916    0.1271
    0.9979    0.1258

};

\addplot [mark=none,cyan,line width=1.5] table{
               0   -0.0000
    0.0011    0.0000
    0.0081    0.0003
    0.0208    0.0012
    0.0437    0.0014
    0.0690    0.0005
    0.1271   -0.0007
    0.1915    0.0057
    0.2241    0.0131
    0.2461    0.0158
    0.2696    0.0159
    0.2905    0.0155
    0.3030    0.0157
    0.3141    0.0159
    0.3346    0.0188
    0.3517    0.0251
    0.3649    0.0304
    0.3776    0.0341
    0.3922    0.0364
    0.4073    0.0371
    0.4213    0.0372
    0.4334    0.0369
    0.4438    0.0362
    0.4589    0.0353
    0.4810    0.0398
    0.4926    0.0453
    0.5027    0.0503
    0.5133    0.0544
    0.5230    0.0569
    0.5327    0.0584
    0.5427    0.0591
    0.5535    0.0593
    0.5643    0.0593
    0.5741    0.0591
    0.5839    0.0581
    0.5975    0.0566
    0.6176    0.0594
    0.6291    0.0649
    0.6384    0.0701
    0.6474    0.0745
    0.6562    0.0778
    0.6664    0.0803
    0.6773    0.0818
    0.6884    0.0823
    0.6990    0.0824
    0.7092    0.0822
    0.7188    0.0815
    0.7291    0.0798
    0.7448    0.0784
    0.7619    0.0822
    0.7716    0.0876
    0.7797    0.0924
    0.7879    0.0967
    0.7958    0.0999
    0.8050    0.1025
    0.8150    0.1042
    0.8254    0.1049
    0.8355    0.1051
    0.8464    0.1050
    0.8560    0.1046
    0.8653    0.1032
    0.8769    0.1010
    0.8935    0.1009
    0.9059    0.1054
    0.9139    0.1105
    0.9209    0.1151
    0.9283    0.1191
    0.9355    0.1223
    0.9438    0.1250
    0.9530    0.1269
    0.9627    0.1280
    0.9723    0.1283
    0.9825    0.1283
    0.9923    0.1280
    1.0016    0.1269
    1.0114    0.1249
    1.0252    0.1227
    1.0405    0.1247
    1.0496    0.1295
    1.0566    0.1345
    1.0630    0.1387
    1.0699    0.1425
    1.0768    0.1456
    1.0847    0.1482
    1.0936    0.1500
    1.1029    0.1511
    1.1122    0.1515
    1.1220    0.1516
    1.1315    0.1514
    1.1408    0.1505
    1.1503    0.1487
    1.1634    0.1465
    1.1793    0.1480
    1.1894    0.1529
    1.1969    0.1580
    1.2036    0.1623
    1.2110    0.1663
    1.2179    0.1693
    1.2258    0.1718
    1.2345    0.1736
    1.2436    0.1746
    1.2528    0.1751
    1.2627    0.1752
    1.2721    0.1751
    1.2815    0.1744
    1.2912    0.1727
    1.3055    0.1712
    1.3226    0.1746
    1.3323    0.1799
    1.3403    0.1849
    1.3483    0.1894
    1.3562    0.1929
    1.3654    0.1959
    1.3758    0.1979
    1.3866    0.1989
    1.3972    0.1993
    1.4072    0.1993
    1.4171    0.1990
    1.4264    0.1981
    1.4382    0.1963
    1.4577    0.1977
    1.4707    0.2032
    1.4798    0.2085
    1.4887    0.2133
    1.4969    0.2169
    1.5059    0.2197
};

\addplot [mark=none,red,dashed,line width=1.5] table{
   0   -0.0000
    0.0066    0.0003
    0.0347    0.0016
    0.0845   -0.0001
    0.2040    0.0089
    0.2504    0.0161
    0.2908    0.0156
    0.3124    0.0160
    0.3469    0.0233
    0.3710    0.0325
    0.3972    0.0368
    0.4235    0.0372
    0.4437    0.0362
    0.4774    0.0384
    0.4991    0.0486
    0.5176    0.0556
    0.5350    0.0586
    0.5542    0.0593
    0.5728    0.0591
    0.5924    0.0569
    0.6245    0.0623
    0.6413    0.0716
    0.6573    0.0781
    0.6765    0.0817
    0.6964    0.0824
    0.7148    0.0818
    0.7340    0.0789
    0.7634    0.0826
    0.7788    0.0917
    0.7931    0.0988
    0.8098    0.1034
    0.8285    0.1049
    0.8474    0.1049
    0.8644    0.1032
    0.8882    0.0999
    0.9100    0.1074
    0.9229    0.1159
    0.9358    0.1222
    0.9515    0.1265
    0.9690    0.1281
    0.9871    0.1280
    1.0040    0.1262
    1.0255    0.1221
    1.0475    0.1273
    1.0594    0.1358
    1.0711    0.1428
    1.0846    0.1479
    1.1008    0.1507
    1.1174    0.1514
    1.1349    0.1509
    1.1520    0.1478
    1.1772    0.1462
    1.1939    0.1549
    1.2055    0.1627
    1.2187    0.1692
    1.2313    0.1727
    1.2454    0.1745
    1.2615    0.1749
    1.2788    0.1743
    1.2971    0.1710
    1.3245    0.1739
    1.3393    0.1831
    1.3525    0.1906
    1.3674    0.1959
    1.3813    0.1982
    1.3969    0.1989
    1.4149    0.1987
    1.4325    0.1963
    1.4616    0.1971
    1.4787    0.2066
    1.4932    0.2146
    1.5074    0.2195
    1.5236    0.2220
    1.5408    0.2225
    1.5588    0.2221
    1.5772    0.2193
    1.6064    0.2219
    1.6219    0.2311
    1.6358    0.2385
    1.6518    0.2435
    1.6702    0.2456
    1.6891    0.2457
    1.7066    0.2442
    1.7283    0.2402
    1.7519    0.2455
    1.7642    0.2541
    1.7765    0.2610
    1.7910    0.2660
    1.8082    0.2684
    1.8257    0.2687
    1.8432    0.2675
    1.8612    0.2633
    1.8845    0.2620
    1.8978    0.2700
    1.9083    0.2780
    1.9189    0.2840
    1.9319    0.2886
    1.9437    0.2908
    1.9571    0.2916
    1.9738    0.2915
    1.9900    0.2896   

};

\end{axis}

\end{tikzpicture} & \begin{tikzpicture}

\begin{axis}[
  width = 0.45\textwidth,
  title = {$\nu = 10$},
  xmin = 0,
  xmax = 2, 
  ymin = -0.1,
  ymax = 0.2,
  xtick = {0,0.5,1,1.5,2},
  ytick = {-0.1,0,0.1,0.2},
  ylabel = {$y/L_p$},
  xlabel = {$x/L_p$},
  label style={font=\normalsize},
  legend style={anchor=east,draw=none},
  legend pos = south east,
  ]

\addplot [mark=none,black,line width=1.5] table{
         0   -0.0000
    0.0016    0.0002
    0.0049    0.0005
    0.0088    0.0008
    0.0134    0.0010
    0.0194    0.0013
    0.0270    0.0015
    0.0368    0.0015
    0.0481    0.0014
    0.0600    0.0013
    0.0708    0.0010
    0.0842    0.0006
    0.1123   -0.0005
    0.1493   -0.0012
    0.1727    0.0012
    0.1920    0.0070
    0.2185    0.0160
    0.2474    0.0185
    0.2596    0.0177
    0.2685    0.0169
    0.2743    0.0164
    0.2801    0.0159
    0.2854    0.0157
    0.2904    0.0156
    0.2957    0.0156
    0.3016    0.0156
    0.3084    0.0157
    0.3157    0.0160
    0.3233    0.0166
    0.3306    0.0178
    0.3373    0.0195
    0.3424    0.0211
    0.3478    0.0231
    0.3537    0.0256
    0.3581    0.0274
    0.3631    0.0294
    0.3687    0.0313
    0.3747    0.0331
    0.3812    0.0346
    0.3882    0.0357
    0.3969    0.0364
    0.4083    0.0363
    0.4324    0.0341
    0.4511    0.0324
    0.4659    0.0330
    0.4829    0.0386
    0.4924    0.0437
    0.4988    0.0471
    0.5047    0.0499
    0.5113    0.0525
    0.5187    0.0546
    0.5267    0.0562
    0.5350    0.0569
    0.5434    0.0568
    0.5521    0.0560
    0.5636    0.0541
    0.5861    0.0493
    0.5988    0.0475
    0.6079    0.0475
    0.6156    0.0490
    0.6250    0.0540
    0.6321    0.0596
    0.6368    0.0632
    0.6412    0.0662
    0.6455    0.0688
    0.6504    0.0713
    0.6547    0.0731
    0.6589    0.0745
    0.6634    0.0757
    0.6681    0.0765
    0.6730    0.0770
    0.6779    0.0772
    0.6830    0.0770
    0.6886    0.0764
    0.6946    0.0754
    0.7014    0.0738
    0.7101    0.0711
    0.7246    0.0656
    0.7353    0.0608
    0.7416    0.0571
    0.7450    0.0537
    0.7469    0.0502
    0.7483    0.0465
    0.7494    0.0426
    0.7506    0.0387
    0.7520    0.0349
    0.7535    0.0317
    0.7551    0.0293
    0.7567    0.0274
    0.7584    0.0258
    0.7601    0.0245
    0.7619    0.0232
    0.7639    0.0217
    0.7659    0.0201
    0.7679    0.0181
    0.7700    0.0159
    0.7718    0.0140
    0.7738    0.0118
    0.7762    0.0095
    0.7791    0.0069
    0.7821    0.0045
    0.7850    0.0025
    0.7883    0.0005
    0.7921   -0.0015
    0.7963   -0.0033
    0.8010   -0.0049
    0.8062   -0.0062
    0.8117   -0.0071
    0.8175   -0.0075
    0.8234   -0.0074
    0.8293   -0.0068
    0.8350   -0.0059
    0.8413   -0.0043
    0.8488   -0.0018
    0.8623    0.0037
    0.8809    0.0110
    0.9016    0.0135
    0.9286    0.0105
    0.9465    0.0086
    0.9597    0.0082
    0.9752    0.0092       

};

\addplot [mark=none,cyan,line width=1.5] table{
          0   -0.0000
    0.0016    0.0002
    0.0051    0.0004
    0.0090    0.0007
    0.0138    0.0009
    0.0200    0.0011
    0.0280    0.0012
    0.0384    0.0011
    0.0504    0.0009
    0.0615    0.0008
    0.0729    0.0006
    0.0872    0.0001
    0.1281   -0.0014
    0.1546   -0.0015
    0.1743    0.0010
    0.1903    0.0058
    0.2176    0.0159
    0.2533    0.0191
    0.2629    0.0184
    0.2705    0.0177
    0.2761    0.0173
    0.2815    0.0171
    0.2865    0.0170
    0.2922    0.0169
    0.2989    0.0169
    0.3067    0.0169
    0.3154    0.0172
    0.3243    0.0180
    0.3322    0.0192
    0.3383    0.0207
    0.3445    0.0227
    0.3492    0.0244
    0.3542    0.0264
    0.3598    0.0286
    0.3660    0.0309
    0.3728    0.0330
    0.3803    0.0347
    0.3883    0.0360
    0.3971    0.0367
    0.4090    0.0366
    0.4356    0.0341
    0.4544    0.0326
    0.4695    0.0338
    0.4883    0.0412
    0.4952    0.0450
    0.5013    0.0482
    0.5070    0.0507
    0.5133    0.0530
    0.5203    0.0549
    0.5277    0.0562
    0.5354    0.0568
    0.5431    0.0567
    0.5515    0.0560
    0.5624    0.0541
    0.5853    0.0489
    0.5986    0.0463
    0.6064    0.0455
    0.6128    0.0457
    0.6188    0.0474
    0.6276    0.0535
    0.6322    0.0576
    0.6362    0.0611
    0.6398    0.0639
    0.6439    0.0667
    0.6474    0.0688
    0.6513    0.0708
    0.6557    0.0726
    0.6604    0.0742
    0.6655    0.0754
    0.6708    0.0762
    0.6763    0.0765
    0.6817    0.0765
    0.6871    0.0760
    0.6922    0.0752
    0.6980    0.0739
    0.7045    0.0720
    0.7124    0.0691
    0.7243    0.0636
    0.7393    0.0530
    0.7450    0.0421
    0.7488    0.0308
    0.7524    0.0222
    0.7580    0.0145
    0.7702    0.0054
    0.7810    0.0005
    0.7888   -0.0021
    0.7955   -0.0038
    0.8009   -0.0048
    0.8067   -0.0055
    0.8118   -0.0058
    0.8162   -0.0057
    0.8207   -0.0053
    0.8251   -0.0047
    0.8293   -0.0038
    0.8334   -0.0026
    0.8373   -0.0013
    0.8415    0.0005
    0.8455    0.0025
    0.8500    0.0050
    0.8551    0.0082
    0.8646    0.0141
    0.8721    0.0174
    0.8778    0.0189
    0.8843    0.0198
    0.9047    0.0182
    0.9298    0.0131
    0.9423    0.0112
    0.9532    0.0105
    0.9631    0.0107
    0.9727    0.0118
    0.9821    0.0136
    0.9931    0.0166 
};

\addplot [mark=none,red,dashed,line width=1.5] table{
          0   -0.0000
    0.0043    0.0004
    0.0118    0.0008
    0.0228    0.0012
    0.0407    0.0010
    0.0618    0.0008
    0.0842    0.0002
    0.1429   -0.0017
    0.1780    0.0018
    0.2103    0.0135
    0.2542    0.0191
    0.2689    0.0179
    0.2808    0.0172
    0.2914    0.0168
    0.3048    0.0164
    0.3214    0.0167
    0.3359    0.0192
    0.3471    0.0229
    0.3563    0.0268
    0.3672    0.0309
    0.3803    0.0344
    0.3952    0.0363
    0.4203    0.0353
    0.4584    0.0317
    0.4856    0.0389
    0.4987    0.0462
    0.5093    0.0511
    0.5201    0.0545
    0.5305    0.0561
    0.5432    0.0563
    0.5592    0.0543
    0.5916    0.0464
    0.6076    0.0436
    0.6174    0.0442
    0.6277    0.0516
    0.6345    0.0584
    0.6407    0.0637
    0.6471    0.0679
    0.6541    0.0714
    0.6607    0.0737
    0.6680    0.0754
    0.6757    0.0761
    0.6835    0.0759
    0.6924    0.0747
    0.7026    0.0721
    0.7163    0.0667
    0.7339    0.0559
    0.7418    0.0454
    0.7465    0.0345
    0.7515    0.0236
    0.7600    0.0130
    0.7778    0.0023
    0.7915   -0.0027
    0.8021   -0.0051
    0.8129   -0.0059
    0.8211   -0.0055
    0.8288   -0.0042
    0.8366   -0.0020
    0.8446    0.0014
    0.8537    0.0064
    0.8709    0.0158
    0.8824    0.0184
    0.9115    0.0159
    0.9432    0.0106
    0.9667    0.0105
    0.9881    0.0144
    1.0263    0.0247
    1.0915    0.0268
    1.1476    0.0317
    1.2037    0.0420
    1.2717    0.0444
    1.3294    0.0536
    1.3579    0.0604
    1.3822    0.0622
    1.4188    0.0591
    1.4566    0.0614
    1.4841    0.0724
    1.4977    0.0770
    1.5117    0.0793
    1.5262    0.0793
    1.5397    0.0770
    1.5543    0.0726
    1.5705    0.0664
    1.5810    0.0621
    1.5864    0.0597
    1.5906    0.0580
    1.5944    0.0564
    1.5977    0.0550
    1.6006    0.0536
    1.6027    0.0519
    1.6038    0.0502
    1.6045    0.0480
    1.6050    0.0450
    1.6057    0.0410
    1.6070    0.0353
    1.6090    0.0292
    1.6114    0.0238
    1.6144    0.0188
    1.6184    0.0136
    1.6224    0.0093
    1.6279    0.0048
    1.6337    0.0013
    1.6401   -0.0017
    1.6478   -0.0041
    1.6566   -0.0056
    1.6659   -0.0058
    1.6752   -0.0047
    1.6847   -0.0023
    1.6971    0.0025
    1.7216    0.0129
    1.7551    0.0144
    1.7887    0.0096
    1.8131    0.0097  

};

\end{axis}

\end{tikzpicture} \\
\end{tabular}
\end{center}
\mcaption{Effects of the side walls on RBC dynamics. We considered the same RBCs
in~\figref{f:rowConv}. The DLD device also has the same period $n_p = 6$. However,
we, now, fixed the number of rows to $n_{\mathrm{row}} = 12$ and performed the
simulations in a single period ($n_{\mathrm{col}} = p$) and two periods
($n_{\mathrm{col}} = 2p$) of the device. We plotted RBCs' inclination angles (on
the top) and trajectories (on the bottom). The RBC with $\nu = 0.1$ (on the
left) displaces in a single period of the device with a steady mean inclination
angle. In two periods the transition and the equilibrium behaviors did not
change. The RBC with $\nu = 10$ (on the right) zig-zags once in a single period
and twice in two periods. However, if our domain includes only $n_{\mathrm{col}} = p$ number of columns, the RBC flips near the end of the first period. In two periods the RBC flips again but near the end of the second period not the first
one. If our domain includes 
$n_{\mathrm{col}} = \left \lceil 1.5 p \right \rceil$ number of columns and we simulate the flow within a period, we can avoid
the flip near the end of the first period. So that, the RBC shows the same
dynamics in a single period in both devices with 
$n_{\mathrm{col}} = \left \lceil 1.5 p \right \rceil$ and 
$n_{\mathrm{col}} = 2p$. $x$ and $y$ coordinates
are normalized by the length of one period of the device $L_p$.}{f:colConv}
\end{figure}

Dynamics of RBC given by a simulation in a single period of a DLD device must
repeat itself as the simulation domain includes more periods if the simulation
in a single period is not spoiled by the side wall effects. In order to
determine if our simulations in a single period is accurate, we fixed the number
of rows to $n_{\mathrm{row}} = 12$ and considered the number of columns
$n_{\mathrm{col}} = (p, 2p)$ for the same examples above. We present the results
in~\figref{f:colConv}. Here again, the results on the left are for the
displacing RBC ($\nu = 0.1$) and those on the right are for the zig-zagging RBC
($\nu = 10$). The displacing RBC goes through a transition within one period and
follows a trajectory along the pillars. As one more period is included in the
domain, the trends in neither the inclination angle nor the trajectory changes.
The zig-zagging RBC falls down a lower lane once within one period and repeats
this motion if the device includes one more period. However, the RBC in the
device with $n_{\mathrm{col}} = p$ columns flips near the end of the first
period (see the top right figure on~\figref{f:colConv}). The RBC in the device
with two periods also flips but near the end of the second period not the first
one. This can be avoided by having 
$n_{\mathrm{col}} = \left \lceil 1.5p \right \rceil$ columns in the domain. It seems that it suffices to have
$n_{\mathrm{col}} = \left \lceil 1.5p \right \rceil$ number of columns in our
domain and perform simulations within one period to alleviate the wall effects.

In addition to those precautions, we also discussed in~\secref{s:dldModel} that
it is essential to replace the circular pillars which, if placed, would cross
the top and bottom walls with non-circular pillars in such a way that a similar
hydraulic resistance is maintained in the entire domain. Consequently, we have a homogeneous flow almost everywhere in our DLD domain when RBCs are not present (see~\figref{f:N6WholeDLD}).

\begin{figure}
\begin{center}
\begin{tabular}{c c}
 \input{figs/N6VC1Angles.tikz} & \input{figs/N6VC10Angles.tikz} \\
\input{figs/N6VC1Trajs.tikz} & \input{figs/N6VC10Trajs.tikz} \\
\end{tabular}
\end{center}
\mcaption{ Angular orientations and trajectories of RBCs initialized with
different inclination angles and at lateral positions. We consider
RBCs with low ($\nu = 1$) and high ($\nu = 10$) viscosity contrasts in
the DLD device with the row-shift fraction $\epsilon = 0.1667$. Under
these conditions the less viscous displaces and the more viscous one
zig-zags. The plots on the left are for $\nu = 1$ and those on the
right are for $\nu = 10$. At the top we show the inclination angles
$\alpha$ as the RBC moves along the flow direction $x$. At the bottom
we show the positions of the RBCs' centroids. $x$ and $y$ coordinates
are normalized by the length of a period $L$. The scales for $x$ and $y$ 
coordinates are chosen differently for better visualization.}{f:IAsTrajsN6}
\end{figure}
{\em Sensitivity to the initial angular orientation and lateral position.}
We investigated whether our DLD model is sensitive to the
initial lateral position and inclination angle of an RBC. 
We performed several simulations with RBCs
initialized at two different lateral positions and with various inclination angles. 
We chose a DLD device with the row-shift fraction $\epsilon = 0.1667$ and
considered the viscosity contrasts $\nu = (1, 10)$.
In~\figref{f:IAsTrajsN6} we show the trajectories and inclination
angles ($\alpha$) of the RBCs in those simulations. The less viscous
RBC displaces (on the left in~\figref{f:IAsTrajsN6}) while the more
viscous one zig-zags (on the right). For all initial inclination
angles $\alpha_0$ the displacing RBCs reach an equilibrium angular
orientation alternating between two angles (see the top left).  They
also follow the same trajectories after different transients (see bottom
left). Additionally, the RBC with $\alpha_0 = \upi/2$ is initialized
at a different lateral position and it also attains the equilibrium
orientation and the trajectory. In the zig-zag mode there is no
equilibrium in dynamics. Although trajectories are sensitive to the
RBC's initial position and inclination angle~\citep{quek-chiam-e11},
the transport mode is supposed to persist. For the zig-zagging case 
(high viscosity contrast) we observe different trends in the inclination angles
(top right) and trajectories (bottom right) depending on the
initialization. The RBCs with $\alpha_0 = (0, \upi/2)$ and those with
$\alpha = (\upi/6, \upi/4, \upi/3)$ show the same variations in the
inclination angles and trajectories. Ultimately all of the more
viscous RBCs zig-zags. Therefore, we can conclude that our model is
insensitive to how an RBC is initialized.

\section{Convergence study\label{a:appendix1}}
\begin{figure}
\begin{center}
\begin{tabular}{c c}
 \begin{tikzpicture}

\begin{axis}[
  width = 0.45\textwidth,
  title = {$\nu = 1$},
  xmin = 0,
  xmax = 1, 
  ymin = -2,
  ymax = 2,
  xtick = {0,0.2,0.4,0.6,0.8,1},
  ytick = {-2,-1.5,-1,-0.5,0,0.5,1,1.5,2},
  ylabel = {Inclination angle ($\alpha$) [rad]},
  xlabel = {$x/L$},
  label style={font=\normalsize},
  ]

\addplot [mark=none,cyan,line width=1.5] table{
  0    1.5708
    0.0004   -1.5525
    0.0014   -1.5102
    0.0039   -1.4132
    0.0084   -1.2623
    0.0144   -1.1073
    0.0227   -0.9541
    0.0346   -0.7978
    0.0500   -0.6125
    0.0619   -0.4596
    0.0778   -0.2878
    0.1033   -0.1475
    0.1455   -0.1649
    0.1773   -0.2176
    0.2032   -0.1391
    0.2313   -0.0681
    0.2456   -0.0928
    0.2579   -0.1342
    0.2698   -0.1863
    0.2824   -0.2441
    0.2970   -0.3029
    0.3138   -0.3717
    0.3270   -0.4812
    0.3343   -0.6278
    0.3390   -0.7787
    0.3434   -0.9500
    0.3474   -1.1059
    0.3516   -1.2797
    0.3559   -1.4607
    0.3596    1.5211
    0.3636    1.3426
    0.3678    1.1612
    0.3721    0.9934
    0.3766    0.8410
    0.3820    0.7005
    0.3876    0.5899
    0.3936    0.5051
    0.3997    0.4451
    0.4059    0.4085
    0.4126    0.3966
    0.4217    0.4217
    0.4322    0.5059
    0.4416    0.6169
    0.4523    0.7253
    0.4672    0.8216
    0.4806    0.8504
    0.4868    0.8102
    0.4920    0.7484
    0.4963    0.6868
    0.5012    0.6120
    0.5049    0.5567
    0.5093    0.4913
    0.5143    0.4240
    0.5196    0.3577
    0.5254    0.2947
    0.5315    0.2413
    0.5377    0.1987
    0.5440    0.1699
    0.5501    0.1565
    0.5567    0.1593
    0.5655    0.1981
    0.5740    0.2798
    0.5818    0.3943
    0.5891    0.5213
    0.5963    0.6350
    0.6059    0.7510
    0.6179    0.8415
    0.6265    0.8241
    0.6309    0.7812
    0.6352    0.7247
    0.6395    0.6599
    0.6433    0.6021
    0.6473    0.5418
    0.6517    0.4786
    0.6566    0.4131
    0.6619    0.3493
    0.6676    0.2906
    0.6735    0.2401
    0.6795    0.2010
    0.6856    0.1742
    0.6915    0.1623
    0.6979    0.1673
    0.7064    0.2053
    0.7145    0.2828
    0.7225    0.3983
    0.7296    0.5184
    0.7369    0.6351
    0.7463    0.7529
    0.7574    0.8536
    0.7667    0.8525
    0.7713    0.8081
    0.7751    0.7559
    0.7791    0.6944
    0.7827    0.6378
    0.7869    0.5718
    0.7911    0.5076
    0.7947    0.4554
    0.7987    0.4027
    0.8030    0.3510
    0.8075    0.3027
    0.8122    0.2591
    0.8171    0.2205
    0.8220    0.1906
    0.8268    0.1695
    0.8322    0.1569
    0.8382    0.1579
    0.8457    0.1842
    0.8537    0.2489
    0.8618    0.3531
    0.8688    0.4696
    0.8757    0.5850
    0.8839    0.7024
    0.8932    0.8187
    0.9024    0.8898
    0.9088    0.8658
    0.9122    0.8270
    0.9159    0.7735
    0.9192    0.7201
    0.9228    0.6609
    0.9260    0.6094
    0.9296    0.5515
    0.9337    0.4901
    0.9382    0.4276
    0.9419    0.3806
    0.9456    0.3380
    0.9495    0.2980
    0.9535    0.2615
    0.9576    0.2296
    0.9617    0.2028
    0.9664    0.1797
    0.9712    0.1644
    0.9765    0.1593
    0.9830    0.1713
    0.9908    0.2159
    0.9983    0.2959

};

\addplot [mark=none,green,line width=1.5] table{

  0   -1.5708
    0.0003   -1.5551
    0.0016   -1.5025
    0.0047   -1.3812
    0.0092   -1.2357
    0.0149   -1.0898
    0.0224   -0.9498
    0.0326   -0.8059
    0.0456   -0.6405
    0.0572   -0.4877
    0.0700   -0.3362
    0.0872   -0.1965
    0.1354   -0.1311
    0.1923   -0.1403
    0.2121   -0.0472
    0.2282   -0.0247
    0.2394   -0.0401
    0.2494   -0.0667
    0.2585   -0.0966
    0.2676   -0.1280
    0.2776   -0.1573
    0.2896   -0.1716
    0.3068   -0.1274
    0.3189   -0.0096
    0.3255    0.1106
    0.3322    0.2539
    0.3388    0.3804
    0.3455    0.4626
    0.3525    0.4863
    0.3577    0.4686
    0.3618    0.4410
    0.3664    0.4013
    0.3711    0.3565
    0.3748    0.3206
    0.3789    0.2834
    0.3831    0.2472
    0.3874    0.2134
    0.3918    0.1833
    0.3963    0.1579
    0.4007    0.1382
    0.4052    0.1252
    0.4103    0.1191
    0.4159    0.1258
    0.4226    0.1568
    0.4289    0.2108
    0.4347    0.2843
    0.4407    0.3835
    0.4462    0.4851
    0.4520    0.5900
    0.4587    0.6901
    0.4663    0.7777
    0.4748    0.8409
    0.4817    0.8476
    0.4864    0.8210
    0.4904    0.7801
    0.4939    0.7343
    0.4974    0.6831
    0.5014    0.6230
    0.5044    0.5780
    0.5078    0.5276
    0.5115    0.4750
    0.5155    0.4217
    0.5198    0.3695
    0.5243    0.3202
    0.5290    0.2756
    0.5339    0.2357
    0.5389    0.2060
    0.5438    0.1852
    0.5487    0.1745
    0.5537    0.1747
    0.5596    0.1911
    0.5657    0.2297
    0.5717    0.2904
    0.5777    0.3751
    0.5833    0.4689
    0.5890    0.5689
    0.5955    0.6651
    0.6029    0.7523
    0.6113    0.8273
    0.6191    0.8619
    0.6245    0.8490
    0.6286    0.8160
    0.6324    0.7710
    0.6357    0.7254
    0.6391    0.6742
    0.6430    0.6155
    0.6463    0.5648
    0.6496    0.5164
    0.6531    0.4665
    0.6570    0.4163
    0.6611    0.3673
    0.6654    0.3212
    0.6699    0.2795
    0.6746    0.2435
    0.6793    0.2145
    0.6840    0.1934
    0.6886    0.1812
    0.6934    0.1786
    0.6989    0.1897
    0.7048    0.2207
    0.7111    0.2775
    0.7170    0.3535
    0.7227    0.4455
    0.7283    0.5423
    0.7345    0.6393
    0.7416    0.7290
    0.7495    0.8093
    0.7574    0.8645
    0.7638    0.8692
    0.7681    0.8441
    0.7718    0.8048
    0.7749    0.7639
    0.7784    0.7118
    0.7815    0.6650
    0.7846    0.6162
    0.7880    0.5634
    0.7918    0.5077
    0.7959    0.4505
    0.7989    0.4116
    0.8022    0.3717
    0.8057    0.3336
    0.8093    0.2982
    0.8130    0.2648
    0.8168    0.2369
    0.8206    0.2138
    0.8245    0.1960
    0.8290    0.1823
    0.8334    0.1774
    0.8387    0.1836
    0.8446    0.2081
    0.8508    0.2564
    0.8567    0.3253
    0.8624    0.4120
    0.8679    0.5057
    0.8738    0.6028
    0.8805    0.6944
    0.8876    0.7786
    0.8948    0.8504
    0.9021    0.8880
    0.9066    0.8776
    0.9100    0.8502
    0.9132    0.8123
    0.9161    0.7709
    0.9194    0.7199
    0.9221    0.6778
    0.9250    0.6307
    0.9283    0.5801
    0.9318    0.5263
    0.9353    0.4773
    0.9383    0.4372
    0.9414    0.3974
    0.9448    0.3588
    0.9483    0.3222
    0.9518    0.2884
    0.9555    0.2582
    0.9592    0.2320
    0.9630    0.2106
    0.9667    0.1945
    0.9710    0.1828
    0.9754    0.1793
    0.9805    0.1880
    0.9862    0.2137
    0.9922    0.2625
    0.9979    0.3304

};

\addplot [mark=none,red,dashed,line width=1.5] table{

 0    1.5708
    0.0006   -1.5451
    0.0032   -1.4392
    0.0085   -1.2552
    0.0158   -1.0695
    0.0265   -0.8859
    0.0398   -0.7101
    0.0554   -0.5045
    0.0750   -0.2806
    0.1104   -0.1199
    0.1910   -0.1471
    0.2153   -0.0350
    0.2316   -0.0231
    0.2443   -0.0472
    0.2562   -0.0825
    0.2678   -0.1201
    0.2802   -0.1498
    0.2971   -0.1380
    0.3138   -0.0170
    0.3228    0.1337
    0.3303    0.2920
    0.3376    0.4288
    0.3456    0.5158
    0.3533    0.5212
    0.3592    0.4858
    0.3650    0.4332
    0.3707    0.3763
    0.3757    0.3260
    0.3811    0.2764
    0.3868    0.2308
    0.3926    0.1920
    0.3985    0.1626
    0.4044    0.1446
    0.4110    0.1402
    0.4186    0.1605
    0.4269    0.2200
    0.4345    0.3140
    0.4416    0.4332
    0.4488    0.5633
    0.4567    0.6854
    0.4663    0.7897
    0.4763    0.8469
    0.4840    0.8304
    0.4892    0.7846
    0.4937    0.7275
    0.4986    0.6572
    0.5031    0.5882
    0.5076    0.5233
    0.5125    0.4550
    0.5180    0.3866
    0.5239    0.3223
    0.5302    0.2666
    0.5367    0.2228
    0.5432    0.1940
    0.5496    0.1826
    0.5565    0.1918
    0.5645    0.2355
    0.5720    0.3126
    0.5793    0.4225
    0.5863    0.5439
    0.5939    0.6631
    0.6031    0.7683
    0.6129    0.8446
    0.6220    0.8569
    0.6276    0.8204
    0.6321    0.7697
    0.6367    0.7053
    0.6408    0.6437
    0.6454    0.5737
    0.6506    0.4977
    0.6547    0.4429
    0.6592    0.3869
    0.6641    0.3338
    0.6692    0.2859
    0.6744    0.2454
    0.6797    0.2140
    0.6851    0.1931
    0.6910    0.1835
    0.6976    0.1927
    0.7056    0.2360
    0.7129    0.3102
    0.7203    0.4197
    0.7272    0.5388
    0.7347    0.6557
    0.7435    0.7602
    0.7527    0.8422
    0.7617    0.8726
    0.7674    0.8460
    0.7719    0.7990
    0.7764    0.7373
    0.7802    0.6805
    0.7843    0.6159
    0.7890    0.5445
    0.7932    0.4834
    0.7974    0.4280
    0.8018    0.3721
    0.8066    0.3215
    0.8116    0.2763
    0.8167    0.2383
    0.8219    0.2090
    0.8271    0.1898
    0.8329    0.1815
    0.8394    0.1911
    0.8471    0.2328
    0.8542    0.3043
    0.8615    0.4103
    0.8683    0.5263
    0.8754    0.6401
    0.8836    0.7439
    0.8923    0.8345
    0.8998    0.8856
    0.9059    0.8814
    0.9102    0.8467
    0.9144    0.7945
    0.9180    0.7418
    0.9222    0.6765
    0.9257    0.6203
    0.9294    0.5627
    0.9335    0.5015
    0.9382    0.4381
    0.9432    0.3769
    0.9486    0.3188
    0.9543    0.2681
    0.9601    0.2272
    0.9660    0.1983
    0.9718    0.1835
    0.9776    0.1842
    0.9846    0.2086
    0.9916    0.2628
    0.9989    0.3522
};

\end{axis}

\end{tikzpicture} & \begin{tikzpicture}

\begin{axis}[
  width = 0.45\textwidth,
  title = {$\nu = 10$},
  xmin = 0,
  xmax = 1, 
  ymin = -2,
  ymax = 2,
  xtick = {0,0.2,0.4,0.6,0.8,1},
  ytick = {-2,-1.5,-1,-0.5,0,0.5,1,1.5,2},
  ylabel = {Inclination angle ($\alpha$) [rad]},
  xlabel = {$x/L$},
  label style={font=\normalsize},
  legend style={anchor=east,draw=none},
  legend pos = south east,
  ]

\addplot [mark=none,cyan,line width=1.5] table{
 0    1.5708
    0.0016   -1.5127
    0.0049   -1.4092
    0.0088   -1.3043
    0.0136   -1.2008
    0.0197   -1.0983
    0.0278   -0.9947
    0.0387   -0.8836
    0.0508   -0.7640
    0.0649   -0.6147
    0.0757   -0.4995
    0.0912   -0.3727
    0.1247   -0.3033
    0.1558   -0.4407
    0.1730   -0.5631
    0.1903   -0.6738
    0.2120   -0.7106
    0.2413   -0.6614
    0.2601   -0.7459
    0.2691   -0.8567
    0.2765   -0.9832
    0.2833   -1.1214
    0.2897   -1.2537
    0.2956   -1.3689
    0.3021   -1.4773
    0.3096    1.5569
    0.3180    1.4485
    0.3258    1.3438
    0.3332    1.2304
    0.3398    1.1139
    0.3450    1.0110
    0.3506    0.8948
    0.3561    0.7826
    0.3608    0.6916
    0.3661    0.5962
    0.3719    0.5022
    0.3782    0.4131
    0.3850    0.3328
    0.3933    0.2563
    0.4031    0.1955
    0.4183    0.1633
    0.4421    0.2731
    0.4574    0.4369
    0.4714    0.5908
    0.4878    0.6525
    0.4955    0.5913
    0.5017    0.5167
    0.5080    0.4327
    0.5152    0.3405
    0.5231    0.2458
    0.5316    0.1569
    0.5405    0.0796
    0.5491    0.0217
    0.5597   -0.0221
    0.5773   -0.0110
    0.5960    0.1390
    0.6050    0.2958
    0.6123    0.4734
    0.6189    0.6322
    0.6267    0.7376
    0.6339    0.7078
    0.6383    0.6493
    0.6428    0.5811
    0.6469    0.5160
    0.6517    0.4392
    0.6558    0.3752
    0.6601    0.3124
    0.6646    0.2470
    0.6694    0.1839
    0.6743    0.1230
    0.6794    0.0656
    0.6844    0.0130
    0.6902   -0.0422
    0.6963   -0.0909
    0.7030   -0.1339
    0.7114   -0.1630
    0.7243   -0.1548
    0.7349   -0.1126
    0.7413   -0.0759
    0.7447   -0.0389
    0.7464    0.0065
    0.7473    0.0630
    0.7481    0.1281
    0.7488    0.1979
    0.7497    0.2616
    0.7507    0.3187
    0.7520    0.3671
    0.7536    0.4203
    0.7559    0.4833
    0.7580    0.5586
    0.7598    0.6346
    0.7620    0.7220
    0.7640    0.8174
    0.7660    0.9171
    0.7682    1.0371
    0.7702    1.1416
    0.7726    1.2644
    0.7747    1.3712
    0.7768    1.4805
    0.7791   -1.5453
    0.7815   -1.4249
    0.7839   -1.3029
    0.7862   -1.1831
    0.7884   -1.0665
    0.7908   -0.9608
    0.7934   -0.8603
    0.7962   -0.7667
    0.7993   -0.6805
    0.8030   -0.5925
    0.8074   -0.5065
    0.8121   -0.4265
    0.8171   -0.3532
    0.8224   -0.2869
    0.8281   -0.2267
    0.8345   -0.1696
    0.8418   -0.1202
    0.8517   -0.0820
    0.8672   -0.0982
    0.8851   -0.1970
    0.9061   -0.2264
    0.9329   -0.1775
    0.9561   -0.0883
    0.9804   -0.0049

};

\addplot [mark=none,green,line width=1.5] table{

 0   -1.5708
    0.0009   -1.5383
    0.0030   -1.4658
    0.0056   -1.3882
    0.0085   -1.3112
    0.0117   -1.2360
    0.0155   -1.1621
    0.0200   -1.0870
    0.0255   -1.0124
    0.0321   -0.9365
    0.0399   -0.8554
    0.0482   -0.7676
    0.0575   -0.6656
    0.0648   -0.5822
    0.0730   -0.4942
    0.0829   -0.4045
    0.0976   -0.3145
    0.1351   -0.3174
    0.1554   -0.4258
    0.1719   -0.5408
    0.1856   -0.6281
    0.2055   -0.6694
    0.2431   -0.5984
    0.2549   -0.6531
    0.2625   -0.7251
    0.2681   -0.7987
    0.2735   -0.8874
    0.2786   -0.9861
    0.2834   -1.0888
    0.2880   -1.1892
    0.2923   -1.2820
    0.2964   -1.3648
    0.3007   -1.4426
    0.3053   -1.5179
    0.3105    1.5451
    0.3158    1.4696
    0.3214    1.3901
    0.3268    1.3102
    0.3320    1.2263
    0.3368    1.1415
    0.3404    1.0722
    0.3442    0.9968
    0.3481    0.9169
    0.3523    0.8326
    0.3558    0.7636
    0.3592    0.6981
    0.3628    0.6316
    0.3667    0.5647
    0.3708    0.4984
    0.3753    0.4342
    0.3799    0.3734
    0.3848    0.3177
    0.3902    0.2649
    0.3964    0.2166
    0.4040    0.1738
    0.4141    0.1460
    0.4343    0.1998
    0.4477    0.3223
    0.4583    0.4512
    0.4678    0.5651
    0.4819    0.6638
    0.4907    0.6339
    0.4957    0.5842
    0.5001    0.5309
    0.5041    0.4753
    0.5086    0.4161
    0.5135    0.3524
    0.5183    0.2909
    0.5222    0.2447
    0.5263    0.1988
    0.5307    0.1540
    0.5358    0.1056
    0.5409    0.0623
    0.5465    0.0218
    0.5526   -0.0145
    0.5599   -0.0437
    0.5710   -0.0521
    0.5895    0.0487
    0.5985    0.1679
    0.6041    0.2848
    0.6086    0.4074
    0.6126    0.5267
    0.6166    0.6407
    0.6215    0.7490
    0.6283    0.7914
    0.6315    0.7651
    0.6344    0.7277
    0.6373    0.6844
    0.6398    0.6426
    0.6425    0.5982
    0.6454    0.5470
    0.6483    0.4988
    0.6509    0.4569
    0.6537    0.4128
    0.6566    0.3672
    0.6597    0.3205
    0.6629    0.2732
    0.6663    0.2261
    0.6698    0.1795
    0.6733    0.1341
    0.6769    0.0905
    0.6805    0.0489
    0.6841    0.0100
    0.6876   -0.0259
    0.6914   -0.0615
    0.6954   -0.0952
    0.6996   -0.1263
    0.7043   -0.1536
    0.7095   -0.1744
    0.7164   -0.1859
    0.7271   -0.1709
    0.7385   -0.1475
    0.7438   -0.1602
    0.7465   -0.1696
    0.7487   -0.1416
    0.7513   -0.0808
    0.7538   -0.0379
    0.7574   -0.0142
    0.7645   -0.0237
    0.7744   -0.0591
    0.7825   -0.0709
    0.7885   -0.0637
    0.7931   -0.0492
    0.7976   -0.0290
    0.8017   -0.0044
    0.8061    0.0264
    0.8096    0.0534
    0.8130    0.0824
    0.8164    0.1138
    0.8198    0.1474
    0.8232    0.1828
    0.8266    0.2193
    0.8299    0.2553
    0.8331    0.2923
    0.8362    0.3285
    0.8392    0.3633
    0.8420    0.3959
    0.8449    0.4292
    0.8478    0.4599
    0.8510    0.4887
    0.8545    0.5124
    0.8589    0.5226
    0.8664    0.4637
    0.8719    0.3470
    0.8760    0.2257
    0.8796    0.1143
    0.8837    0.0044
    0.8912   -0.1044
    0.9160   -0.1854
    0.9295   -0.1584
    0.9389   -0.1152
    0.9477   -0.0641
    0.9558   -0.0110
    0.9625    0.0354
    0.9692    0.0827
    0.9755    0.1291
    0.9824    0.1771
    0.9896    0.2229
    0.9985    0.2619

};

\addplot [mark=none,red,dashed,line width=1.5] table{

 0    1.5708
    0.0010   -1.5344
    0.0033   -1.4564
    0.0061   -1.3743
    0.0092   -1.2936
    0.0128   -1.2136
    0.0170   -1.1344
    0.0222   -1.0555
    0.0285   -0.9756
    0.0362   -0.8917
    0.0450   -0.8002
    0.0546   -0.6946
    0.0623   -0.6065
    0.0710   -0.5115
    0.0809   -0.4177
    0.0959   -0.3194
    0.1349   -0.3211
    0.1540   -0.4269
    0.1676   -0.5271
    0.1792   -0.6144
    0.1929   -0.6881
    0.2236   -0.6647
    0.2484   -0.6629
    0.2577   -0.7295
    0.2647   -0.8108
    0.2703   -0.8989
    0.2756   -1.0004
    0.2802   -1.0983
    0.2837   -1.1767
    0.2876   -1.2661
    0.2915   -1.3480
    0.2958   -1.4319
    0.3005   -1.5130
    0.3057    1.5488
    0.3112    1.4718
    0.3175    1.3895
    0.3234    1.3105
    0.3291    1.2291
    0.3346    1.1417
    0.3393    1.0597
    0.3433    0.9847
    0.3475    0.9041
    0.3520    0.8180
    0.3560    0.7403
    0.3597    0.6722
    0.3636    0.6026
    0.3679    0.5324
    0.3725    0.4633
    0.3773    0.3967
    0.3824    0.3346
    0.3878    0.2789
    0.3942    0.2239
    0.4015    0.1768
    0.4108    0.1408
    0.4309    0.1668
    0.4471    0.3023
    0.4565    0.4185
    0.4660    0.5385
    0.4788    0.6513
    0.4902    0.6329
    0.4956    0.5821
    0.5001    0.5282
    0.5048    0.4671
    0.5088    0.4137
    0.5131    0.3569
    0.5177    0.2985
    0.5226    0.2401
    0.5277    0.1837
    0.5329    0.1308
    0.5383    0.0827
    0.5436    0.0398
    0.5497   -0.0019
    0.5566   -0.0374
    0.5656   -0.0608
    0.5832   -0.0131
    0.5961    0.1200
    0.6031    0.2539
    0.6078    0.3793
    0.6118    0.5044
    0.6158    0.6251
    0.6204    0.7439
    0.6271    0.8124
    0.6309    0.7880
    0.6339    0.7498
    0.6366    0.7091
    0.6394    0.6625
    0.6420    0.6187
    0.6449    0.5691
    0.6481    0.5142
    0.6507    0.4704
    0.6534    0.4256
    0.6564    0.3782
    0.6595    0.3308
    0.6628    0.2826
    0.6662    0.2344
    0.6697    0.1868
    0.6733    0.1403
    0.6769    0.0954
    0.6806    0.0528
    0.6842    0.0128
    0.6878   -0.0241
    0.6917   -0.0617
    0.6957   -0.0962
    0.7002   -0.1290
    0.7050   -0.1570
    0.7106   -0.1785
    0.7178   -0.1874
    0.7284   -0.1722
    0.7387   -0.1645
    0.7437   -0.1982
    0.7477   -0.2195
    0.7511   -0.1651
    0.7542   -0.1123
    0.7578   -0.0854
    0.7647   -0.0883
    0.7751   -0.1175
    0.7827   -0.1223
    0.7886   -0.1114
    0.7938   -0.0914
    0.7984   -0.0661
    0.8029   -0.0351
    0.8066   -0.0059
    0.8105    0.0270
    0.8144    0.0632
    0.8183    0.1022
    0.8222    0.1434
    0.8261    0.1860
    0.8299    0.2292
    0.8335    0.2721
    0.8370    0.3137
    0.8403    0.3530
    0.8434    0.3891
    0.8466    0.4235
    0.8498    0.4550
    0.8534    0.4828
    0.8577    0.4993
    0.8654    0.4581
    0.8721    0.3267
    0.8765    0.2035
    0.8806    0.0808
    0.8859   -0.0383
    0.8997   -0.1539
    0.9253   -0.1693
    0.9358   -0.1296
    0.9449   -0.0802
    0.9531   -0.0289
    0.9615    0.0279
    0.9698    0.0868
    0.9777    0.1424
    0.9851    0.1939
    0.9931    0.2407
};

\end{axis}

\end{tikzpicture} \\
\begin{tikzpicture}

\begin{axis}[
  width = 0.45\textwidth,
  title = {$\nu = 1$},
  xmin = 0,
  xmax = 1, 
  ymin = -0.2,
  ymax = 0.2,
  xtick = {0,0.2,0.4,0.6,0.8,1},
  ytick = {-0.2,-0.1,0,0.1,0.2},
  ylabel = {$y/L$},
  xlabel = {$x/L$},
  label style={font=\normalsize},
  legend style={anchor=east,draw=none},
  legend entries={Resolution \# 1, Resolution \# 2, Resolution \# 3},
  legend pos = south east,
  ]

\addplot [mark=none,cyan,line width=1.5] table{
  0   -0.0000
    0.0004    0.0000
    0.0014    0.0001
    0.0039    0.0002
    0.0084    0.0005
    0.0144    0.0009
    0.0227    0.0013
    0.0346    0.0016
    0.0500    0.0014
    0.0619    0.0010
    0.0778    0.0004
    0.1033   -0.0005
    0.1455   -0.0008
    0.1773    0.0018
    0.2032    0.0086
    0.2313    0.0144
    0.2456    0.0156
    0.2579    0.0156
    0.2698    0.0148
    0.2824    0.0134
    0.2970    0.0116
    0.3138    0.0107
    0.3270    0.0127
    0.3343    0.0166
    0.3390    0.0201
    0.3434    0.0235
    0.3474    0.0260
    0.3516    0.0281
    0.3559    0.0296
    0.3596    0.0305
    0.3636    0.0310
    0.3678    0.0313
    0.3721    0.0317
    0.3766    0.0323
    0.3820    0.0331
    0.3876    0.0340
    0.3936    0.0348
    0.3997    0.0353
    0.4059    0.0356
    0.4126    0.0356
    0.4217    0.0352
    0.4322    0.0342
    0.4416    0.0325
    0.4523    0.0304
    0.4672    0.0298
    0.4806    0.0337
    0.4868    0.0373
    0.4920    0.0407
    0.4963    0.0435
    0.5012    0.0465
    0.5049    0.0485
    0.5093    0.0506
    0.5143    0.0525
    0.5196    0.0542
    0.5254    0.0555
    0.5315    0.0565
    0.5377    0.0570
    0.5440    0.0572
    0.5501    0.0570
    0.5567    0.0567
    0.5655    0.0560
    0.5740    0.0552
    0.5818    0.0542
    0.5891    0.0530
    0.5963    0.0520
    0.6059    0.0516
    0.6179    0.0540
    0.6265    0.0587
    0.6309    0.0618
    0.6352    0.0648
    0.6395    0.0677
    0.6433    0.0700
    0.6473    0.0721
    0.6517    0.0742
    0.6566    0.0760
    0.6619    0.0776
    0.6676    0.0789
    0.6735    0.0798
    0.6795    0.0803
    0.6856    0.0805
    0.6915    0.0803
    0.6979    0.0800
    0.7064    0.0792
    0.7145    0.0783
    0.7225    0.0771
    0.7296    0.0758
    0.7369    0.0745
    0.7463    0.0737
    0.7574    0.0753
    0.7667    0.0801
    0.7713    0.0835
    0.7751    0.0864
    0.7791    0.0893
    0.7827    0.0916
    0.7869    0.0940
    0.7911    0.0962
    0.7947    0.0977
    0.7987    0.0992
    0.8030    0.1005
    0.8075    0.1016
    0.8122    0.1024
    0.8171    0.1029
    0.8220    0.1032
    0.8268    0.1033
    0.8322    0.1031
    0.8382    0.1028
    0.8457    0.1021
    0.8537    0.1012
    0.8618    0.0999
    0.8688    0.0986
    0.8757    0.0970
    0.8839    0.0957
    0.8932    0.0956
    0.9024    0.0985
    0.9088    0.1029
    0.9122    0.1057
    0.9159    0.1088
    0.9192    0.1114
    0.9228    0.1139
    0.9260    0.1159
    0.9296    0.1180
    0.9337    0.1200
    0.9382    0.1218
    0.9419    0.1231
    0.9456    0.1241
    0.9495    0.1249
    0.9535    0.1256
    0.9576    0.1261
    0.9617    0.1264
    0.9664    0.1265
    0.9712    0.1264
    0.9765    0.1262
    0.9830    0.1257
    0.9908    0.1248
    0.9983    0.1237
};

\addplot [mark=none,green,line width=1.5] table{

  0   -0.0000
    0.0003    0.0000
    0.0016    0.0001
    0.0047    0.0003
    0.0092    0.0006
    0.0149    0.0010
    0.0224    0.0015
    0.0326    0.0017
    0.0456    0.0017
    0.0572    0.0014
    0.0700    0.0009
    0.0872    0.0001
    0.1354   -0.0006
    0.1923    0.0058
    0.2121    0.0109
    0.2282    0.0138
    0.2394    0.0150
    0.2494    0.0155
    0.2585    0.0154
    0.2676    0.0149
    0.2776    0.0140
    0.2896    0.0126
    0.3068    0.0112
    0.3189    0.0115
    0.3255    0.0123
    0.3322    0.0137
    0.3388    0.0155
    0.3455    0.0180
    0.3525    0.0212
    0.3577    0.0237
    0.3618    0.0255
    0.3664    0.0274
    0.3711    0.0292
    0.3748    0.0304
    0.3789    0.0314
    0.3831    0.0323
    0.3874    0.0330
    0.3918    0.0336
    0.3963    0.0339
    0.4007    0.0340
    0.4052    0.0340
    0.4103    0.0338
    0.4159    0.0335
    0.4226    0.0330
    0.4289    0.0324
    0.4347    0.0318
    0.4407    0.0311
    0.4462    0.0303
    0.4520    0.0295
    0.4587    0.0288
    0.4663    0.0289
    0.4748    0.0305
    0.4817    0.0335
    0.4864    0.0363
    0.4904    0.0389
    0.4939    0.0413
    0.4974    0.0436
    0.5014    0.0460
    0.5044    0.0476
    0.5078    0.0493
    0.5115    0.0510
    0.5155    0.0525
    0.5198    0.0538
    0.5243    0.0550
    0.5290    0.0559
    0.5339    0.0565
    0.5389    0.0569
    0.5438    0.0571
    0.5487    0.0570
    0.5537    0.0569
    0.5596    0.0566
    0.5657    0.0562
    0.5717    0.0557
    0.5777    0.0550
    0.5833    0.0542
    0.5890    0.0532
    0.5955    0.0522
    0.6029    0.0516
    0.6113    0.0523
    0.6191    0.0548
    0.6245    0.0577
    0.6286    0.0604
    0.6324    0.0631
    0.6357    0.0654
    0.6391    0.0676
    0.6430    0.0699
    0.6463    0.0717
    0.6496    0.0733
    0.6531    0.0748
    0.6570    0.0763
    0.6611    0.0775
    0.6654    0.0786
    0.6699    0.0794
    0.6746    0.0800
    0.6793    0.0804
    0.6840    0.0806
    0.6886    0.0806
    0.6934    0.0805
    0.6989    0.0803
    0.7048    0.0799
    0.7111    0.0793
    0.7170    0.0786
    0.7227    0.0778
    0.7283    0.0767
    0.7345    0.0755
    0.7416    0.0746
    0.7495    0.0745
    0.7574    0.0761
    0.7638    0.0790
    0.7681    0.0818
    0.7718    0.0844
    0.7749    0.0867
    0.7784    0.0891
    0.7815    0.0911
    0.7846    0.0930
    0.7880    0.0949
    0.7918    0.0967
    0.7959    0.0984
    0.7989    0.0995
    0.8022    0.1005
    0.8057    0.1014
    0.8093    0.1022
    0.8130    0.1028
    0.8168    0.1032
    0.8206    0.1035
    0.8245    0.1037
    0.8290    0.1037
    0.8334    0.1036
    0.8387    0.1033
    0.8446    0.1029
    0.8508    0.1024
    0.8567    0.1017
    0.8624    0.1009
    0.8679    0.0998
    0.8738    0.0985
    0.8805    0.0973
    0.8876    0.0967
    0.8948    0.0972
    0.9021    0.0997
    0.9066    0.1024
    0.9100    0.1049
    0.9132    0.1073
    0.9161    0.1096
    0.9194    0.1120
    0.9221    0.1139
    0.9250    0.1157
    0.9283    0.1176
    0.9318    0.1194
    0.9353    0.1209
    0.9383    0.1221
    0.9414    0.1232
    0.9448    0.1242
    0.9483    0.1250
    0.9518    0.1257
    0.9555    0.1262
    0.9592    0.1266
    0.9630    0.1269
    0.9667    0.1270
    0.9710    0.1270
    0.9754    0.1269
    0.9805    0.1266
    0.9862    0.1262
    0.9922    0.1256
    0.9979    0.1248
};

\addplot [mark=none,red,dashed,line width=1.5] table{

 0   -0.0000
    0.0006    0.0000
    0.0032    0.0002
    0.0085    0.0005
    0.0158    0.0010
    0.0265    0.0015
    0.0398    0.0016
    0.0554    0.0013
    0.0750    0.0005
    0.1104   -0.0005
    0.1910    0.0056
    0.2153    0.0116
    0.2316    0.0143
    0.2443    0.0154
    0.2562    0.0155
    0.2678    0.0149
    0.2802    0.0137
    0.2971    0.0120
    0.3138    0.0115
    0.3228    0.0122
    0.3303    0.0135
    0.3376    0.0154
    0.3456    0.0184
    0.3533    0.0220
    0.3592    0.0248
    0.3650    0.0273
    0.3707    0.0295
    0.3757    0.0310
    0.3811    0.0324
    0.3868    0.0334
    0.3926    0.0341
    0.3985    0.0344
    0.4044    0.0345
    0.4110    0.0343
    0.4186    0.0339
    0.4269    0.0333
    0.4345    0.0326
    0.4416    0.0318
    0.4488    0.0309
    0.4567    0.0300
    0.4663    0.0300
    0.4763    0.0323
    0.4840    0.0359
    0.4892    0.0391
    0.4937    0.0420
    0.4986    0.0450
    0.5031    0.0475
    0.5076    0.0498
    0.5125    0.0518
    0.5180    0.0537
    0.5239    0.0552
    0.5302    0.0564
    0.5367    0.0571
    0.5432    0.0574
    0.5496    0.0573
    0.5565    0.0571
    0.5645    0.0567
    0.5720    0.0561
    0.5793    0.0553
    0.5863    0.0543
    0.5939    0.0530
    0.6031    0.0524
    0.6129    0.0536
    0.6220    0.0571
    0.6276    0.0605
    0.6321    0.0635
    0.6367    0.0666
    0.6408    0.0691
    0.6454    0.0717
    0.6506    0.0741
    0.6547    0.0758
    0.6592    0.0773
    0.6641    0.0785
    0.6692    0.0795
    0.6744    0.0803
    0.6797    0.0807
    0.6851    0.0809
    0.6910    0.0809
    0.6976    0.0806
    0.7056    0.0801
    0.7129    0.0795
    0.7203    0.0785
    0.7272    0.0773
    0.7347    0.0760
    0.7435    0.0750
    0.7527    0.0756
    0.7617    0.0787
    0.7674    0.0820
    0.7719    0.0851
    0.7764    0.0882
    0.7802    0.0907
    0.7843    0.0932
    0.7890    0.0957
    0.7932    0.0976
    0.7974    0.0992
    0.8018    0.1006
    0.8066    0.1018
    0.8116    0.1027
    0.8167    0.1034
    0.8219    0.1037
    0.8271    0.1038
    0.8329    0.1037
    0.8394    0.1035
    0.8471    0.1030
    0.8542    0.1023
    0.8615    0.1013
    0.8683    0.1000
    0.8754    0.0985
    0.8836    0.0973
    0.8923    0.0974
    0.8998    0.0992
    0.9059    0.1024
    0.9102    0.1054
    0.9144    0.1086
    0.9180    0.1113
    0.9222    0.1141
    0.9257    0.1163
    0.9294    0.1183
    0.9335    0.1203
    0.9382    0.1222
    0.9432    0.1238
    0.9486    0.1252
    0.9543    0.1262
    0.9601    0.1268
    0.9660    0.1271
    0.9718    0.1271
    0.9776    0.1269
    0.9846    0.1264
    0.9916    0.1258
    0.9989    0.1249

};

\end{axis}

\end{tikzpicture} & \begin{tikzpicture}

\begin{axis}[
  width = 0.45\textwidth,
  title = {$\nu = 10$},
  xmin = 0,
  xmax = 1, 
  ymin = -0.2,
  ymax = 0.2,
  xtick = {0,0.2,0.4,0.6,0.8,1},
  ytick = {-0.2,-0.1,0,0.1,0.2},
  ylabel = {$y/L$},
  xlabel = {$x/L$},
  label style={font=\normalsize},
  legend style={anchor=east,draw=none},
  legend pos = south east,
  ]

\addplot [mark=none,cyan,line width=1.5] table{
 0   -0.0000
    0.0016    0.0002
    0.0049    0.0005
    0.0088    0.0007
    0.0136    0.0010
    0.0197    0.0012
    0.0278    0.0014
    0.0387    0.0013
    0.0508    0.0011
    0.0649    0.0009
    0.0757    0.0007
    0.0912    0.0001
    0.1247   -0.0011
    0.1558   -0.0015
    0.1730    0.0004
    0.1903    0.0053
    0.2120    0.0137
    0.2413    0.0190
    0.2601    0.0183
    0.2691    0.0175
    0.2765    0.0169
    0.2833    0.0165
    0.2897    0.0164
    0.2956    0.0164
    0.3021    0.0164
    0.3096    0.0166
    0.3180    0.0170
    0.3258    0.0177
    0.3332    0.0190
    0.3398    0.0206
    0.3450    0.0223
    0.3506    0.0244
    0.3561    0.0266
    0.3608    0.0285
    0.3661    0.0305
    0.3719    0.0324
    0.3782    0.0340
    0.3850    0.0354
    0.3933    0.0364
    0.4031    0.0368
    0.4183    0.0362
    0.4421    0.0337
    0.4574    0.0328
    0.4714    0.0341
    0.4878    0.0403
    0.4955    0.0446
    0.5017    0.0479
    0.5080    0.0508
    0.5152    0.0534
    0.5231    0.0555
    0.5316    0.0568
    0.5405    0.0572
    0.5491    0.0568
    0.5597    0.0554
    0.5773    0.0519
    0.5960    0.0481
    0.6050    0.0472
    0.6123    0.0474
    0.6189    0.0489
    0.6267    0.0532
    0.6339    0.0591
    0.6383    0.0627
    0.6428    0.0659
    0.6469    0.0685
    0.6517    0.0710
    0.6558    0.0728
    0.6601    0.0743
    0.6646    0.0755
    0.6694    0.0764
    0.6743    0.0770
    0.6794    0.0772
    0.6844    0.0770
    0.6902    0.0764
    0.6963    0.0754
    0.7030    0.0737
    0.7114    0.0711
    0.7243    0.0660
    0.7349    0.0608
    0.7413    0.0565
    0.7447    0.0526
    0.7464    0.0487
    0.7473    0.0447
    0.7481    0.0406
    0.7488    0.0364
    0.7497    0.0323
    0.7507    0.0282
    0.7520    0.0242
    0.7536    0.0202
    0.7559    0.0161
    0.7580    0.0128
    0.7598    0.0101
    0.7620    0.0076
    0.7640    0.0056
    0.7660    0.0040
    0.7682    0.0023
    0.7702    0.0010
    0.7726   -0.0002
    0.7747   -0.0010
    0.7768   -0.0017
    0.7791   -0.0022
    0.7815   -0.0026
    0.7839   -0.0028
    0.7862   -0.0029
    0.7884   -0.0031
    0.7908   -0.0035
    0.7934   -0.0040
    0.7962   -0.0047
    0.7993   -0.0056
    0.8030   -0.0065
    0.8074   -0.0075
    0.8121   -0.0084
    0.8171   -0.0089
    0.8224   -0.0092
    0.8281   -0.0090
    0.8345   -0.0084
    0.8418   -0.0071
    0.8517   -0.0045
    0.8672    0.0012
    0.8851    0.0081
    0.9061    0.0104
    0.9329    0.0081
    0.9561    0.0064
    0.9804    0.0075

};

\addplot [mark=none,green,line width=1.5] table{

 0   -0.0000
    0.0013    0.0002
    0.0043    0.0004
    0.0077    0.0007
    0.0117    0.0009
    0.0165    0.0012
    0.0225    0.0014
    0.0302    0.0015
    0.0399    0.0015
    0.0504    0.0014
    0.0610    0.0012
    0.0709    0.0010
    0.0829    0.0006
    0.1039   -0.0002
    0.1457   -0.0013
    0.1681    0.0003
    0.1856    0.0045
    0.2154    0.0152
    0.2495    0.0186
    0.2605    0.0179
    0.2681    0.0172
    0.2748    0.0165
    0.2811    0.0161
    0.2869    0.0158
    0.2923    0.0157
    0.2975    0.0157
    0.3030    0.0158
    0.3092    0.0159
    0.3158    0.0162
    0.3228    0.0167
    0.3294    0.0176
    0.3358    0.0190
    0.3404    0.0203
    0.3452    0.0219
    0.3502    0.0239
    0.3549    0.0259
    0.3592    0.0276
    0.3637    0.0294
    0.3687    0.0312
    0.3741    0.0328
    0.3799    0.0342
    0.3861    0.0353
    0.3933    0.0361
    0.4020    0.0364
    0.4141    0.0360
    0.4391    0.0333
    0.4532    0.0320
    0.4654    0.0323
    0.4819    0.0370
    0.4922    0.0425
    0.4982    0.0459
    0.5031    0.0484
    0.5086    0.0508
    0.5147    0.0530
    0.5202    0.0545
    0.5253    0.0556
    0.5307    0.0563
    0.5371    0.0567
    0.5435    0.0566
    0.5509    0.0559
    0.5599    0.0546
    0.5761    0.0509
    0.5940    0.0467
    0.6028    0.0453
    0.6086    0.0449
    0.6136    0.0451
    0.6189    0.0464
    0.6271    0.0519
    0.6315    0.0560
    0.6351    0.0593
    0.6387    0.0623
    0.6418    0.0647
    0.6454    0.0671
    0.6490    0.0691
    0.6523    0.0708
    0.6558    0.0723
    0.6597    0.0737
    0.6637    0.0748
    0.6680    0.0757
    0.6724    0.0763
    0.6769    0.0765
    0.6814    0.0765
    0.6859    0.0762
    0.6904    0.0757
    0.6954    0.0747
    0.7007    0.0734
    0.7068    0.0715
    0.7144    0.0687
    0.7271    0.0627
    0.7402    0.0533
    0.7453    0.0443
    0.7481    0.0350
    0.7513    0.0258
    0.7546    0.0194
    0.7603    0.0127
    0.7717    0.0047
    0.7825   -0.0000
    0.7894   -0.0022
    0.7958   -0.0039
    0.8007   -0.0048
    0.8061   -0.0055
    0.8104   -0.0057
    0.8147   -0.0058
    0.8190   -0.0056
    0.8232   -0.0051
    0.8275   -0.0044
    0.8315   -0.0034
    0.8355   -0.0022
    0.8392   -0.0008
    0.8426    0.0007
    0.8464    0.0026
    0.8502    0.0048
    0.8545    0.0075
    0.8605    0.0116
    0.8693    0.0167
    0.8751    0.0187
    0.8796    0.0197
    0.8851    0.0205
    0.8999    0.0197
    0.9275    0.0142
    0.9389    0.0121
    0.9499    0.0110
    0.9592    0.0108
    0.9675    0.0113
    0.9755    0.0124
    0.9841    0.0142
    0.9937    0.0169

};

\addplot [mark=none,red,dashed,line width=1.5] table{

0   -0.0000
    0.0027    0.0003
    0.0077    0.0007
    0.0139    0.0011
    0.0225    0.0014
    0.0350    0.0015
    0.0504    0.0014
    0.0658    0.0011
    0.0829    0.0006
    0.1310   -0.0011
    0.1681    0.0003
    0.1952    0.0080
    0.2495    0.0186
    0.2646    0.0175
    0.2748    0.0165
    0.2840    0.0159
    0.2923    0.0157
    0.3001    0.0158
    0.3092    0.0159
    0.3193    0.0164
    0.3294    0.0176
    0.3381    0.0196
    0.3452    0.0219
    0.3528    0.0250
    0.3592    0.0276
    0.3662    0.0303
    0.3741    0.0328
    0.3830    0.0348
    0.3933    0.0361
    0.4073    0.0364
    0.4391    0.0333
    0.4595    0.0319
    0.4819    0.0370
    0.4951    0.0442
    0.5031    0.0484
    0.5116    0.0520
    0.5202    0.0545
    0.5279    0.0560
    0.5371    0.0567
    0.5472    0.0563
    0.5599    0.0546
    0.5876    0.0481
    0.6028    0.0453
    0.6112    0.0449
    0.6189    0.0464
    0.6295    0.0541
    0.6351    0.0593
    0.6401    0.0634
    0.6454    0.0671
    0.6506    0.0700
    0.6558    0.0723
    0.6617    0.0743
    0.6680    0.0757
    0.6747    0.0764
    0.6814    0.0765
    0.6881    0.0760
    0.6954    0.0747
    0.7037    0.0725
    0.7144    0.0687
    0.7350    0.0579
    0.7453    0.0443
    0.7496    0.0304
    0.7546    0.0194
    0.7657    0.0083
    0.7825   -0.0000
    0.7925   -0.0031
    0.8007   -0.0048
    0.8083   -0.0056
    0.8147   -0.0058
    0.8211   -0.0054
    0.8275   -0.0044
    0.8335   -0.0028
    0.8392   -0.0008
    0.8445    0.0016
    0.8502    0.0048
    0.8571    0.0093
    0.8693    0.0167
    0.8773    0.0193
    0.8851    0.0205
    0.9184    0.0161
    0.9389    0.0121
    0.9551    0.0108
    0.9675    0.0113
    0.9797    0.0132
    0.9937    0.0169

};

\end{axis}

\end{tikzpicture} \\
\end{tabular}
\end{center}
\mcaption{ Convergence in trajectories and inclination angles of RBCs with different
viscosity contrasts. In order to show that the spatio-temporal resolution we
used is sufficient for the convergence in physics, we consider a DLD device with
a row-shift fraction of $\epsilon = 0.1667$ and RBCs with the capillary number
$C_a = 0.648$ and two different viscosity contrasts, $\nu = 1$ (on the left) and
$\nu = 10$. We performed simulations with three different
resolutions~\tabref{t:convResolutions}. On the top figures we show the
inclination angles of RBCs during their motion and on the bottom figures we show
the positions of the RBCs centroids. $x$ and $y$ positions are normalized by the
length of a period of the device. Among these resolutions we have used the
second one in our simulations because it is computationally more efficient than
the third one and delivers as accurate results as the third one does.}{f:resolConv}
\end{figure}

We conducted a convergence study to find sufficiently fine spatial and temporal
resolution to ensure that the underlying physics is not spoiled by the numerical
errors. We considered only one DLD device with the row-shift fraction $\epsilon = 0.1667$. The device consists of 8 rows and 
$\left \lceil 1.5 (1/0.1667) \right \rceil = 9$ columns of pillars. Since RBC shows two different transport modes in
a DLD device depending on its viscosity contrast and capillary number, we wanted
to run our numerical tests for both transport modes. That's why, we simulated
the flow of two RBCs with viscosity contrasts of $\nu = 1$ and $\nu = 10$ with
the same capillary number $C_a = 0.648$. It turns out that under these
conditions the RBC with $\nu = 1$ displaces and the one with $\nu = 10$ zig-
zags.
\begin{table}
\mcaption{List of spatio-temporal resolutions used in the convergence study. $N_{\Gamma}$
is the number of points on the exterior wall, $N_{\gamma}$ is the number of
points per pillar, $N$ is the number of points on RBC, $\rho_{\mathrm{AL}}$ is
the error tolerance for the adaptive time stepping and sets the temporal
resolution (see~\citet{kabacaoglu-biros-e17b} for the numerical scheme). The
results of the convergence study are on~\figref{f:resolConv}.}{t:convResolutions}
\centering
\begin{tabular}{c  c  c  c  c}

 Resolution \# & $N_{\Gamma}$ &  $N_{\gamma}$ & $N$ & $\rho_{\mathrm{AL}}$ \\ 
  \hline
  1 & 2784 & 48 & 48 & 5E-4 \\

  2 & 3712 & 64 & 64 & 1E-4 \\

  3 & 5568 & 96 & 96 & 5E-5 \\

\end{tabular} 
\end{table} 

We present the numerical scheme for the simulation of RBCs in DLD devices
in~\citet{kabacaoglu-biros-e17b}. In this scheme, the spatial resolution is set
by the number of points discretizing the boundary of the exterior wall,
$N_{\Gamma}$, the number of points discretizing the boundary of each pillar
$N_{\gamma}$ and the number of points discretizing the RBC's membrane $N$. The
temporal resolution is determined by the error tolerance $\rho_{\mathrm{AL}}$,
i.e. the smaller the tolerance is, the smaller the time step sizes are (so the
finer the temporal resolution is). We started with a certain spatial and
temporal resolution, then increased both together
(see~\tabref{t:convResolutions} for the resolutions). Here, $N_{\Gamma}$ and
$N_{\gamma}$ are chosen such that the maximum arc-length spacing is the same for
the exterior wall and the pillars. We present the inclination angles and
trajectories of the RBCs in~\figref{f:resolConv}. As expected, the RBC with 
$\nu = 1$ displaces (see the figure on the bottom left) and the one with 
$\nu = 10$ zig-zags (see the figure on the bottom right). This is captured with all the
resolutions in~\tabref{t:convResolutions}. However, the resolution \#1 leads the
displacing RBC to flip at $x/L \approx 0.38$. The zig-zagging RBC also flips at
$x/L \approx 0.78$ with this resolution. At the higher resolutions the RBCs do
not flip at these locations. Additionally, both the trajectories and the
inclination angles given by the last two resolutions
in~\tabref{t:convResolutions} agree well. Therefore, we chose the resolution \#
2, i.e. $N_{\Gamma} = 3712$, $N_{\gamma} = 64$, $N = 64$ and 
$\rho_{\mathrm{AL}} = 1\mathrm{E}-4$. As we changed the row-shift fraction, we scaled the number of
points for the exterior wall such that the maximum arc-length spacings of a
pillar and the exterior wall are the same.

\bibliographystyle{plainnat} 
\bibliography{separation17}
\biboptions{sort&compress}
\end{document}